\documentclass[apj,revtex4]{emulateapj_2013}

\usepackage{amsmath,amssymb,pdflscape,mathptmx,graphicx,natbib,rotating,rotfloat,multirow}
                                                        
\newcommand{\todo}{\ifmmode {\Huge \bullet} \else {\Huge$\bullet$}\fi}
\newcommand{\ltsim}{\raisebox{-.5ex}{$\;\stackrel{<}{\sim}\;$}}

\def\B{\textit{B}}
\def\V{\textit{V}}

\def\U{\textit{u$^\prime$}}
\def\G{\textit{g$^\prime$}}
\def\R{\textit{r$^\prime$}}
\def\I{\textit{i$^\prime$}}
\def\Z{\textit{z$^\prime$}}
\def\hst{\textit{HST}}
\def\FJ{F125W}
\def\FH{F160W}
\def\FW{F350LP}
\def\FX{F775W}
\def\FZ{F850LP}
\def\snana{{\scriptsize SNANA}} 
\def\sex{SE{\scriptsize XTRACTOR}}
\def\bpz{{\scriptsize BPZ}}
\def\salt{{\scriptsize SALT2}}
\def\stardust{{\scriptsize STARDUST}}
\def\snid{{\scriptsize SNID}}

\def\Nl{2.4}
\def\Nm{6.4}
\def\Nh{4.1}
\def\Nld{2.4}
\def\Nmd{7.0}
\def\Nhd{8.0}

\def\Rl{0.46^{+0.42,+0.10}_{-0.32,-0.13}}
\def\Rm{0.45^{+0.22,+0.13}_{-0.19,-0.06}}
\def\Rh{0.45^{+0.34,+0.05}_{-0.22,-0.09}}
\def\Rul{1.7}
\def\Rdl{$0.45^{+0.42,+0.10}_{-0.32,-0.13}$}
\def\Rdm{$0.42^{+0.21,+0.12}_{-0.18,-0.05}$}
\def\Rdh{$0.27^{+0.21,+0.03}_{-0.13,-0.06}$}
\def\Rdul{$0.6$}
\def\plind{$-1.00^{+0.06(0.09)}_{-0.06(0.10)}~{\rm (statistical)}~^{+0.12}_{-0.08}~{\rm (systematic)}$}

\newcommand\nar{{NewAR}}%
\newcommand\pasa{{PASA}}%

\newcommand{\TAU}{School of Physics and Astronomy, Tel-Aviv University, Tel-Aviv 69978, Israel}
\newcommand{\AMNH}{Department of Astrophysics, American Museum of Natural History, New York, NY 10024, USA}
\newcommand{\JHU}{Department of Physics and Astronomy, The Johns Hopkins University, Baltimore, MD 21218, USA}
\newcommand{\Rutgers}{Department of Physics and Astronomy, Rutgers, The State University of New Jersey, Piscataway, NJ 08854, USA}
\newcommand{\STSCI}{Space Telescope Science Institute, Baltimore, MD 21218, USA}

\newcommand{\Andalucia}{Instituto de Astrof\'{\i}sica de Andaluc\'{\i}a (CSIC), Granada, Spain}
\newcommand{\Barcelona}{Institut de Ciencies de l'Espai, (IEEC-CSIC), E-08193 Bellaterra (Barcelona), Spain}
\newcommand{\Trieste}{INAF - Osservatorio Astronomico di Trieste, Trieste, Italy}
\newcommand{\Napoli}{INAF - Osservatorio Astronomico di Capodimonte, Via Moiariello 16, I-80131 Napoli, Italy}

\newcommand{\Berkeley}{Department of Astronomy, University of California, Berkeley, CA 94720-3411, USA}
\newcommand{\UTA}{Department of Astronomy, University of Texas, Austin, TX 78712-0259, USA}
\newcommand{\NOAO}{National Optical Astronomy Observatory, 950 North Cherry Avenue, Tucson, AZ 85719, USA} 
\newcommand{\Bohr}{Dark Cosmology Centre, Niels Bohr Institute, University of Copenhagen, Juliane Maries Vej 30, DK-2100 Copenhagen, Denmark}
\newcommand{\ESO}{European Southern Observatory, Garching bei M\"{u}nchen, Germany}

\newcommand{\Dame}{Physics Department, University of Notre Dame, Notre Dame, IN 46556, USA}

\newcommand{\UCR}{Department of Physics and Astronomy, University of California, Riverside, CA 92521, USA}
\newcommand{\CCPP}{CCPP, New York University, 4 Washington Place, New York, NY 10003, USA}
\newcommand{\Bruno}{Excellence Cluster Universe, Technische Universitaet Muenchen, 85748 Garching, Germany}
\newcommand{\Goddard}{Astrophysics Science Division, NASA Goddard Space Flight Center, Mail Code 661, Greenbelt, MD 20771, USA}

\shorttitle{CLASH SN Ia Rates to $z=2.4$}
\shortauthors{Graur et al.}

\begin{document}

\title{Type-Ia Supernova Rates to Redshift $2.4$ from CLASH: the Cluster Lensing And Supernova survey with Hubble}

\author{O.~Graur\altaffilmark{1,2,3,4},
S.~A.~Rodney\altaffilmark{1,20},
D.~Maoz\altaffilmark{2},
A.~G.~Riess\altaffilmark{1,5},
S.~W.~Jha\altaffilmark{6},
M.~Postman\altaffilmark{5},
T.~Dahlen\altaffilmark{5},
T.~W.-S.~Holoien\altaffilmark{6},
C.~McCully\altaffilmark{6},
B.~Patel\altaffilmark{6},
L.-G.~Strolger\altaffilmark{5},
N.~Ben\'{i}tez\altaffilmark{7},
D.~Coe\altaffilmark{5},
S.~Jouvel\altaffilmark{8},
E.~Medezinski\altaffilmark{1},
A.~Molino\altaffilmark{7},
M.~Nonino\altaffilmark{9},
L.~Bradley\altaffilmark{5},
A.~Koekemoer\altaffilmark{5},
I.~Balestra\altaffilmark{9,10},
S.~B.~Cenko\altaffilmark{11,12},
K.~I.~Clubb\altaffilmark{12},
M.~E.~Dickinson\altaffilmark{13},
A.~V.~Filippenko\altaffilmark{12},
T.~F.~Frederiksen\altaffilmark{14},
P.~Garnavich\altaffilmark{15},
J.~Hjorth\altaffilmark{14},
D.~O.~Jones\altaffilmark{1},
B.~Leibundgut\altaffilmark{16,17},
T.~Matheson\altaffilmark{13},
B.~Mobasher\altaffilmark{18},
P.~Rosati\altaffilmark{16},
J.~M.~Silverman\altaffilmark{19,21},
V.~U\altaffilmark{18},
K.~Jedruszczuk\altaffilmark{3},
C.~Li\altaffilmark{3},
K.~Lin\altaffilmark{3},
M.~Mirmelstein\altaffilmark{2},
J.~Neustadt\altaffilmark{3},
A.~Ovadia\altaffilmark{3}, and
E.~H.~Rogers\altaffilmark{3}
}

\altaffiltext{1}{\JHU}
\altaffiltext{2}{\TAU}
\altaffiltext{3}{\AMNH}
\altaffiltext{4}{\CCPP}
\altaffiltext{5}{\STSCI}
\altaffiltext{6}{\Rutgers}
\altaffiltext{7}{\Andalucia}
\altaffiltext{8}{\Barcelona}
\altaffiltext{9}{\Trieste}
\altaffiltext{10}{\Napoli}
\altaffiltext{11}{\Goddard}
\altaffiltext{12}{\Berkeley}
\altaffiltext{13}{\NOAO}
\altaffiltext{14}{\Bohr}
\altaffiltext{15}{\Dame}
\altaffiltext{16}{\ESO}
\altaffiltext{17}{\Bruno}
\altaffiltext{18}{\UCR}
\altaffiltext{19}{\UTA}
\altaffiltext{20}{Hubble Fellow}
\altaffiltext{21}{NSF Astronomy and Astrophysics Postdoctoral Fellow}

\email{orgraur@jhu.edu}




\begin{abstract}
\noindent
We present the supernova (SN) sample and Type-Ia SN (SN~Ia) rates from the Cluster Lensing And Supernova survey with Hubble (CLASH). Using the Advanced Camera for Surveys and the Wide Field Camera 3 on the {\it Hubble Space Telescope} (\hst), we have imaged 25 galaxy-cluster fields and parallel fields of non-cluster galaxies. We report a sample of 27 SNe discovered in the parallel fields. Of these SNe, $\sim13$ are classified as SN~Ia candidates, including four SN~Ia candidates at redshifts $z>1.2$. We measure volumetric SN~Ia rates to redshift $1.8$ and add the first upper limit on the SN~Ia rate in the range $1.8<z<2.4$. The results are consistent with the rates measured by the \hst/GOODS and Subaru Deep Field SN surveys. We model these results together with previous measurements at $z<1$ from the literature. The best-fitting SN~Ia delay-time distribution (DTD; the distribution of times that elapse between a short burst of star formation and subsequent SN~Ia explosions) is a power law with an index of \plind, where the statistical uncertainty is a result of the 68\% and 95\% (in parentheses) statistical uncertainties reported for the various SN~Ia rates (from this work and from the literature), and the systematic uncertainty reflects the range of possible cosmic star-formation histories. We also test DTD models produced by an assortment of published binary population synthesis (BPS) simulations. The shapes of all BPS double-degenerate DTDs are consistent with the volumetric SN~Ia measurements, when the DTD models are scaled up by factors of 3--9. In contrast, all BPS single-degenerate DTDs are ruled out by the measurements at $>99$\% significance level.

\end{abstract}



\section{Introduction}
\label{sec:clash_intro}

Although Type-Ia supernovae (SNe~Ia) have been used to measure extragalactic distances and thus reveal the accelerating expansion of the universe \citep{1998AJ....116.1009R,1998ApJ...507...46S,1999ApJ...517..565P}, the nature of the stellar system that leads to these explosions remains unclear (see review by \citealt{Howell2011}).
The current consensus is that the progenitor is a carbon-oxygen white dwarf (CO WD) that accretes matter from a binary companion until the pressure or temperature somewhere in the WD become high enough to ignite the carbon and lead to a thermonuclear explosion of the WD \citep{Leibundgut2000}.
Different scenarios have been proposed to explain the nature of the binary companion and the process of mass accretion.
The leading scenarios are the single-degenerate scenario (SD; \citealt{Whelan1973}), in which the binary companion is either a main-sequence star, a subgiant just leaving the main sequence, a red giant, or a stripped ``He star,'' and the WD accretes mass from the secondary through Roche-lobe overflow or a stellar wind.
In the double-degenerate scenario (DD; \citealt{Iben1984,Webbink1984}), the companion is a second CO WD and the two WDs merge due to loss of energy and angular momentum to gravitational waves.

Each of these scenarios predicts a different form of the distribution of times that elapse between a short burst of star formation and any subsequent SN~Ia events, known as the delay-time distribution (DTD; see \citealt{2012NewAR..56..122W} and \citealt{2013FrPhy...8..116H} for recent reviews).
The DTD can be thought of as a transfer function connecting the star-formation history (SFH) of a specific stellar environment and that environment's SN~Ia rate.
Thus, by measuring the SN~Ia rate and comparing it to the SFH, one might reconstruct the DTD.
The SN~Ia DTD has been recovered using several techniques applied to different SN samples collected from different types of stellar environments (see review by \citealt{Maoz2012review}).
The emerging picture is that of a power-law DTD with an index of $\sim -1$, a form that arises naturally from the DD scenario, although combinations of DTDs from a DD channel and a SD channel cannot be ruled out. 
One method to recover the DTD, $\Psi(t)$, is to measure the SN~Ia rate, $R_{\rm Ia}(t)$, as a function of cosmic time $t$ in field galaxies, and compare them to the cosmic SFH, $S(t)$:
\begin{equation}
\label{eq:clash_intro_rates}
R_{\textrm{Ia}}(t) = \int_0^t S(t-\tau) \Psi(\tau)d\tau.
\end{equation}

Measurements of the volumetric SN~Ia rates (i.e., the SN~Ia rates per unit volume) in field galaxies agree out to $z \approx 1$.
\citet[G11]{Graur2011} provide a compilation of all SN~Ia rates measured up to 2011, and later measurements are presented by \citet{Krughoff2011}, \citet{perrett2012}, \citet{2012ApJ...745...31B}, \citet{melinder2012}, and \citet{GraurMaoz2013}.
Volumetric SN~Ia rate measurements were first extended to $z>1$ by \citet{dahlen2004}, with additional data analyzed by \citet*[D08]{dahlen2008}, using the {\it Hubble Space Telescope} (\hst) to survey the GOODS fields (\citealt{2004ApJ...600L.103G,Riess2004}).
The \hst/GOODS survey discovered 20 SNe~Ia at $1<z<1.4$ and 3 at $1.4<z<1.8$.
G11 conducted a SN survey in the Subaru Deep Field (SDF) using the 8.2-m Subaru telescope and discovered 27 SN~Ia candidates at $1<z<1.5$ and 10 at $1.5<z<2$.

The SN~Ia rate uncertainties at $z>1$, and especially at $z>1.5$, are dominated by small-number statistics.
The three $z>1.4$ \hst/GOODS SNe~Ia were discovered in host galaxies having a spectroscopic redshift (spec-$z$) and no active galactic nucleus (AGN) activity.
On the other hand, the classification of the larger SDF SN sample from G11 relies on photometric redshift (photo-$z$) measurements that might be systematically biased toward high redshifts (see their Section 4.2).
While G11 also used several methods to weed out interloping AGNs, 
there could still be some AGN contamination because each 
SN in the SDF sample was only observed on one epoch.
The \hst/GOODS sample, while smaller than the SDF sample, suffers from lower systematic uncertainties owing to the spectroscopic classification of the SN host galaxies and measurements of their redshifts, and a better sampling of the SN light curves.
Of the 10 $z>1.5$ SN host galaxies in the SDF sample, only one galaxy has so far had its redshift and lack of AGN activity confirmed spectroscopically \citep{Frederiksen2013sdf}.

Although the GOODS and SDF $z>1$ SN~Ia rates are consistent, their interpretation differs between the two groups.
Based on the GOODS data, \citet{dahlen2004,dahlen2008} argued that the SN~Ia rate declined at $z>0.8$.
Fitting this declining SN~Ia rate evolution, \citet{Strolger2004} and \citet*{Strolger2010} surmised that the DTD is confined to delay times of 3--4 Gyr.
In contrast, based on the SDF data, G11 found that the SN~Ia rate evolution does not decline at high redshifts, but rather levels off, as would be expected of a power-law DTD.

Two new SN surveys are attempting to resolve this conflict.
These surveys are components of two three-year \hst\ Multi-Cycle Treasury programs that use the Advanced Camera for Surveys (ACS) and the new Wide Field Camera 3 (WFC3).
Results from the Cosmic Assembly Near-infrared Deep Extragalactic Legacy Survey (CANDELS; \citealt{Grogin2011,Koekemoer2011}) will be reported by Rodney et al. (in preparation).
Here, we describe results from the Cluster Lensing And Supernova survey with Hubble (CLASH; \citealt{Postman2012}).
CLASH imaged 25 galaxy clusters in 16 broad-band filters from the near-ultraviolet (NUV) to the near-infrared (NIR) with the ACS and WFC3 cameras working in parallel mode.
While one camera was pointed at the galaxy cluster, the other one was used to observe a parallel field far enough from the galaxy cluster so as not to be significantly affected by strong lensing.

In this work, we report a sample of 27 SNe discovered in the parallel fields of the 25 CLASH galaxy clusters.
In Section \ref{sec:clash_obs}, we describe the CLASH observations and our imaging and spectroscopic follow-up program.
We report our SN sample in Section \ref{sec:clash_sne}, where we also conduct detection-efficiency simulations and classify the SNe.
Using our SN~Ia sample, we measure SN~Ia rates out to $z \approx 2.4$ in Section \ref{sec:clash_rates} and use them to test different forms of the DTD in Section \ref{sec:clash_dtd}.
Finally, we summarize our results in Section \ref{sec:clash_conclude}.
Throughout this work, we assume a $\Lambda$-cold-dark-matter cosmological model with parameters $\Omega_{\Lambda} = 0.7$, $\Omega_m = 0.3$, and $H_0 = 70$~km~s$^{-1}$~Mpc$^{-1}$.
Unless noted otherwise, all magnitudes are on the Vega system.

We designate our SN candidates according to the cluster and year in which they were discovered and the first three letters of the nickname given to them for internal tracking purposes.
For example, CLI11Had is a CLASH (CL) SN that was discovered in one of the parallel fields around the ninth (or Ith) cluster, MACS0717.5+3745, in 2011, and was nicknamed ``Hadrian.''
For the sake of brevity, we will henceforth refer to our SN candidates simply as SNe.


\section{Observations}
\label{sec:clash_obs}

\subsection{Imaging}
\label{subsec:clash_imaging}

The CLASH observation strategy is described in detail by \citet{Postman2012}.
During Cycles 18--20, CLASH observed 25 galaxy clusters in the redshift range 0.187--0.890.
The central region of each galaxy cluster was imaged with 16 broad-band filters from the NUV to the NIR using the ACS and WFC3 cameras on \hst.
In ACS, we used the Wide Field Channel (WFC), with a field of view of $202'' \times 202''$ and a pixel scale of $0.05''$ pixel$^{-1}$. 
WFC3 includes two detectors: an infrared channel (WFC3-IR) with a field of view of $123'' \times 136''$ and a scale of $0.13''$ pixel$^{-1}$; and an ultraviolet--visible channel (WFC3-UVIS) with a field of view of $162'' \times 162''$ and a scale of $0.04''$ pixel$^{-1}$.

The orientation of \hst\ and the cadence between succeeding visits to the galaxy cluster (``prime'') field were chosen so that two ACS and two WFC3 parallel fields would each be observed on four separate occasions, with a median cadence of 18 days.
Each visit to a WFC3 parallel field consisted of one orbit comprising two \FH\ filter exposures and one exposure in filters \FJ\ and \FW\ each (filter+system central wavelengths $\lambda_0 \approx$ 15,369, 12,486, and 5846~\AA{}, respectively).
Visits to the ACS parallel fields consisted of one orbit when the prime field was imaged with either the ACS or WFC3-IR cameras and two orbits when the prime field was imaged with the WFC3-UVIS camera.
During single-orbit visits, the parallel ACS orbit comprised four \FZ\ filter exposures and one \FX\ filter exposure (filter+system central wavelengths $\lambda_0 \approx 9445$ and $7764$~\AA{}, respectively).
When the ACS parallel fields were imaged over two orbits, they consisted of six \FZ\ and two \FX\ exposures.
These filters, the reddest in each camera, were chosen to detect high-redshift SNe.
The \FW\ band was added to the WFC3 observations for additional color information to aid in the classification of any SNe discovered in those fields.
The \hst\ angular resolution in our search bands is $\sim 0.10''$ and $\sim 0.17''$ in \FZ\ and \FH, respectively, slightly larger than the pixel scales of their respective cameras.
Table~\ref{table:clash_exptime} lists the typical exposure times and $5\sigma$ limiting magnitudes reached in each of these filters.

The limiting magnitude in each filter was calculated using the method outlined in \citet{Kashikawa2004}: we conducted aperture photometry on hundreds of blank regions in the image, fit a Gaussian to the negative side of the resultant histogram (as the positive tail could be contaminated by light from the sources in the image), and treated the standard deviation of the fit as an estimate of the average noise in the image.
We used circular apertures with radii of 4, 3, and 5 pixels, which correspond to $0.20''$, $0.27''$, and $0.20''$ in the pixel scale to which we drizzle the images taken with ACS, WFC3-IR, and WFC3-UVIS: 0.05, 0.09, and 0.04 arcsec pixel$^{-1}$, respectively.

An additional 52 \hst\ orbits were allocated for follow-up imaging or slitless spectroscopy of targets of opportunity, such as high-redshift SNe~Ia.
This cache of orbits was added to the 150 similar \hst\ orbits allocated to the CANDELS program, for a sum of 202 follow-up orbits for the combined CLASH+CANDELS SN survey (PI: A. Riess). 

Our \hst\ reduction and image-subtraction pipeline is described in detail by Rodney et al. (in preparation).
Briefly, the raw \hst\ images were first calibrated using the {\scriptsize STSDAS}\footnote{http://www.stsci.edu/institute/software\_hardware/pyraf/stsdas} calibration tools.
The calibrations include bias correction, dark subtraction, and flat fielding. 
In the case of WFC3-IR images, ``up-the-ramp'' fitting was used to remove cosmic ray (CR) events.
Charge-transfer efficiency losses in the ACS images were corrected using the algorithm of \citet{2010PASP..122.1035A}.
Next, the subexposures in each filter were combined using {\scriptsize\sc MultiDrizzle} (\citealt{2002hstc.conf..337K}).
This stage also removed the geometrical distortion of the \hst\ focal plane.
For each filter, we created ``template'' images comprised of all previous observations in the same filter.
This means that some SN light may be included in the template images, which we take into account in Section \ref{sec:clash_rates}.
Finally, we subtracted the template images from the drizzled ``target'' images to produce the difference images that were then searched for SNe.
Owing to the stable point-spread-function (PSF) of \hst, we did not need to degrade the PSF of either the target or template images to match the PSF of the images, as done in ground-based SN surveys (e.g., G11).

\begin{table}
 \center
 \caption{Typical Exposure Times for CLASH Parallel Fields}
 \begin{tabular}{llccc}
  \hline
  \hline
  Camera & Filter & Exposures & Total Time  & $5\sigma$ Limiting Magnitude \\
         &        &           & (s)         & (Vega mag)             \\
  \hline
  \multirow{2}{*}{ACS-WFC} & \multirow{2}{*}{F850LP} & 4 & 1500 & 25.0 \\
                           &                         & 6 & 3600 & 25.4 \\
  \multirow{2}{*}{ACS-WFC} & \multirow{2}{*}{F775W}  & 1 & 400  & 25.7 \\
	                   &	                      & 2 & 700  & 25.9 \\
  WFC3-IR                  &                 F160W   & 2 & 1200 & 25.4 \\
  WFC3-IR                  &                 F125W   & 1 & 700  & 25.7 \\
  WFC3-UVIS                &                 F350LP  & 1 & 650  & 27.5 \\
  \hline

 \end{tabular}
 \label{table:clash_exptime}
\end{table}

Most of the CLASH galaxy clusters were observed in the \B, \V, $R_c$, $I_c$, and \Z\ bands with Suprime-Cam (\citealt{2002PASJ...54..833M}) at the prime focus of the 8.2-m Subaru telescope, for the purpose of measuring the amount of shear induced on background galaxies by weak lensing from the galaxy cluster, and for deriving the photometric redshifts of the galaxies in the CLASH parallel fields.
For an example of such observations of the CLASH galaxy cluster MACS1206, and a description of their reduction, see \citet{2012ApJ...755...56U}.

\subsection{Spectroscopy}
\label{subsec:clash_spectro}

\begin{figure}
 \includegraphics[width=0.5\textwidth]{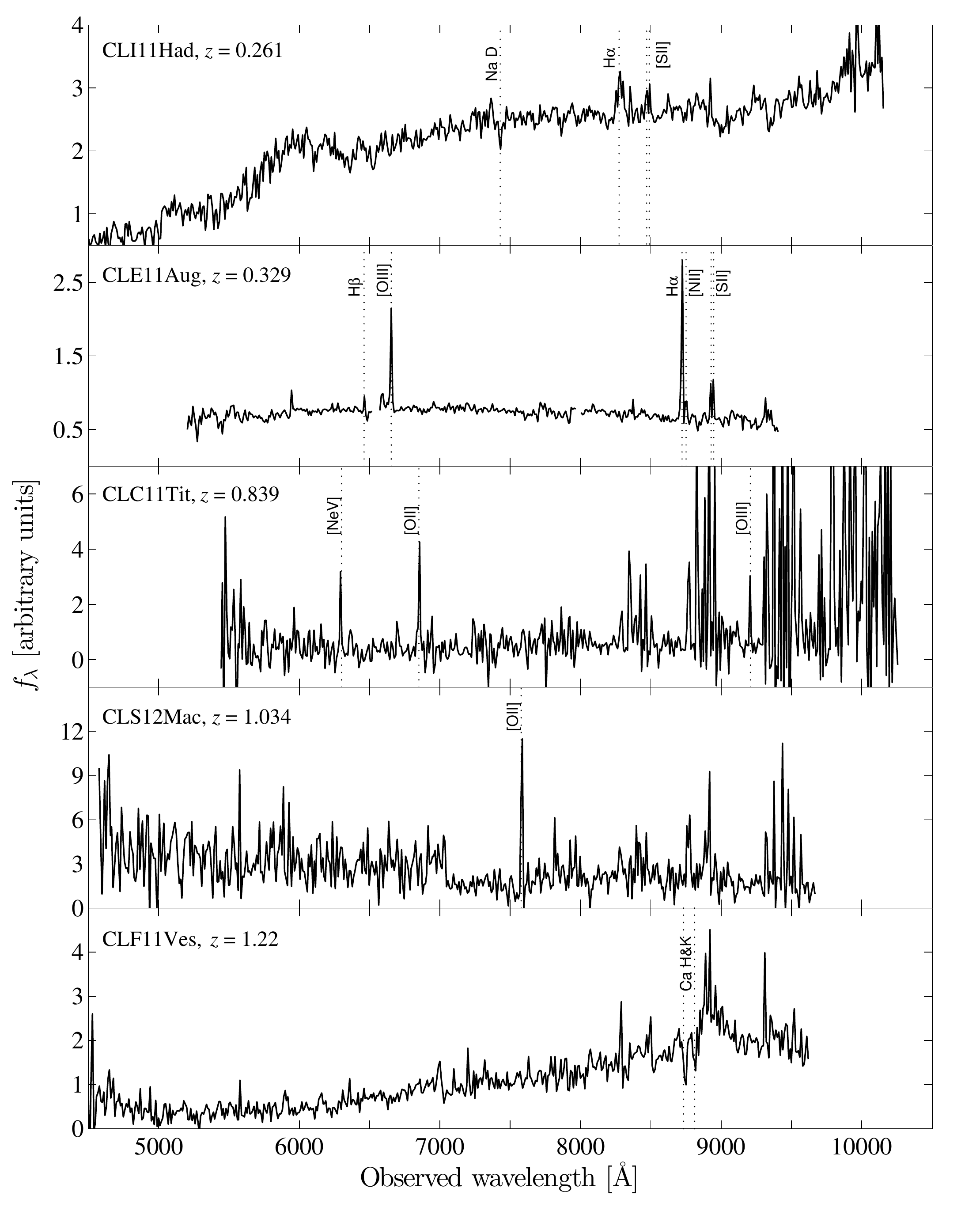}
 \caption[Examples of CLASH SN host-galaxy spectra]{Examples of SN host-galaxy spectra. From top to bottom, we present spectra of the SN host galaxies CLI11Had, at $z=0.261$, taken with Keck+LRIS; CLE11Aug, at $z=0.329$, taken with GeminiS+GMOS (the gaps in this spectrum are the result of physical gaps between GMOS chips); CLC11Tit, at $z=0.839$, obtained with Keck+LRIS; CLS12Mac, at $z=1.034$, obtained with Keck+DEIMOS; and CLF11Ves, at $z=1.22$, obtained with Keck+DEIMOS. All spectra have been binned into 10~\AA{}-wide bins. All flux units and have been arbitrarily scaled.}
 \label{fig:clash_spectra}
\end{figure}

\begin{figure}
 \includegraphics[width=0.5\textwidth]{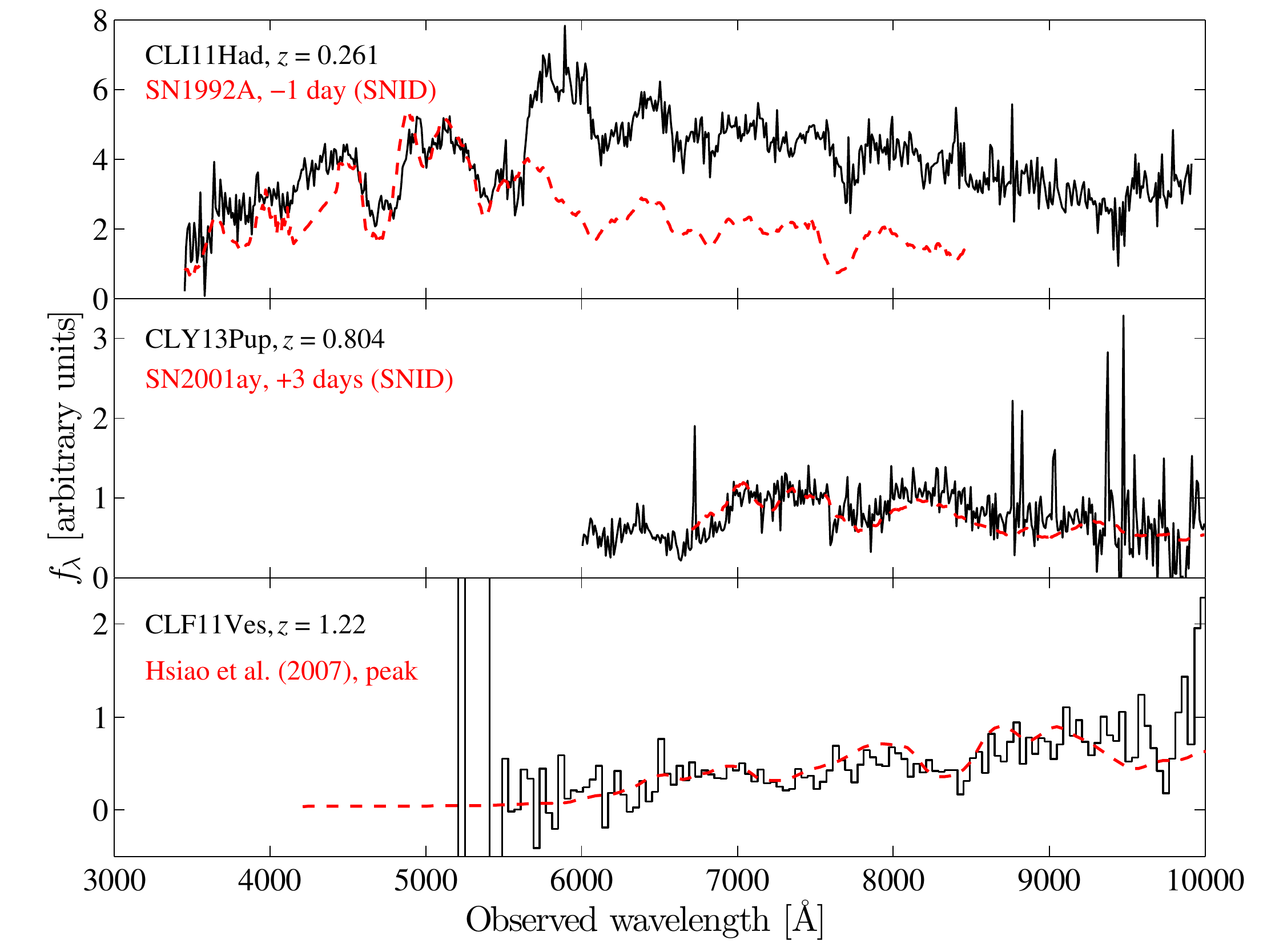}
 \caption{SN spectra. In black solid curves, we show the spectra of the SNe CLI11Had (top), at $z=0.261$, taken with Keck+LRIS; CLY13Pup (center), at $z=0.804$, taken with VLT+FORS2; and CLF11Ves (bottom), at $z=1.22$, taken with the \hst\ G800L ACS grism spectrograph. Overlaid on the spectra as red dashed curves are examples of SNe~Ia from the literature shifted to the same redshift and scaled so as to fit the data. The SN designation, along with that of the example SN~Ia, are noted on the top left of each panel. The top and center spectra have been binned into 10~\AA{}-wide bins, while the bottom spectrum has been binned into 80~\AA{}-wide bins. All flux units are arbitrary. Beyond 5500~\AA{}, the spectrum of CLI11Had may be dominated by host-galaxy light, as no correction for host-galaxy light was performed during reductions.}
 \label{fig:clash_sn_spectra}
\end{figure}

The host galaxies of all SN candidates, and in several cases the SNe themselves, were followed up with spectroscopic observations from several ground-based observatories, as detailed below, or with \hst\ slitless spectroscopy, using the G800L ACS grism spectrograph.
The ground-based observatories and instruments used for this work were the Low Resolution Imaging Spectrometer (LRIS; \citealt{Oke95}) and the DEep Imaging Multi-Object Spectrograph (DEIMOS; \citealt{Faber03}) on the Keck I and II 10-m telescopes, respectively; the Gemini Multi-Object Spectrograph (GMOS; \citealt{2003SPIE.4841.1645H}) on the Gemini North and South telescopes (GeminiN and GeminiS, respectively); the Multi-Object Double Spectrograph (MODS; \citealt{2010SPIE.7735E...9P}) on the Large Binocular Telescope (LBT); and the FOcal Reducer and low dispersion Spectrograph (FORS; \citealt{1998Msngr..94....1A}), the VIsible MultiObject Spectrograph (VIMOS; \citealt{2003SPIE.4841.1670L}), and the X-shooter spectrograph (\citealt{2011A&A...536A.105V}) on the Very Large Telescope (VLT).
Table~\ref{table:clash_SNe} details which instruments were used to obtain spectra of each SN host galaxy.
Several examples of SN host-galaxy spectra are shown in Figure~\ref{fig:clash_spectra}.
The spectra of the SNe CLF11Ves, CLI11Had, and CLY13Pup are shown in Figure~\ref{fig:clash_sn_spectra}.

\section{Supernova Sample}
\label{sec:clash_sne}

In our survey, SNe can be discovered if they either brighten or decline between one search epoch and the next.
By a ``brightening'' SN we do not mean that the SN is necessarily caught on the rising part of its light curve, but rather any case in which the discovery flux is
higher than the template flux.
As a result of the cadence of our survey, it is easier to discover SNe either on the rise or near peak, as in these cases the template image will contain no SN flux.
In contrast, SNe caught while on the decline will invariably have some flux in all our images, thus reducing the flux in the difference image and consequently their probability of detection.
We have discovered a total of 20 brightening and 7 declining SNe in the parallel fields of the 25 CLASH clusters.
Of these, 18 were discovered in the ACS and 9 in the WFC3 fields.
Nineteen (or 70\%) of the SN host galaxies have spectroscopic redshifts, as detailed below in Section \ref{subsec:clash_redshift}.
We classify half of this sample as SNe~Ia, four of which are at $z>1.2$.
Our SN sample is summarized in Table~\ref{table:clash_SNe} and displayed in Figure~\ref{fig:clash_sne}.

We have discovered 12 additional SN candidates in the prime fields.
However, as the effects of gravitational lensing must be taken into account to properly classify any SNe discovered behind the galaxy clusters, we leave their treatment to a future paper.
The complete photometry of all 39 SNe in our sample will appear in a future paper by Graur et al. (in preparation).

\subsection{Candidate Selection}
\label{subsec:clash_cands}

The \FZ- and \FH-band subtraction images were simultaneously searched by eye and scanned with the source-identifying software SE{\scriptsize XTRAC\-TOR} (\citealt{sextractor}) to identify variable objects.
We set \sex\ to locate all objects that had at least four connected pixels with flux $3\sigma$ above the local background level in both the regular subtraction image and in its negative, the latter in order to search for declining SNe.
To increase the detection efficiency in the \FH\ band, the \FH- and \FJ-band subtraction images were searched by eye simultaneously (by toggling between them), as some SNe may appear brighter in the \FJ\ band (see, for example, the light-curve fit of CLI11Had in Figure~\ref{fig:clash_classify_1}).
The \FZ- and \FX-band subtraction images were not similarly toggled due to the high CR contamination in the \FX-band subtraction image.

To be regarded as SN candidates, the variable objects had to pass the following criteria.
\begin{enumerate}
 \item All objects with suspect residual shapes, such as the subtraction residuals of bright galaxy cores or objects with non-PSF shapes, were rejected.
 \item The \FZ- and \FH-band images were comprised of several subexposures (four or six in the \FZ\ band and two in the \FH\ band).
 We used these subexposures to create separate subtraction images, which were then compared to the main subtraction image. 
 The object had to appear in all of the subtraction images to be considered a likely candidate.
 \item To be considered a declining SN, the object had to have a negative flux in the subtraction image and appear in both the target and template images.
 Objects that only appeared in the template images were discarded as either CRs or noise spikes.
 \item Objects with suspect residual shapes in the \FX- or \FJ-band subtraction images were flagged for inspection in the next search epoch.
 As a result of the cadence of the survey, any SNe detected in one of the first two search epochs would be visible in the other search epochs as well, so if the object under consideration did not reappear in a later epoch, it was rejected.
 No candidates were rejected if they did not appear in the \FX- or \FJ-band images.
\end{enumerate}

\subsection{Detection-efficiency Simulations}
\label{subsec:clash_eff}

In our survey, SNe can be missed because of many factors.
Generally, the fainter the SN, the less likely it is to be detected above the background.
On average, \FX-band images suffer from a background (composed of zodiacal light, earthshine, and airglow) level twice as high as \FZ-band images \citep*{2012acs..rept....4S}.
The main sources of background for WFC3-IR observations are earthshine and zodiacal light, with the latter being the dominant source.
Both sources contribute less background at longer wavelengths \citep*{2002wfc..rept...12G}, but as our \FH\ exposures are roughly twice as long as the \FJ\ exposures, they both display roughly the same number of background counts \citep{2012wfci.book.....D}.
There are other factors that affect the discovery probability of a SN, such as its proximity to the core of its host galaxy (SNe that explode close to the cores of their host galaxies are harder to discover due to the noise from the higher background and the residuals from imperfect image subtractions).

To test the effect of these and other factors on our detection efficiency, we planted $\sim 1000$ fake point sources in the raw images at the start of our reduction pipeline.
The fake SNe were planted in random locations around galaxies chosen from \sex\ catalogs of the images following a Gaussian distribution centered on the center of the galaxy, as measured by \sex, with a standard deviation of $\sigma=2R_{50}$, where $R_{50}$ is the radius that contains 50\% of the galaxy light.
This distribution assured that the fake SNe approximately followed the light of the galaxy and that a large number were planted in galaxy cores (e.g., \citealt{2008MNRAS.388L..74F}).
Near the center of a bright galaxy, a SN could be obscured due to the increased Poisson noise and residual subtraction artifacts from small inter-epoch registration errors.    
This is evaluated in Rodney et al. (in preparation), where we find this effect to be negligible, with less than 2\% of galaxies above $z=0.2$ exhibiting core residuals that could obscure a SN.
The magnitudes of the fake SNe were drawn from flat distributions in \FZ\ and \FH\ in the range 22--28 mag.
To simulate the appearance of real SNe~Ia, the \FX\ and \FJ\ magnitudes, respectively, were randomly chosen from a SN~Ia simulation done with the SuperNova ANAlysis (\snana\footnote{http://sdssdp62.fnal.gov/sdsssn/SNANA-PUBLIC/}; \citealt{Kessler2009snana}) software package, which was constructed to reflect a realistic spread of SN~Ia colors in the redshift range $z=0$--3, with host-galaxy extinction according to values chosen from an exponential of the form $P(A_V)=e^{(-A_V/\tau_V)}$, with $\tau_V=0.7$, chosen to approximate the host-galaxy extinction model of \citet{2005MNRAS.362..671R}.
This was done to ensure that the fake SNe resembled the colors of their real counterparts as close as possible, and was of importance mainly in the WFC3 fields, where the SN searchers toggled between the \FH\ and \FJ\ difference images. 
The PSF was simulated using {\scriptsize\sc Tiny Tim}\footnote{http://www.stsci.edu/hst/observatory/focus/TinyTim} (\citealt*{2011SPIE.8127E..16K}).

Figure~\ref{fig:clash_effm} shows our detection efficiency, as a function of the brightness of the fake SNe, in the \FZ\ and \FH\ bands.
The uncertainties of the measurements represent the 68\% binomial confidence intervals.
We follow \citet{2010ApJ...718..876S} and fit the efficiency measurements with the function
\begin{equation}\label{eq:eff}
\eta(m;m_{1/2},s_1,s_2) = \left\{ \begin{array}{ll}
      \left(1 + e^{\frac{m-m_{1/2}}{s_1}}\right)^{-1}, & \mbox{$m\le m_{1/2}$}\\
      \left(1 + e^{\frac{m-m_{1/2}}{s_2}}\right)^{-1}, & \mbox{$m>m_{1/2}$},\\
      \end{array} \right.
\end{equation}
where $m$ is the magnitude in the \FZ\ band; $m_{1/2}$ is the magnitude at which the efficiency drops to 50\%; and $s_1$ and $s_2$ determine the range over which the efficiency drops from 100\% to 50\%, and from 50\% to 0, respectively.
Our detection efficiency drops to 50\% at 25.2 and 25.0 mag in the \FZ\ and \FH\ bands, respectively.

\begin{figure*}
\begin{minipage}{\textwidth}
\center
\includegraphics[width=\textwidth]{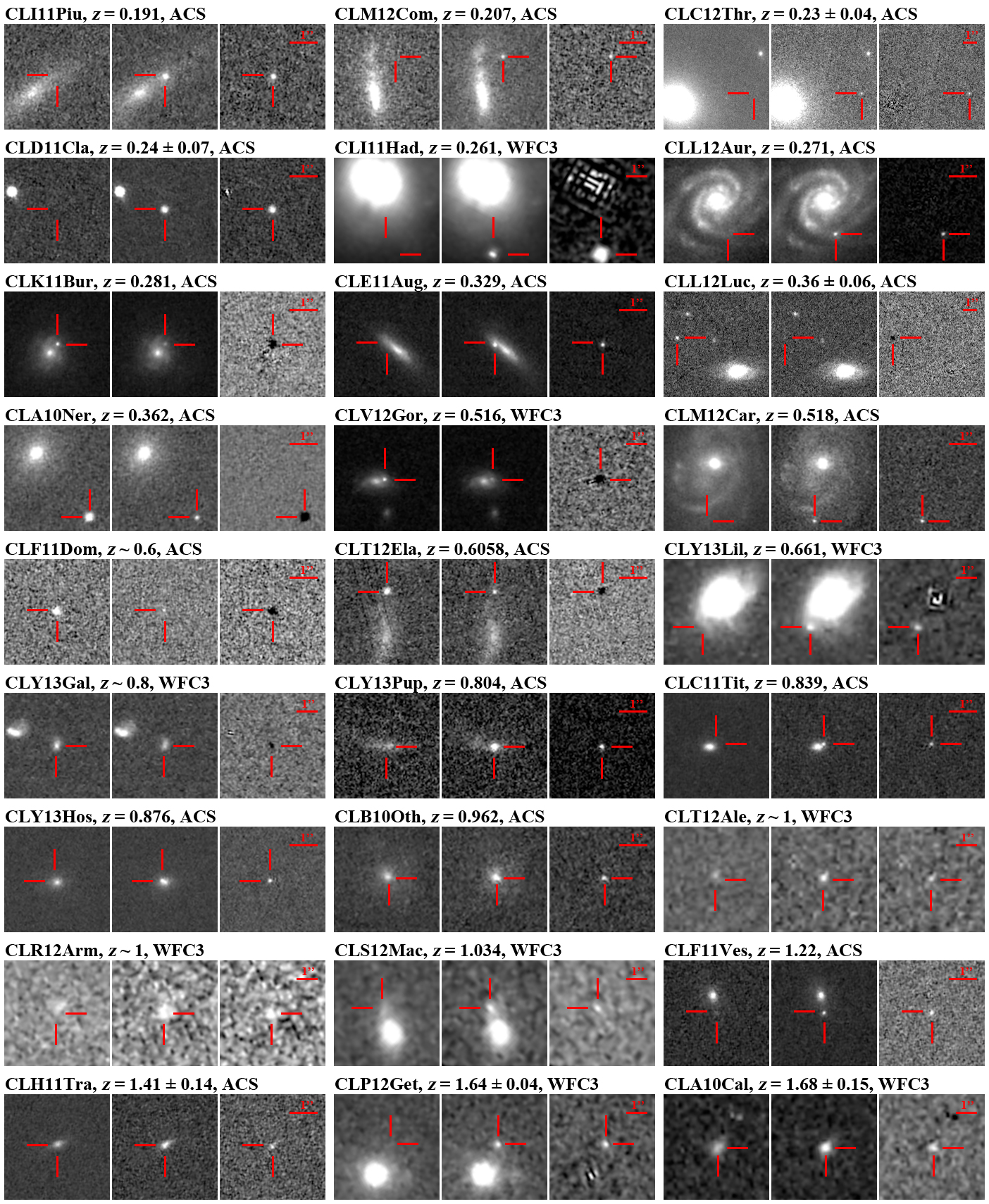}
\caption[SNe discovered in the CLASH parallel fields]{SNe discovered in the parallel fields of the CLASH clusters. North is up and east is left. In the triplet of tiles for each event, the left-hand tiles show the SN host galaxies without any SN light, whereas the center tiles display the SN host galaxy as imaged when the SN was first discovered. For the declining SNe CLK11Bur, CLL12Luc, CLA10Ner, CLV12Gor, CLF11Dom, CLT12Ela, and CLY13Gal, the left-hand and center tiles show the SN and host galaxy on the first and last visits to the field, respectively. The right-hand tiles show the subtraction in the \FZ\ or \FH\ bands for SNe discovered in the ACS or WFC3 parallel fields, respectively. The stretch of the images and the location of the SN differ from panel to panel in order to highlight host-galaxy properties. The header of each panel gives the designation of the SN along with its redshift and camera. Spectroscopic redshifts (cases with no uncertainties in $z$ noted) are given to three significant digits. Photometric redshifts are shown with their uncertainty; in cases where the photometric redshift is not well constrained, we note the approximate peak of the probability density function.}
\label{fig:clash_sne}
\end{minipage}
\end{figure*}

\begin{sidewaystable*}
\vspace{9.0cm}
\centering
\caption{SNe Discovered in the Parallel Fields of the 25 CLASH Galaxy Clusters.}
\begin{tabular}{llccccccl}
\hline
\hline
\multicolumn{1}{c}{ID} & \multicolumn{1}{c}{Nickname} & $\alpha$ (h~m~s) & $\delta$ ($^{\circ}~'~''$) & $P({\rm Ia})_{wp}$ & $P({\rm Ia})_{np}$ & Photo-$z$ & Spec-$z$ & \multicolumn{1}{c}{Spec-$z$ Source} \\
\multicolumn{1}{c}{(1)} & \multicolumn{1}{c}{(2)} & (3) & (4) & (5) & (6) & (7) & (8) & \multicolumn{1}{c}{(9)} \\
\hline

CLA10Cal  & Caligula           & $02:48:25.740$ & $-03:33:08.37$ & $0.95^{+0.03}_{-0.14}$ & $0.98^{+0.00}_{-0.00}$ & $1.68^{+0.15}_{-0.15}$ & $\cdots$ & (VLT+X-shooter)  \\
CLA10Ner  & Nero               & $02:47:40.180$ & $-03:32:53.29$ & $0.76^{+0.09}_{-0.26}$ & $0.82^{+0.01}_{-0.08}$ & $0.32^{+0.08}_{-0.01}$ & 0.362    & Keck+DEIMOS      \\
CLB11Oth  & Otho               & $11:49:56.745$ & $+22:18:42.87$ & $0.79^{+0.06}_{-0.14}$ & $0.89^{+0.00}_{-0.01}$ & $0.86^{+0.15}_{-0.20}$ & 0.962    & VLT+FORS2        \\
CLC11Tit  & Titus              & $17:22:52.985$ & $+32:07:25.74$ & $0.89^{+0.03}_{-0.08}$ & $0.94^{+0.01}_{-0.01}$ & $0.70^{+0.05}_{-0.05}$ & 0.839    & Keck+LRIS        \\
CLD11Cla  & Claudius           & $12:06:08.868$ & $-08:42:54.73$ & $0.13^{+0.24}_{-0.13}$ & $0.19^{+0.18}_{-0.19}$ & $0.24^{+0.07}_{-0.04}$ & $\cdots$ & (VLT+FORS2; Keck+LRIS; ACS+G800L) \\
CLE11Aug  & Augustus           & $13:47:12.802$ & $-11:42:28.97$ & $0.03^{+0.06}_{-0.03}$ & $0.06^{+0.05}_{-0.06}$ & $0.34^{+0.10}_{-0.03}$ & 0.329    & GeminiS+GMOS     \\
CLF11Ves  & Vespasian          & $21:29:42.612$ & $-07:41:48.08$ & $0.97^{+0.01}_{-0.04}$ & $0.98^{+0.00}_{-0.01}$ & $1.31^{+0.05}_{-0.10}$ & 1.22     & ACS+G800L; Keck+DEIMOS  \\
CLF11Dom  & Domitian           & $21:29:53.224$ & $-07:40:56.95$ & $0.63^{+0.16}_{-0.40}$ & $0.71^{+0.06}_{-0.24}$ & $0.71^{+0.14}_{-0.09}$ & $\cdots$ & $\cdots$         \\
CLH11Tra  & Trajan             & $21:39:46.036$ & $-23:38:34.71$ & $1.00^{+0.00}_{-0.00}$ & $1.00^{+0.00}_{-0.00}$ & $1.41^{+0.14}_{-0.11}$ & $\cdots$ & (VLT+VIMOS; Keck+DEIMOS) \\
CLI11Had  & Hadrian            & $07:17:20.115$ & $+37:49:54.54$ & $1.00^{+0.00}_{-0.00}$ & $1.00^{+0.00}_{-0.00}$ & $0.21^{+0.03}_{-0.03}$ & 0.261    & Keck+LRIS        \\
CLK11Bur  & Burgundy           & $06:49:13.878$ & $+70:13:17.91$ & $0.00^{+0.00}_{-0.00}$ & $0.00^{+0.00}_{-0.00}$ & $0.27^{+0.05}_{-0.05}$ & 0.281    & Keck+LRIS        \\
CLI11Piu  & Antoninus Pius     & $07:17:59.057$ & $+37:40:51.13$ & $0.00^{+0.00}_{-0.00}$ & $0.00^{+0.00}_{-0.00}$ & $0.18^{+0.10}_{-0.02}$ & 0.191    & GeminiS+GMOS; LBT+MODS \\
CLL12Aur  & Marcus Aurelius    & $11:16:07.758$ & $+01:23:44.60$ & $0.00^{+0.00}_{-0.00}$ & $0.00^{+0.00}_{-0.00}$ & $0.29^{+0.10}_{-0.04}$ & 0.271    & Keck+DEIMOS      \\
CLL12Luc  & Lucius Verus       & $11:15:57.141$ & $+01:23:19.62$ & $0.00^{+0.00}_{-0.00}$ & $0.00^{+0.00}_{-0.00}$ & $0.33^{+0.06}_{-0.03}$ & 0.36     & Keck+DEIMOS      \\
CLM12Com  & Commodus           & $08:00:52.385$ & $+36:07:37.31$ & $0.00^{+0.00}_{-0.00}$ & $0.00^{+0.00}_{-0.00}$ & $0.19^{+0.05}_{-0.07}$ & 0.207    & Keck+DEIMOS      \\
CLM12Car  & Cardinal           & $08:00:42.802$ & $+36:07:13.81$ & $0.22^{+0.27}_{-0.22}$ & $0.34^{+0.16}_{-0.34}$ & $0.47^{+0.08}_{-0.05}$ & 0.518    & Keck+DEIMOS      \\
CLP12Get  & Geta               & $21:29:23.918$ & $+00:08:24.77$ & $1.00^{+0.00}_{-0.00}$ & $1.00^{+0.00}_{-0.00}$ & $1.64^{+0.04}_{-0.04}$ & 1.64     & VLT+X-shooter; (VLT+VIMOS) \\
CLR12Arm  & Neill Armstrong    & $01:31:30.231$ & $-13:34:39.13$ & $0.01^{+0.02}_{-0.01}$ & $0.01^{+0.01}_{-0.01}$ & $1.12^{+0.63}_{-0.36}$ & $\cdots$ & (VLT+VIMOS; Keck+LRIS)     \\
CLS12Mac  & Macrinus           & $04:15:47.671$ & $-24:00:23.77$ & $0.09^{+0.16}_{-0.09}$ & $0.23^{+0.13}_{-0.22}$ & $0.95^{+0.08}_{-0.03}$ & 1.034    & Keck+DEIMOS      \\
CLT12Ela  & Elagabalus         & $22:48:09.132$ & $-44:35:16.63$ & $0.61^{+0.06}_{-0.10}$ & $0.70^{+0.07}_{-0.06}$ & $0.53^{+0.03}_{-0.04}$ & 0.6058   & Keck+LRIS        \\
CLT12Ale  & Alexander Severus  & $22:49:20.961$ & $-44:32:47.94$ & $0.12^{+0.18}_{-0.11}$ & $0.13^{+0.10}_{-0.10}$ & $1.0^{+2.0}_{-0.5}$    & $\cdots$ & (VLT+FORS2)      \\
CLC12Thr  & Thrax              & $17:22:44.529$ & $+32:03:35.96$ & $0.00^{+0.00}_{-0.00}$ & $0.00^{+0.00}_{-0.00}$ & $0.23^{+0.03}_{-0.04}$ & $\cdots$ & $\cdots$         \\
CLV12Gor  & Gordian            & $11:57:10.258$ & $+33:42:19.60$ & $0.00^{+0.00}_{-0.00}$ & $0.00^{+0.00}_{-0.00}$ & $0.37^{+0.18}_{-0.14}$ & 0.516    & Keck+DEIMOS      \\ 
CLY13Lil  & Lilla              & $13:11:03.822$ & $-03:15:50.66$ & $1.00^{+0.00}_{-0.00}$ & $1.00^{+0.00}_{-0.00}$ & $0.675^{+0.001}_{-0.006}$ & 0.661 & GeminiN+GMOS     \\
CLY13Pup  & Pupienus           & $13:10:43.130$ & $-03:06:13.08$ & $1.00^{+0.00}_{-0.00}$ & $1.00^{+0.00}_{-0.00}$ & $0.76^{+0.04}_{-0.04}$ & 0.804    & VLT+FORS2        \\
CLY13Hos  & Hostilian          & $13:10:46.685$ & $-03:05:22.98$ & $0.92^{+0.00}_{-0.08}$ & $0.95^{+0.00}_{-0.03}$ & $0.90^{+0.05}_{-0.08}$ & 0.876    & Keck+DEIMOS; VLT+FORS2 \\
CLY13Gal  & Trebonianus Gallus & $13:11:02.495$ & $-03:16:54.91$ & $0.00^{+0.00}_{-0.00}$ & $0.00^{+0.00}_{-0.00}$ & $0.8$--$1.8^a$         & $\cdots$ & (VLT+FORS2)      \\ 

\hline
\multicolumn{9}{l}{(1) -- SN identification.} \\
\multicolumn{9}{l}{(2) -- SN nickname.} \\
\multicolumn{9}{l}{(3)--(4) -- Right ascensions and declinations (J2000).} \\
\multicolumn{9}{l}{(5)--(6) -- \stardust\ probability of classification as a SN~Ia with and without assuming a prior on the fraction of each SN subtype as a function of redshift.} \\
\multicolumn{9}{l}{(7) -- Photometric redshift of SN host galaxy, as derived with \bpz.} \\
\multicolumn{9}{l}{(8)--(9) --Spectroscopic redshift of SN host galaxy, where available, and its source. Parentheses indicate unsuccessful attempts or as yet unreduced data.} \\
\multicolumn{9}{l}{$^a$ The 68\% confidence region derived from the photo-$z$ PDF of CLY13Gal, which has two prominent peaks at $z \approx 0.8$ and 1.6.}

\end{tabular}
\label{table:clash_SNe}
\end{sidewaystable*}

\begin{figure*}
 \includegraphics[width=\textwidth]{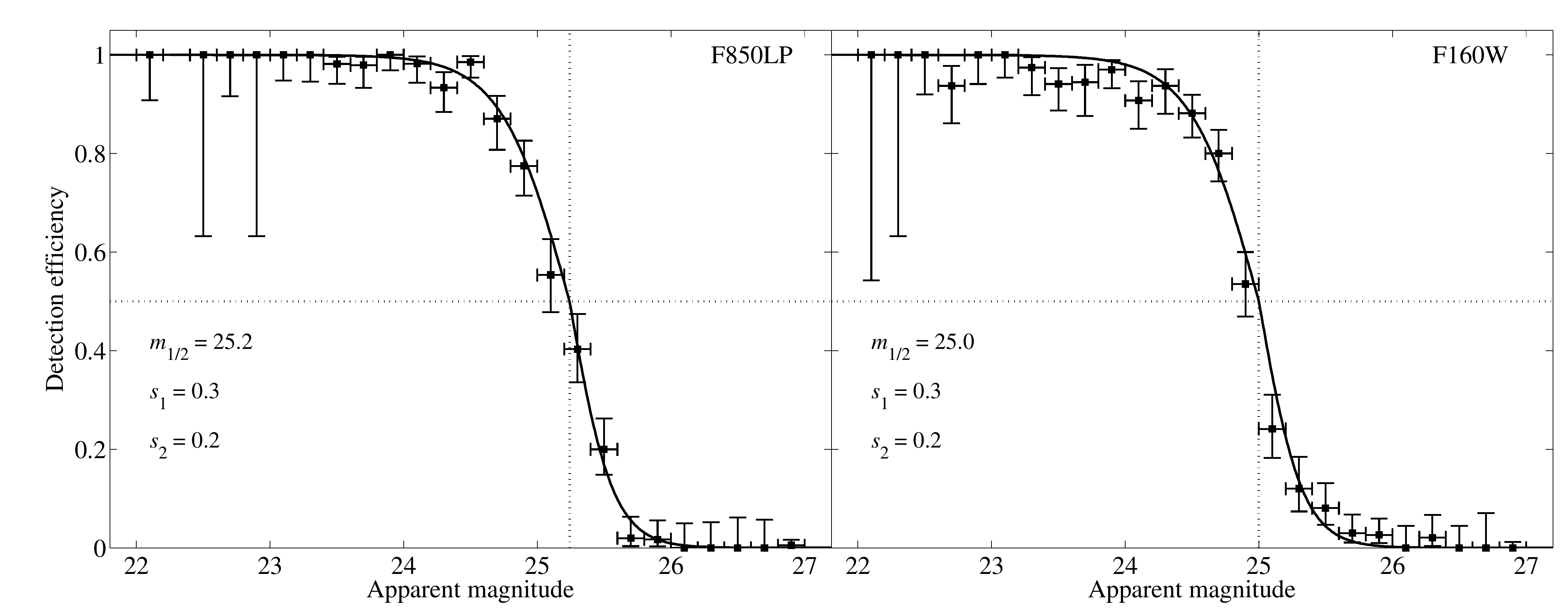}
  \caption[SN detection efficiency vs. magnitude in CLASH]{SN detection efficiency vs. magnitude in the \FZ\ (left) and \FH\ (right) bands. The uncertainties of the measurements are the 68\% binomial confidence intervals. The dotted lines mark where the best-fit efficiency curves drop to 50\%, at 25.2 and 25.0 mag in the \FZ\ and \FH\ bands, respectively.}
  \label{fig:clash_effm}
\end{figure*}

\subsection{Host-galaxy Redshifts}
\label{subsec:clash_redshift}

Our classification method, as with most SN classification techniques, relies on a good knowledge of the redshift of either the SN or its host galaxy.
As part of our survey strategy, we have endeavored to obtain spec-$z$ measurements of the host galaxies of all the SNe in our sample, mostly with ground-based observatories, as described above.
Some of the SN host galaxies suspected of being at $z>1.2$ (CLD11Cla and CLF11Ves) were also followed up with \hst\ slitless spectroscopy using the ACS G800L grism.
At this time, we have acquired and reduced the spectra of 19 of the 27 SN host galaxies in our sample.

\begin{figure}
 \center
 \includegraphics[width=0.5\textwidth]{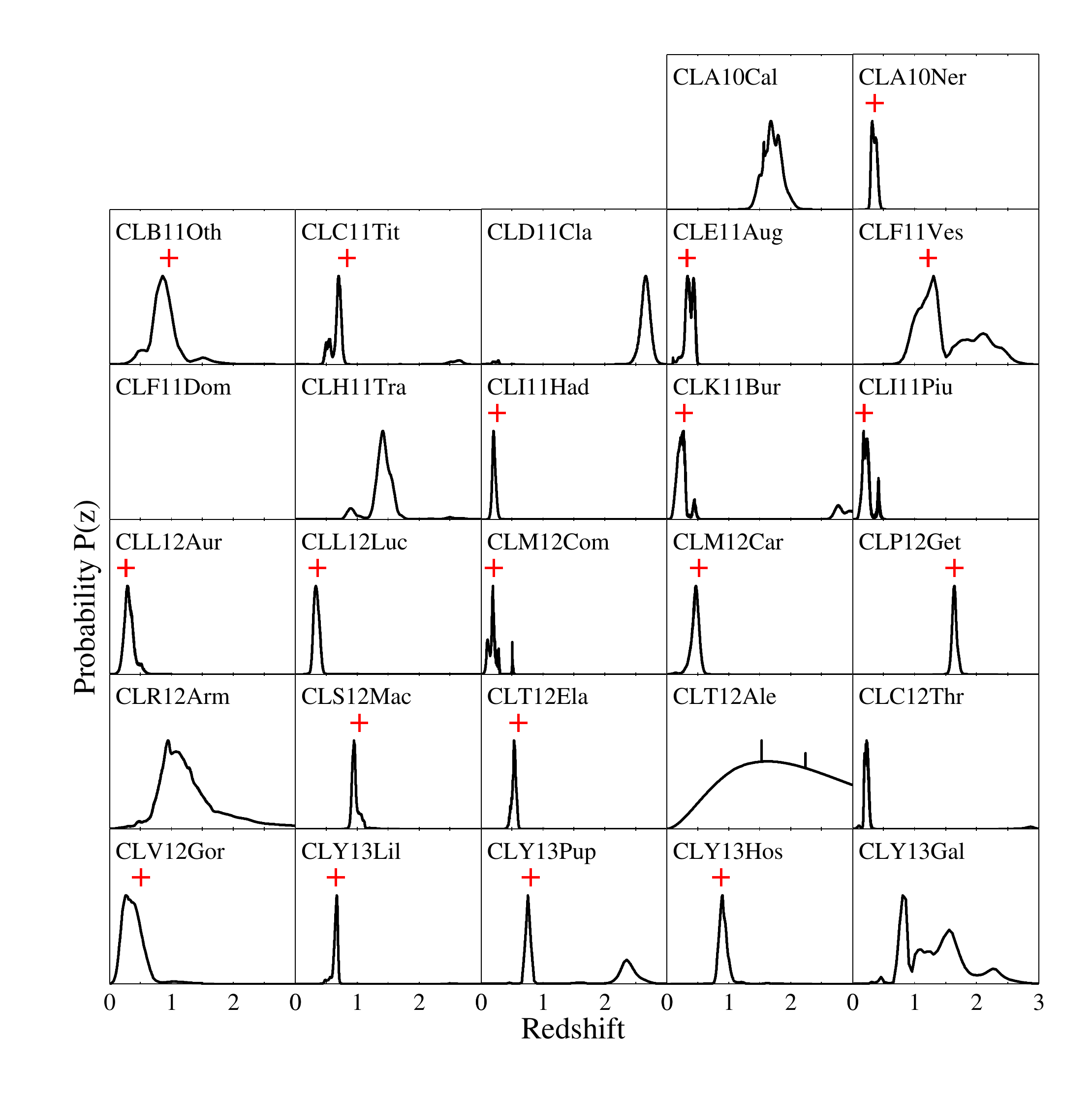}
 \caption[Photometric-redshift fits of the CLASH SN host galaxies]{\bpz\ $z$-PDFs of the SN host galaxies. The $z$-PDFs are the solid curves and the spectroscopic redshifts, where available, are marked by red crosses. The designation of each SN appears in the upper left corner of each panel. All $z$-PDFs have been normalized so that $\int P(z)dz=1$. CLF11Dom is a ``hostless'' SN (see Section \ref{subsubsec:clash_individual}).}
 \label{fig:clash_photoz}
\end{figure}

For the remaining eight SN host galaxies, we rely on photo-$z$ measurements. 
A complete description of our photo-$z$ technique appears in \citet{2013arXiv1308.0063J} and Molino et al. (in preparation).
Here, we give only a brief description of this technique.
The spec-$z$ and photo-$z$ values of the SNe in our sample are listed in Table~\ref{table:clash_SNe} and shown in Figure~\ref{fig:clash_photoz}.

We estimated the redshift and spectral type of all SN host galaxies with photometry obtained from deep Subaru images (in the $B$, $V$, $R_c$, $I_c$, and $z^{\prime}$ bands) and the Bayesian Photometric Redshift code (\bpz; \citealt{Benitez2000}). 
For the host galaxies of SNe that were discovered in the WFC3 parallel fields, we also added galaxy photometry in the \FJ\ and \FH\ bands.
Some host galaxies were previously imaged by the Sloan Digital Sky Survey (SDSS; \citealt{2000AJ....120.1579Y}), allowing us to include photometry in the \U, \G, \R, \I, and \Z\ bands as well.

For each galaxy, \bpz\ calculates a likelihood, $L(z,T)$, as a function of redshift, $z$, and spectral type, $T$, comparing the observed colors of the galaxies with the template library, and then multiplies it by an empirical prior, $p(z,T|m)$, which depends on the galaxy magnitude in some reference band, $m$, yielding a full probability, $p(z,T)$, for each galaxy. 
The new version of \bpz\ (Ben\'{i}tez, in preparation) includes a new template library comprising six spectral energy distribution templates originally from PEGASE (\citealt{1997A&A...326..950F}) and four early-type templates from \citet{2007ApJ...663...81P}.
The PEGASE templates were recalibrated using the \citet{2008ApJ...682..985W} FIREWORKS photometry and spectroscopic redshifts to optimize its performance together with the new early-type galaxy templates. 
In total, we use five templates for early-type galaxies, two for intermediate galaxies, and four for starburst galaxies. 
The prior was calibrated using the GOODS-MUSIC (\citealt{2006A&A...449..951G}), Hubble Ultra Deep Field (\citealt{2006AJ....132..926C}), and COSMOS (\citealt{2009ApJ...690.1236I}) samples. 
Despite its compactness and simplicity, this library produces results which are comparable or slightly better than the best available photo-$z$ methods (see the method comparison in \citealt{2010A&A...523A..31H}), which often include template libraries that are many times larger. 
As a result of the high-quality \hst\ imaging used for its calibration and using an approach similar to that developed by \citet{2006AJ....132..926C}, the representation of typical galaxy colors provided by this library can be used to calibrate ground-based photometry to an accuracy of $\sim 2\%$ (Molino et al., in preparation).

\subsection{Supernova Classification}
\label{subsec:clash_colormag}

We classify our SNe into SNe~Ia, SNe~Ib/c, or SNe~II by fitting light curves to their multi-band photometry using a Bayesian approach first introduced by \citet{2013ApJ...768..166J}, where it was used to classify the CANDELS SN UDS10Wil.
The full description of this classification technique, named the Supernova Taxonomy And Redshift Determination Using \snana\ Templates (\stardust), along with a detailed examination of any systematic biases it might introduce, will appear in a future paper by Rodney et al. (in preparation).
Briefly, for each SN we calculate the probability that it is a SN~Ia, $P({\rm Ia})$, by comparing the observed fluxes (in all available bands and epochs) to light-curve models generated using the \snana\ simulation package.
We classify a SN as a SN~Ia if $P({\rm Ia})\ge0.5$.
However, as detailed below in Section \ref{sec:clash_rates}, when deriving the SN~Ia rates, we sum the $P({\rm Ia})$ values of all the SNe in our sample

The apparent magnitudes of each SN are measured with aperture photometry on the subtraction images of each epoch using the {\scriptsize IRAF} routine {\it apphot} and the same apertures described in Section\ref{subsec:clash_imaging}.
The zero-point magnitudes and aperture corrections for ACS filters are taken from \citet{2005PASP..117.1049S}.
For WFC3-IR and WFC3-UVIS, we use the zero-point magnitudes calculated for a $0.4''$ aperture, as of 2012 March 6, by the Space Telescope Science Institute.\footnote{http://www.stsci.edu/hst/wfc3/phot\_zp\_lbn}
The aperture corrections for the WFC3 filters were calculated by measuring the photometry of several bright stars using different apertures and adopting the correction for a $0.4''$ aperture.

For the SN~Ia simulations, we use the \citet{Guy2007} \salt\footnote{http://supernovae.in2p3.fr/$\sim$guy/salt/} model, with nuisance parameters for the redshift, stretch ($x1$), color ($c$), and time of peak brightness.
The core-collapse (CC) SNe are generated from the \snana\ library of 43 CC~SN templates, taken from the SN samples of the SDSS \citep{2008AJ....135..338F,2008AJ....135..348S,2010ApJ...708..661D}, Supernova Legacy Survey (SNLS; \citealt{2006A&A...447...31A}), and Carnegie Supernova Project \citep{2006PASP..118....2H,2009ApJ...696..713S,2012IAUS..279..361M}.
Each of these CC~SN models also has parameters for redshift, host extinction ($A_V$), date of peak brightness, and peak luminosity.

The remainder of our technique is fundamentally similar to other Bayesian light-curve classifiers (e.g., \citealt{2007ApJ...659..530K}; \citealt*{poznanski2007snabc}; \citealt{2009ApJ...707.1064R,2011ApJ...738..162S}): we compute the likelihood that a given model matches the observable data, multiply it by priors for the model parameters, then marginalize over all models to derive the final posterior classification probability.

The simulated SNe are reddened and dimmed according to the \citet*{1989ApJ...345..245C} reddening law and one of three host-galaxy dust extinction models: ``low,'' ``medium,'' and ``high'' dust models.
For the simulated SNe~Ia, the ``low'' dust model is the \citet{2012ApJ...745...31B} skewed Gaussian fit to the \citet{2006A&A...447...31A} SNLS, while the ``mid'' and ``high'' dust models are taken from \citet{2009ApJS..185...32K} and \citet{neill2006}, respectively.
For CC~SNe, we use models composed of a half Gaussian centered at $A_V=0$ mag and an exponential of the form $e^{-(A_V/\tau_V)}$.
These models have three parameters: the standard distribution of the Gaussian, $\sigma_{A_V}$; the characteristic $A_V$ value, $\tau_V$; and the ratio between the Gaussian and power-law components at $A_V=0$, $A_0$.
For the ``low,'' ``mid,'' and ``high'' dust models, the values of these parameters are $\sigma_{A_V}=0.15$, 0.6, and 0.5 mag; $\tau_V=0.5$, 1.7, and 2.8; and $A_0=1$, 4, and 3 mag.

The peak magnitudes of each SN subtype are chosen according to their observed luminosity functions (LFs), which are detailed in Table~\ref{table:clash_LFs}.
The \citet{li2011LF} LFs, derived from a local sample of SNe observed by the Lick Observatory Supernova Search (LOSS; \citealt{2011MNRAS.412.1419L,li2011rates,li2011LF}), were not originally corrected for host-galaxy extinction.
Here, we adopt ``dust-free'' LFs for SNe~II-P and II-L such that when applying the ``medium'' dust model, the resultant simulated LFs approximate those published by \citet{li2011LF}.

\begin{figure*}
  \begin{tabular}{ccc}
   \includegraphics[width=0.31\textwidth]{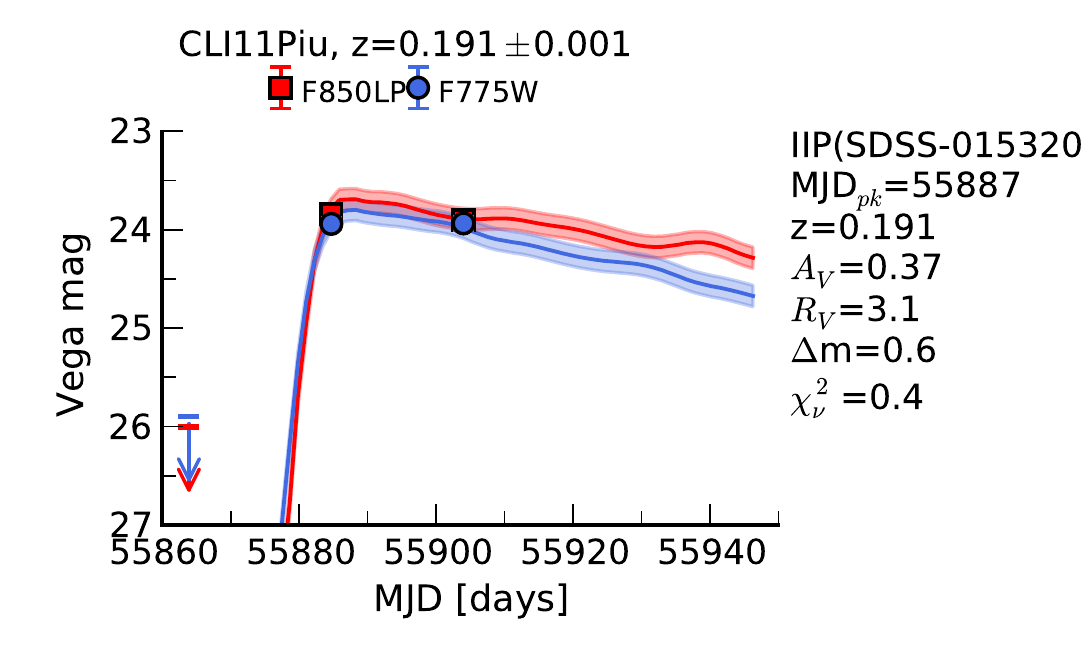} &
   \includegraphics[width=0.31\textwidth]{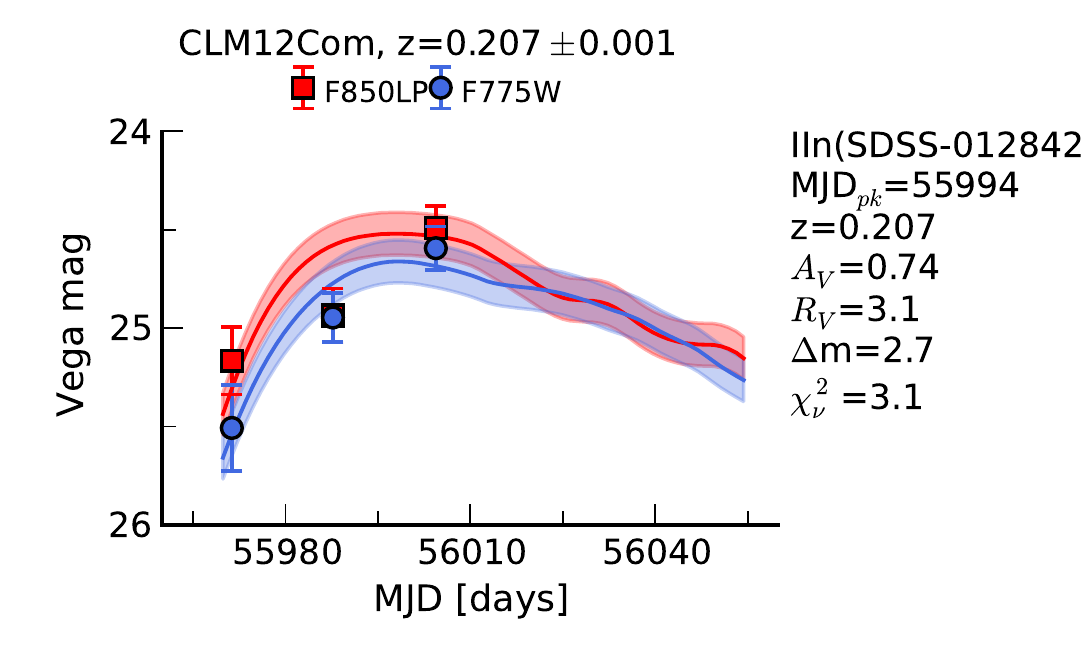} &
   \includegraphics[width=0.31\textwidth]{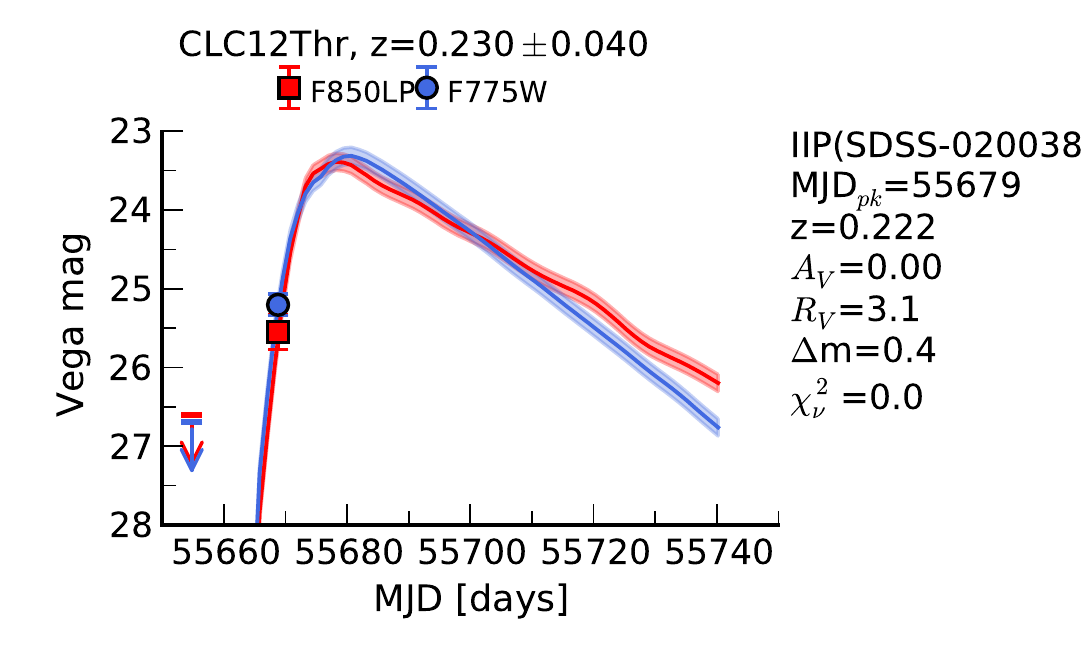} \\
   \includegraphics[width=0.31\textwidth]{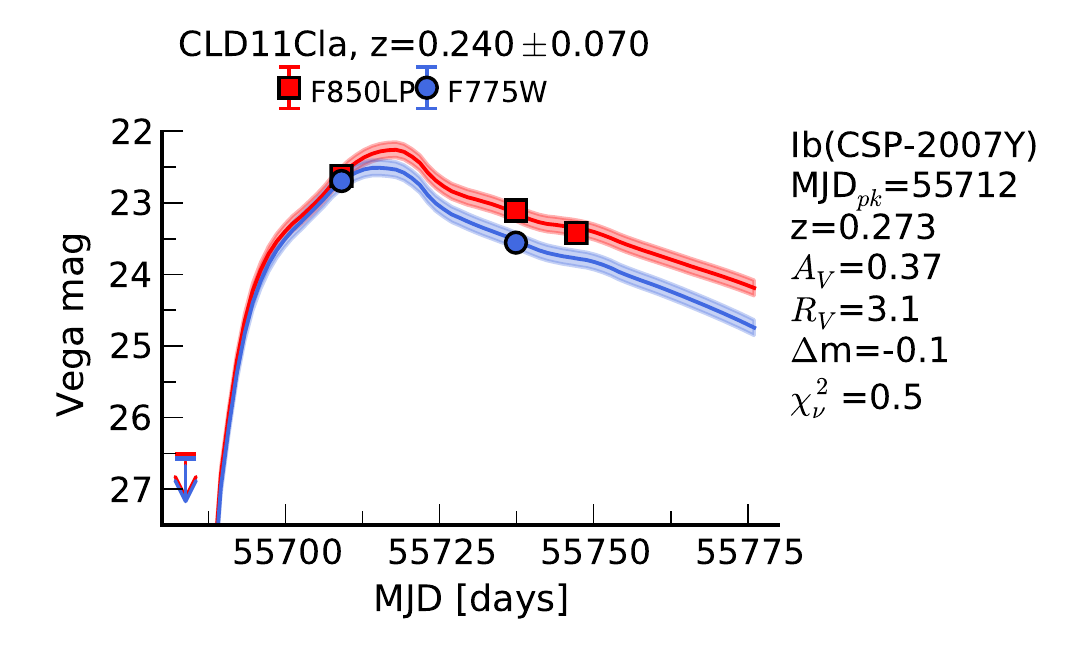} &
   \includegraphics[width=0.31\textwidth]{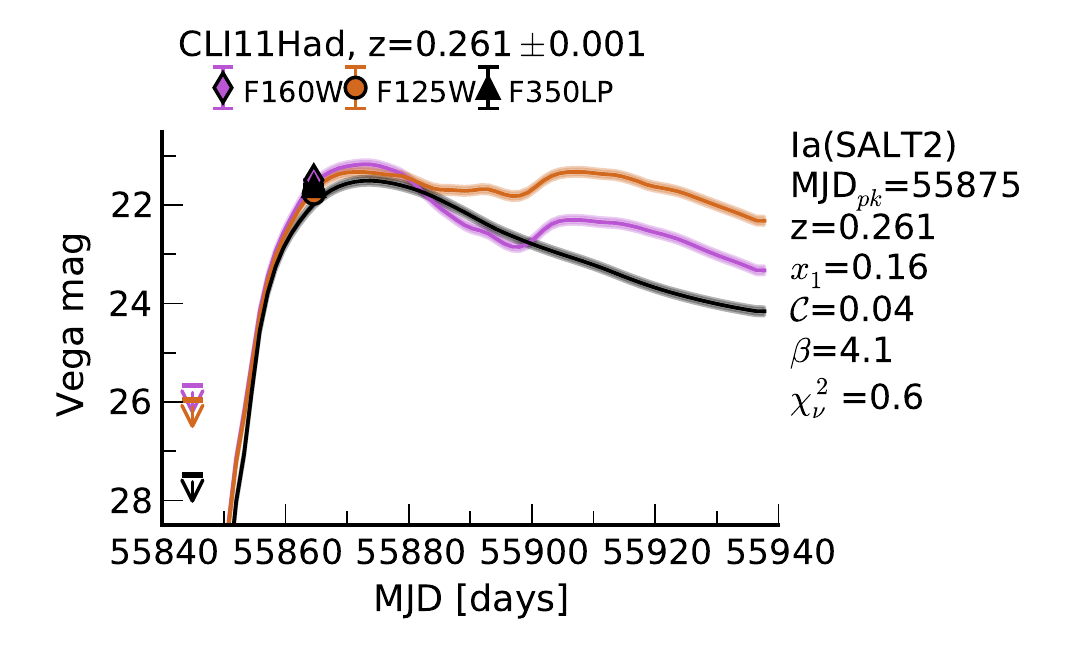} &
   \includegraphics[width=0.31\textwidth]{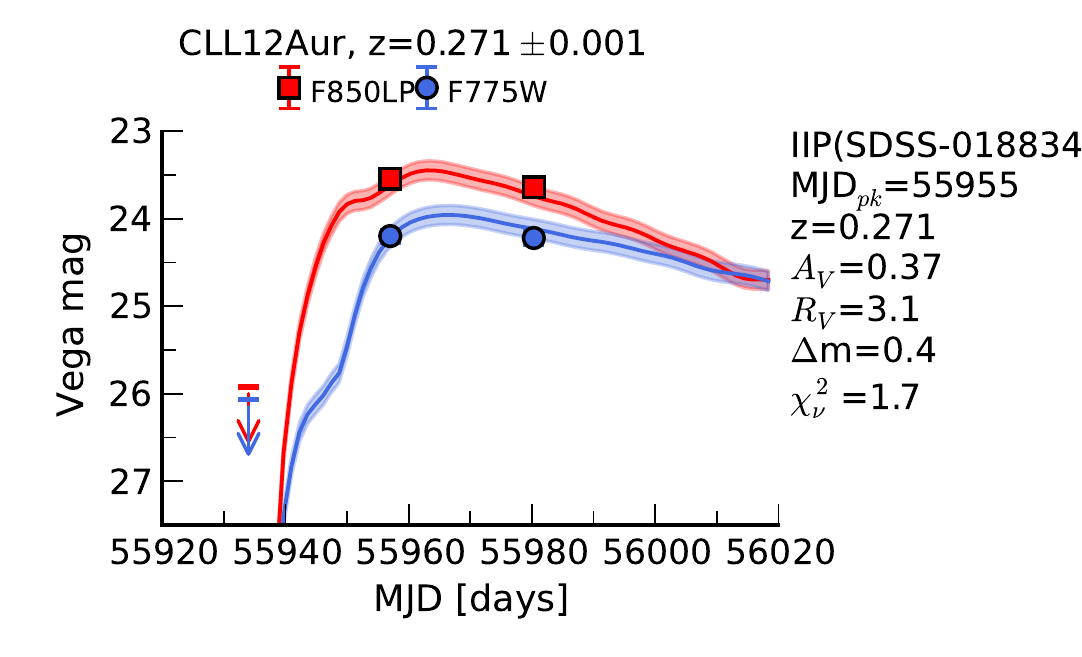} \\
   \includegraphics[width=0.31\textwidth]{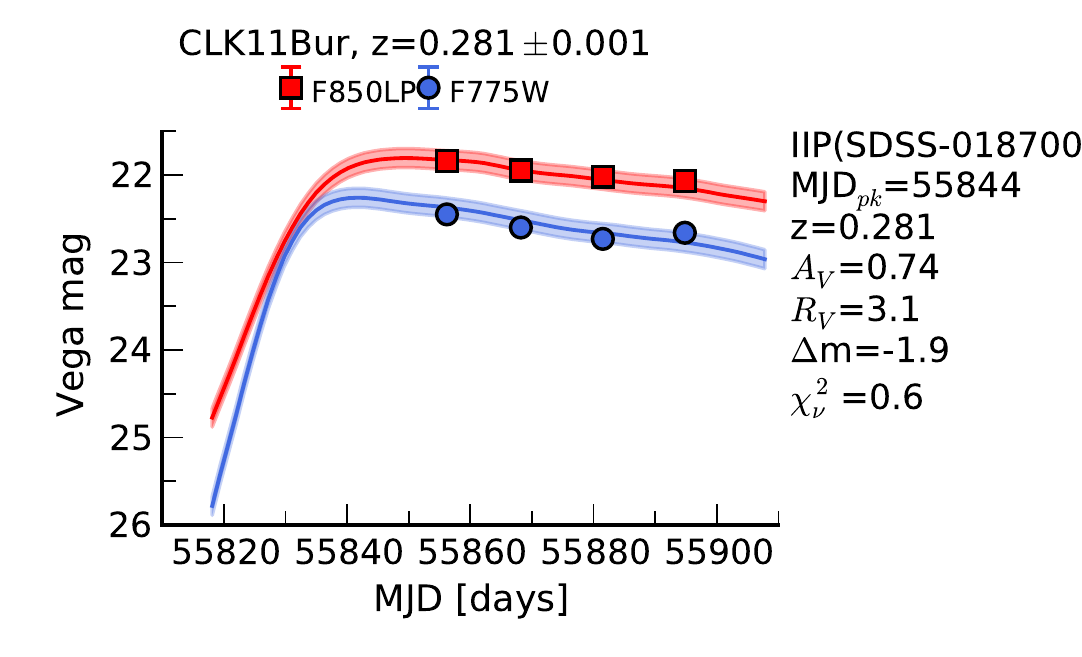} &
   \includegraphics[width=0.31\textwidth]{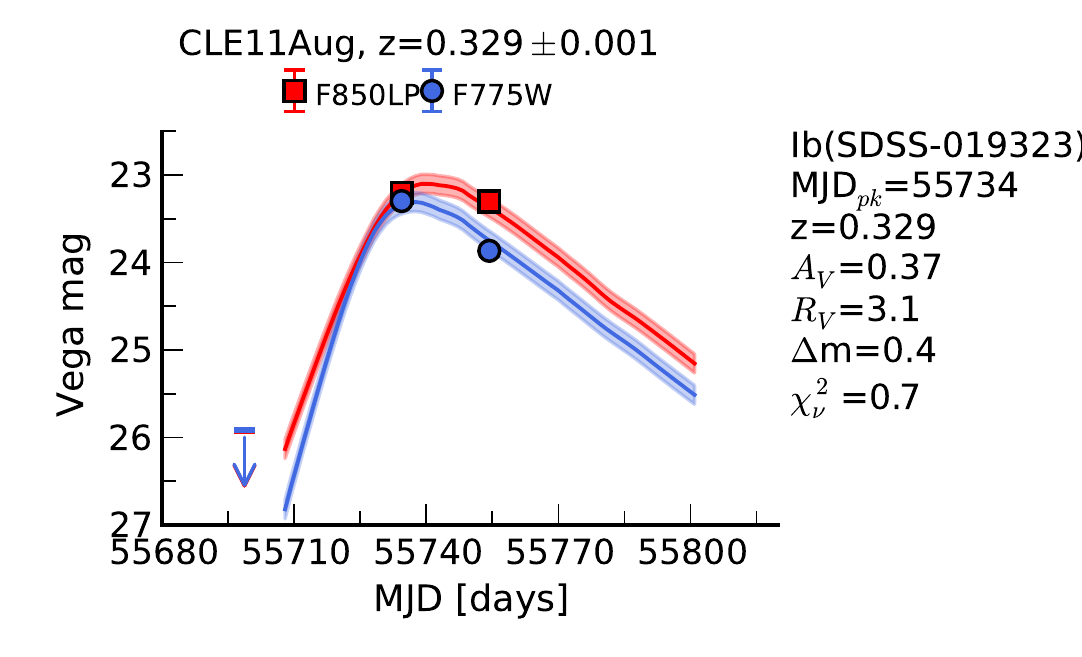} &
   \includegraphics[width=0.31\textwidth]{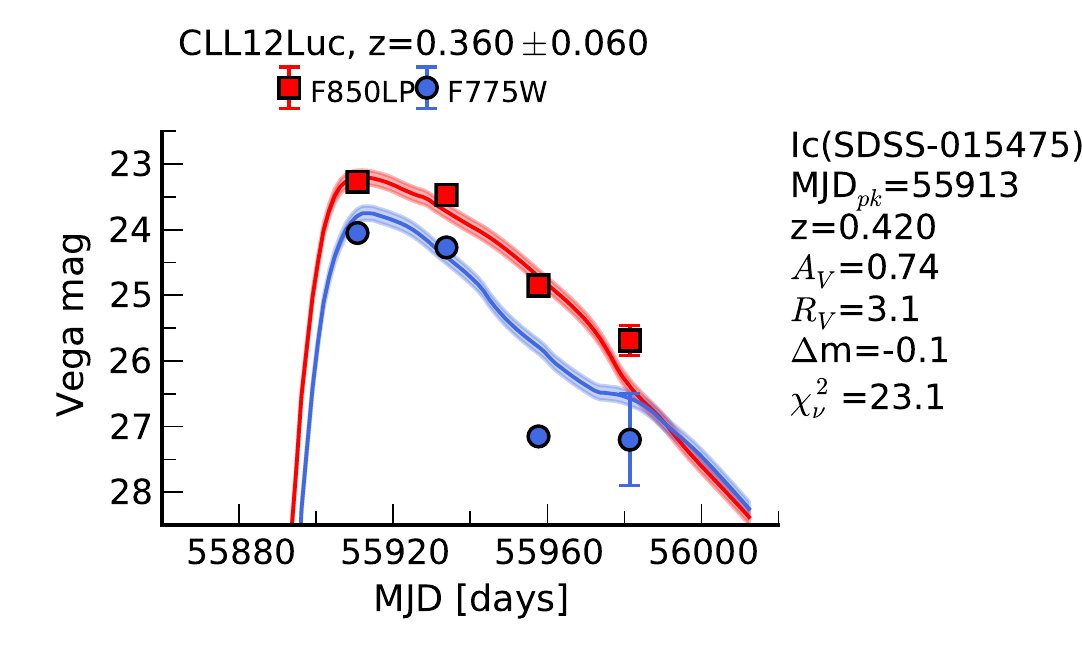} \\
   \includegraphics[width=0.31\textwidth]{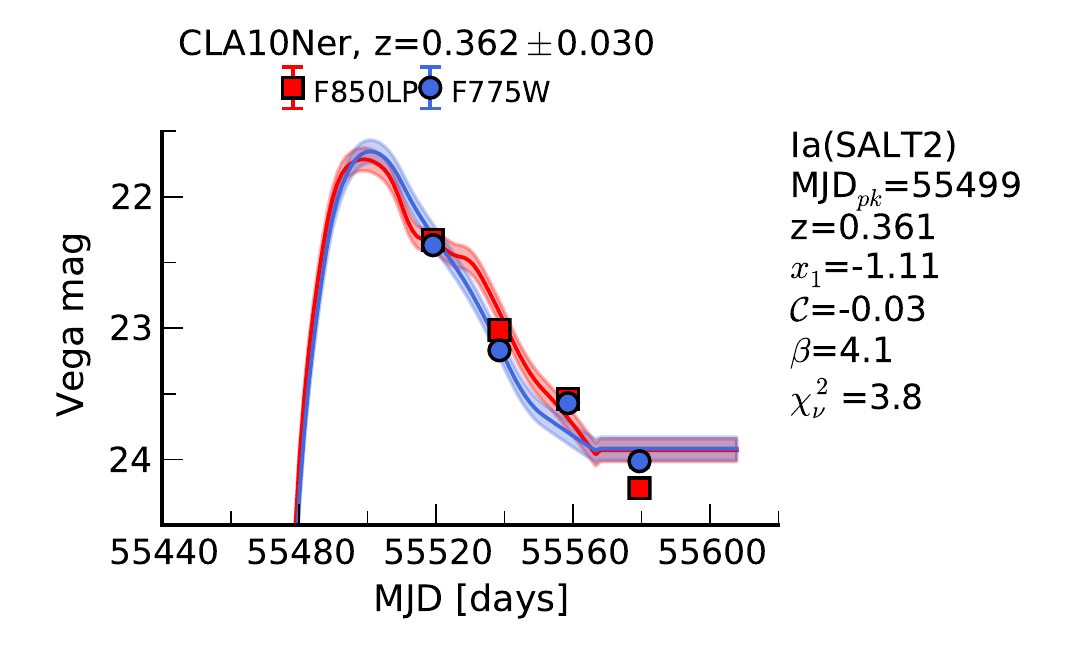} & 
   \includegraphics[width=0.31\textwidth]{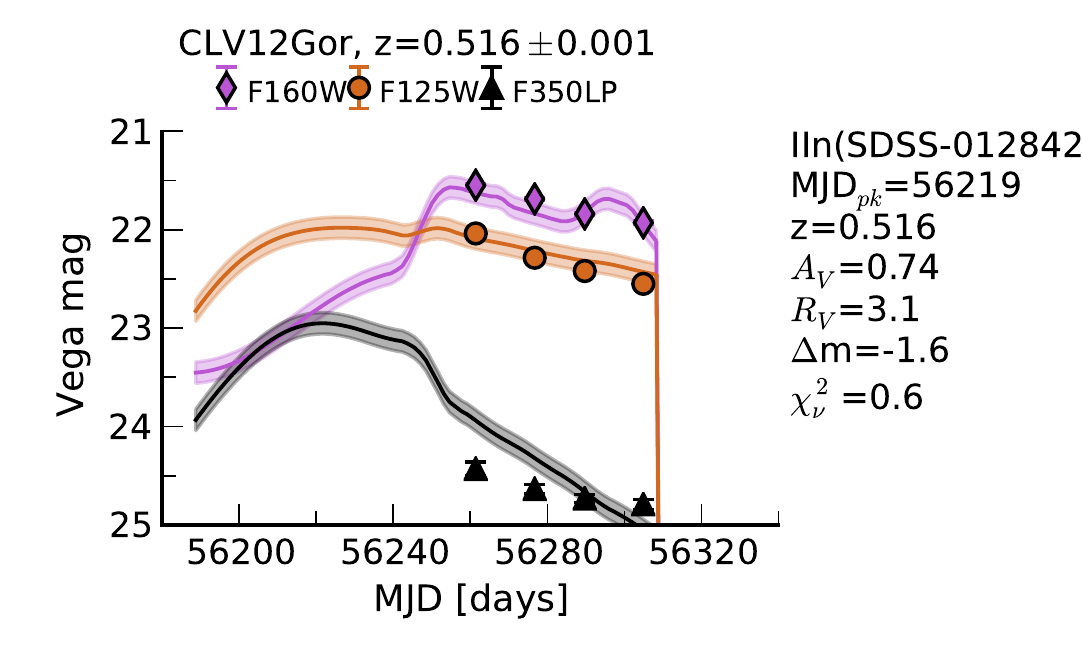} &
   \includegraphics[width=0.31\textwidth]{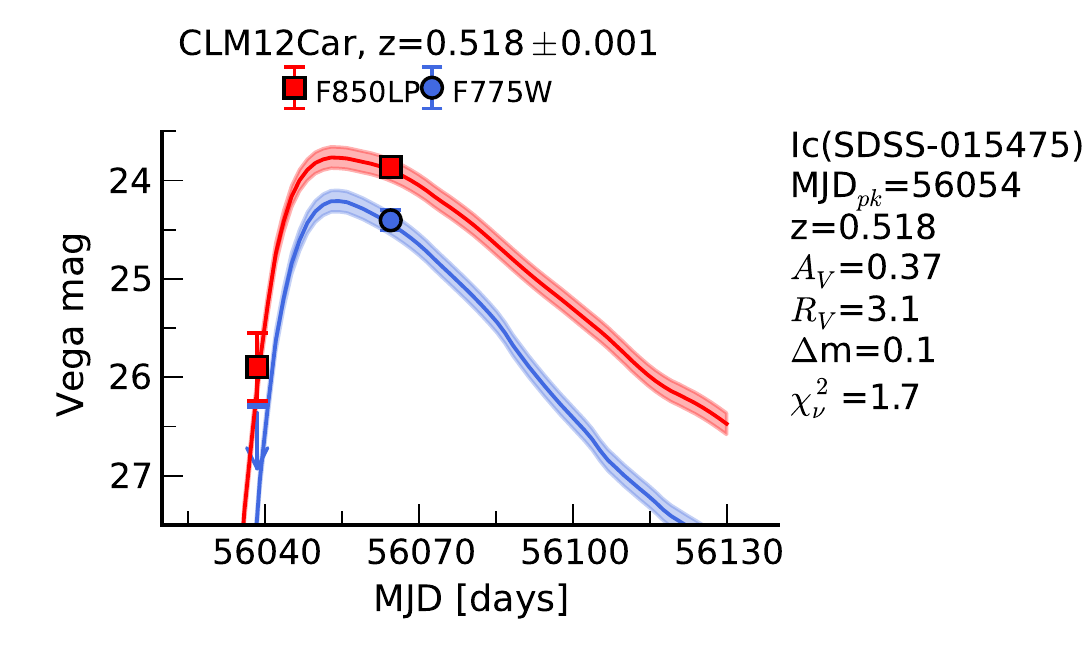} \\
   \includegraphics[width=0.31\textwidth]{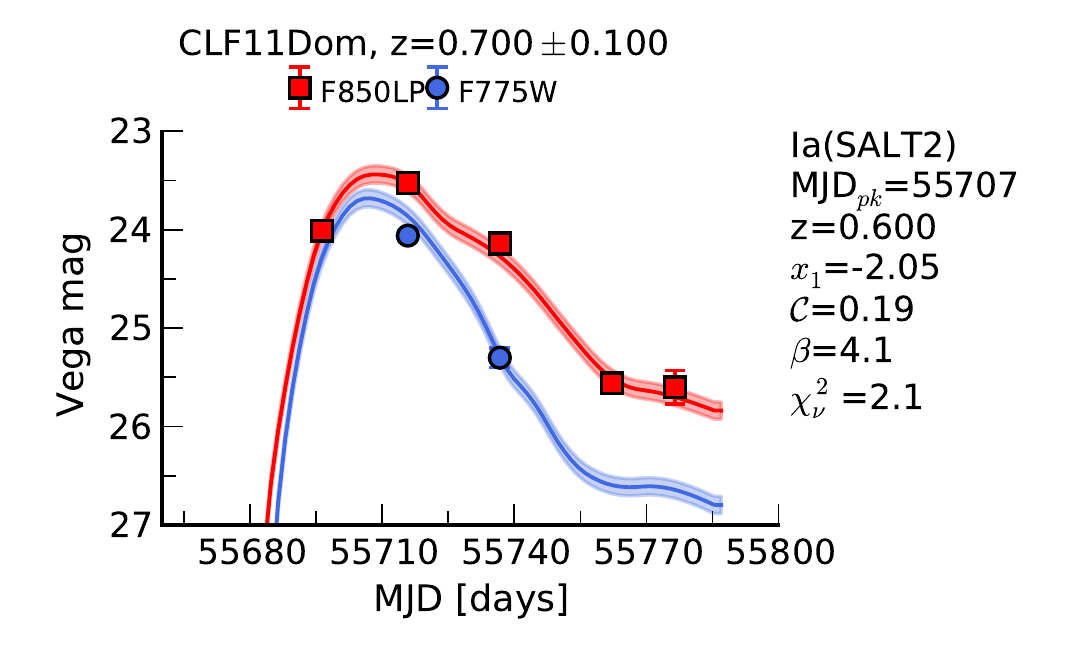} &
   \includegraphics[width=0.31\textwidth]{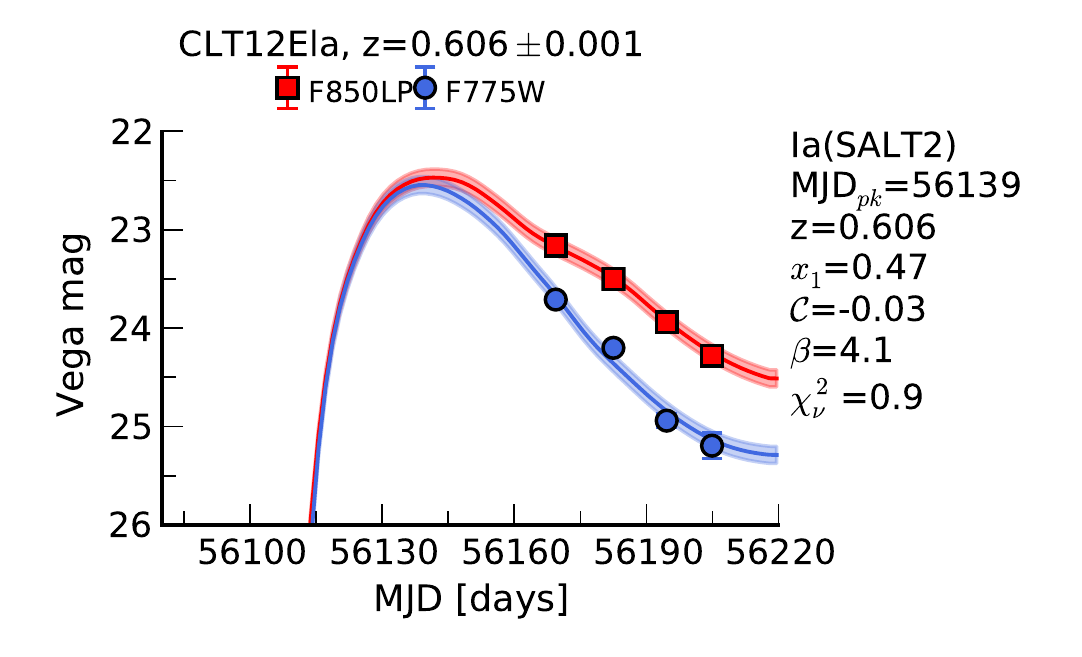} &
   \includegraphics[width=0.31\textwidth]{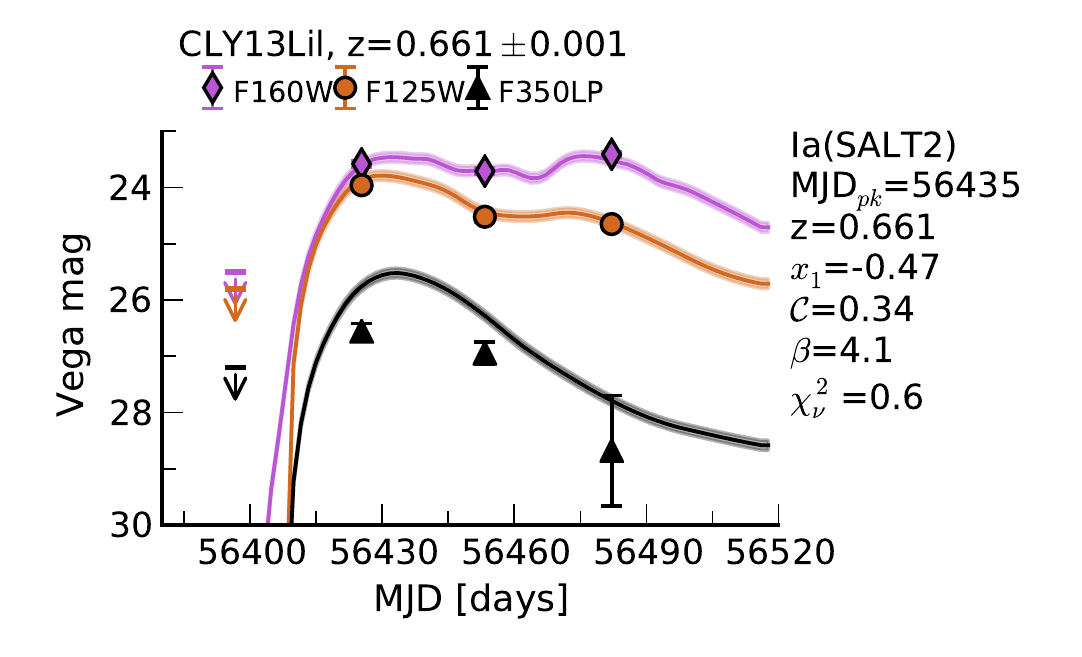} \\
  \end{tabular}
  \caption{Best-fitting \stardust\ light curves for 15 of the 27 SN candidates. Each panel shows the multi-band photometry and best-fitting template light curve of the maximum-likelihood SN type. The title of each panel gives the SN name and its prior redshift, along with a legend detailing the different filters used for photometry. The table to the right of the light curve details the maximum-probability SN template; the posterior values of the light-curve fit parameters, including the date of peak brightness, redshift, and color and shape parameters; and the reduced $\chi^2$ value of the fit, $\chi^2_\nu$. Downturned arrows denote $3\sigma$ upper limits on the flux.}
  \label{fig:clash_classify_1}
\end{figure*}

\begin{figure*}
 \begin{tabular}{ccc}
  \includegraphics[width=0.31\textwidth]{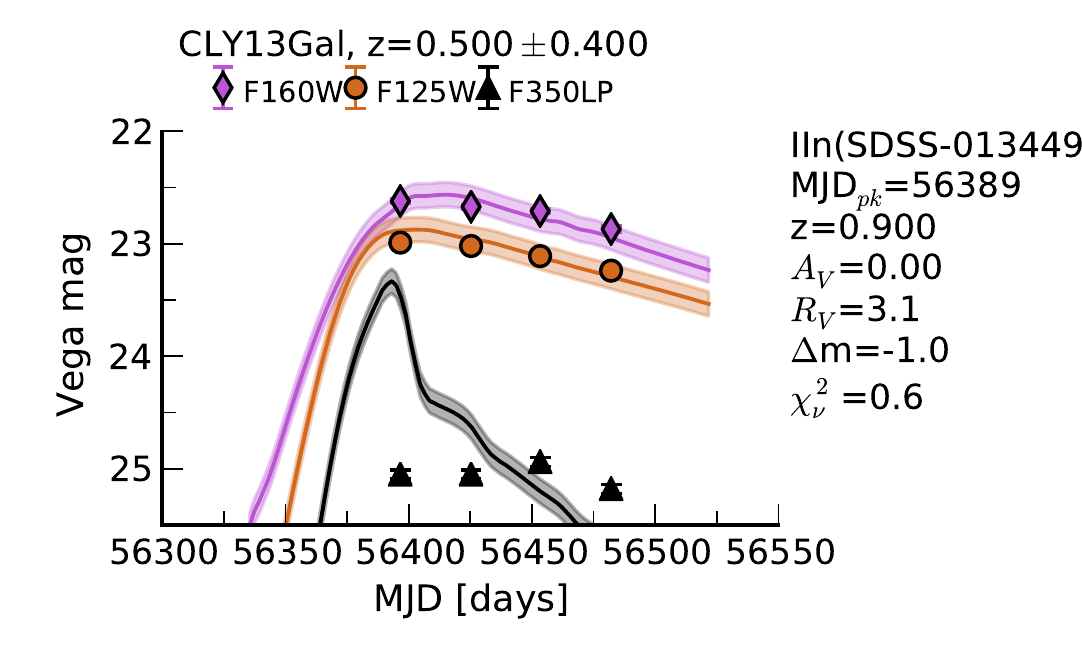} &
  \includegraphics[width=0.31\textwidth]{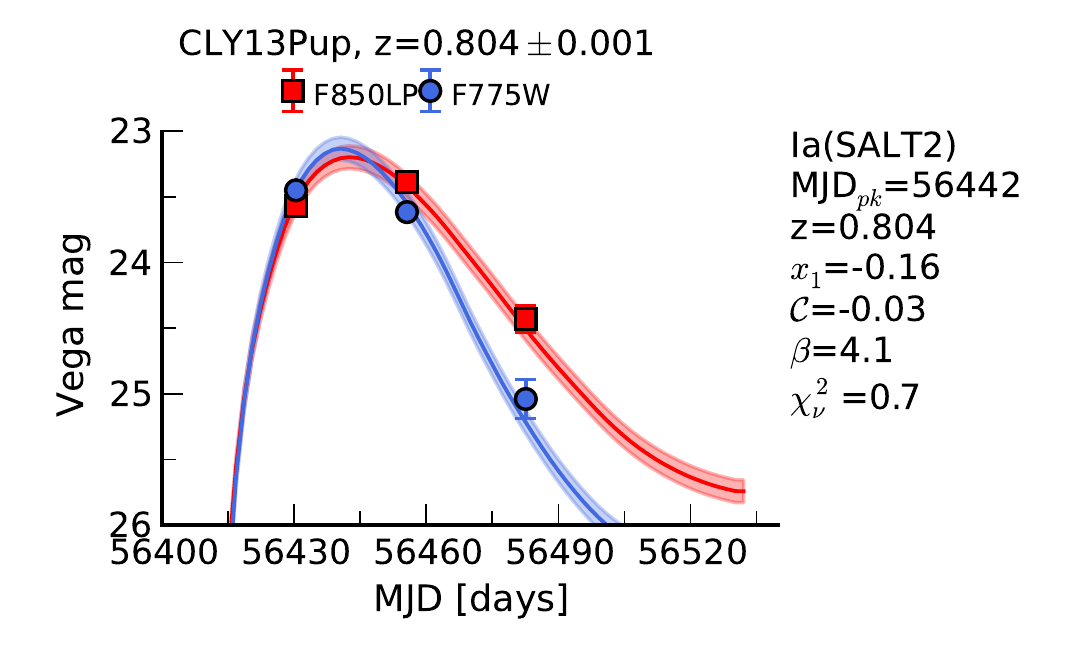} &
  \includegraphics[width=0.31\textwidth]{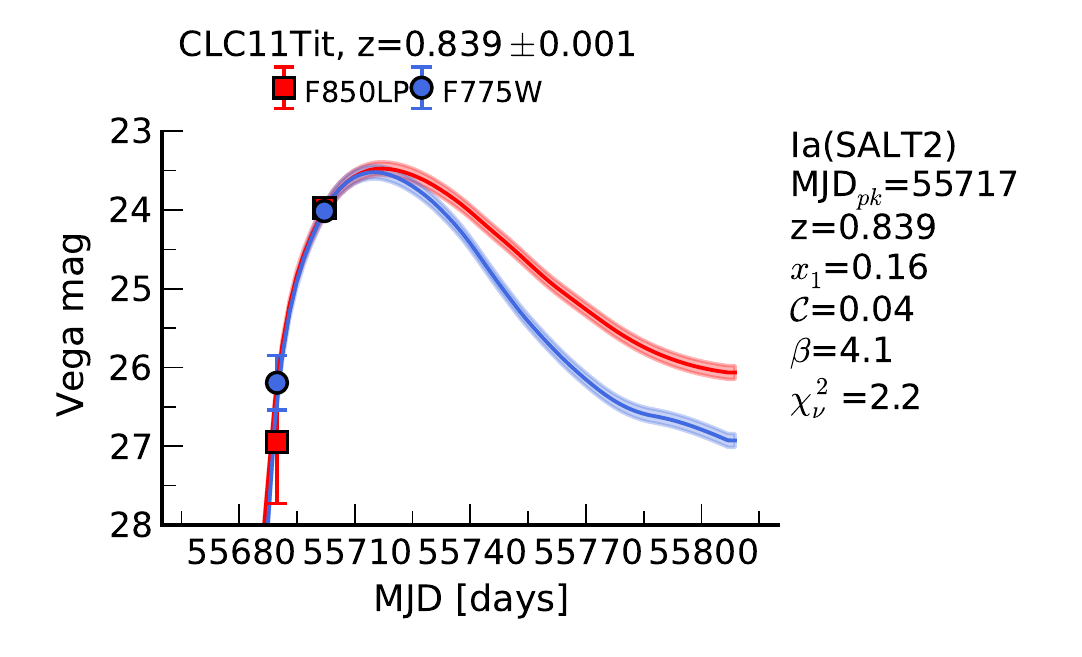} \\
  \includegraphics[width=0.31\textwidth]{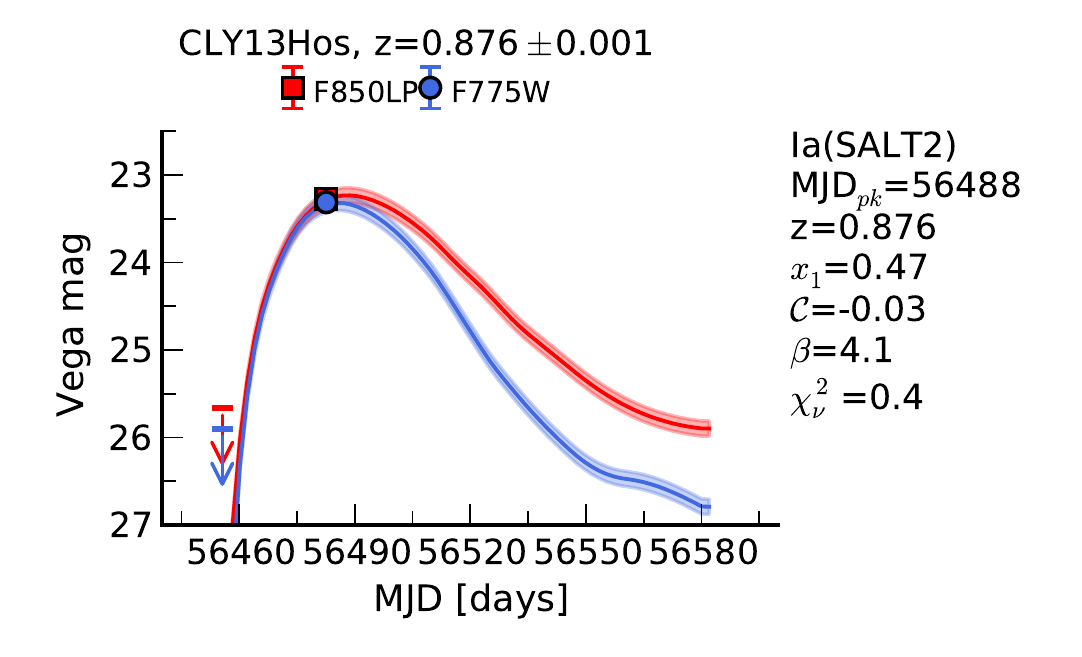} &
  \includegraphics[width=0.31\textwidth]{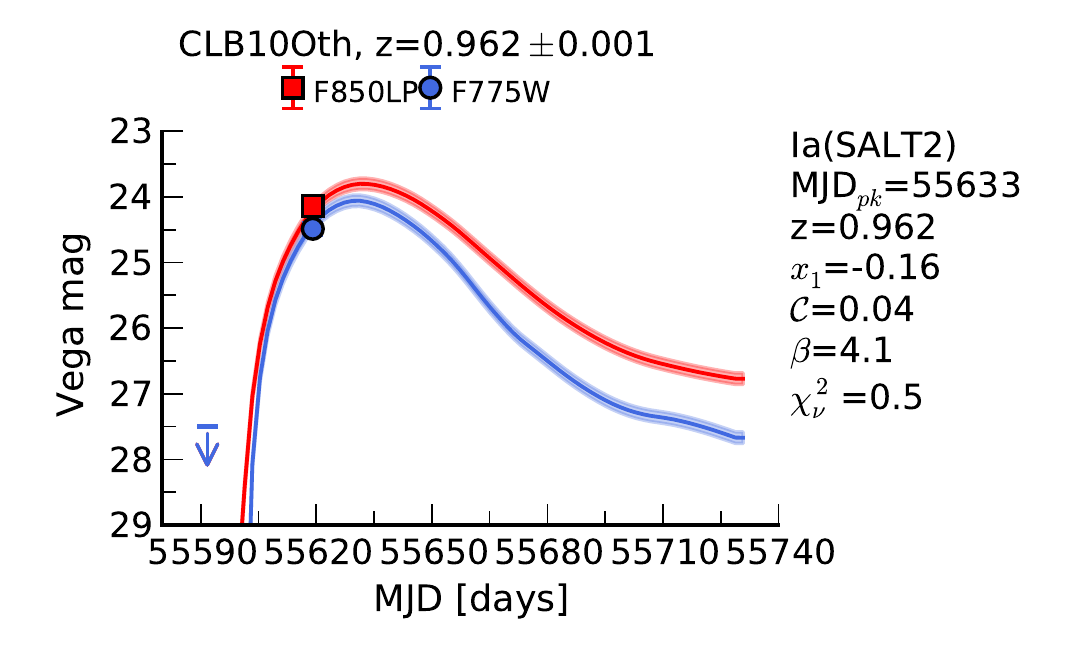} &
  \includegraphics[width=0.31\textwidth]{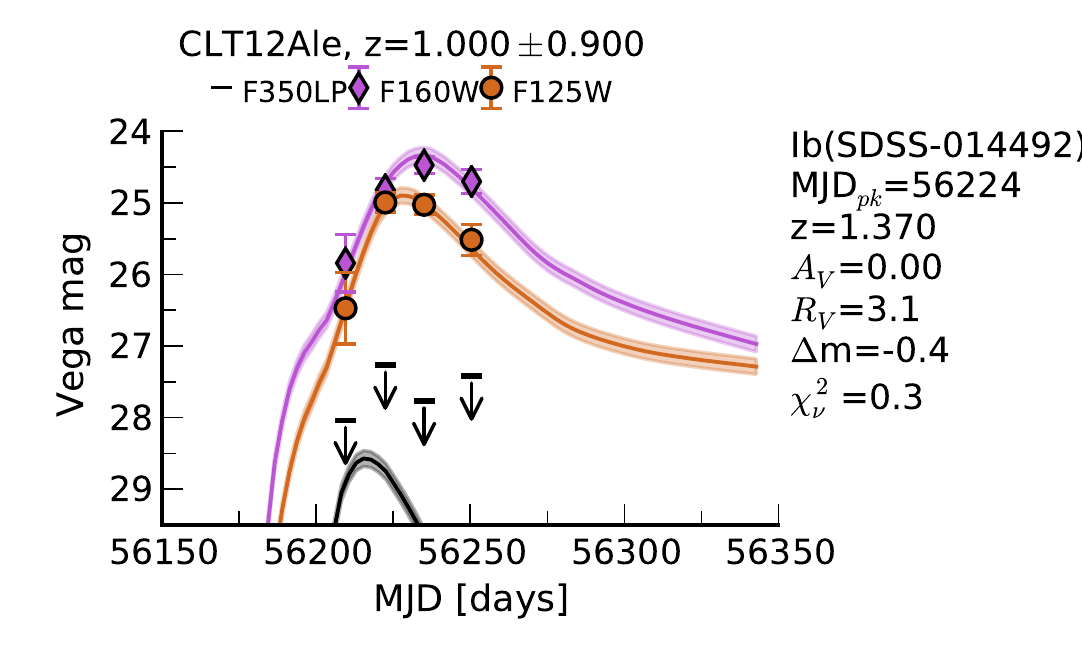} \\
  \includegraphics[width=0.31\textwidth]{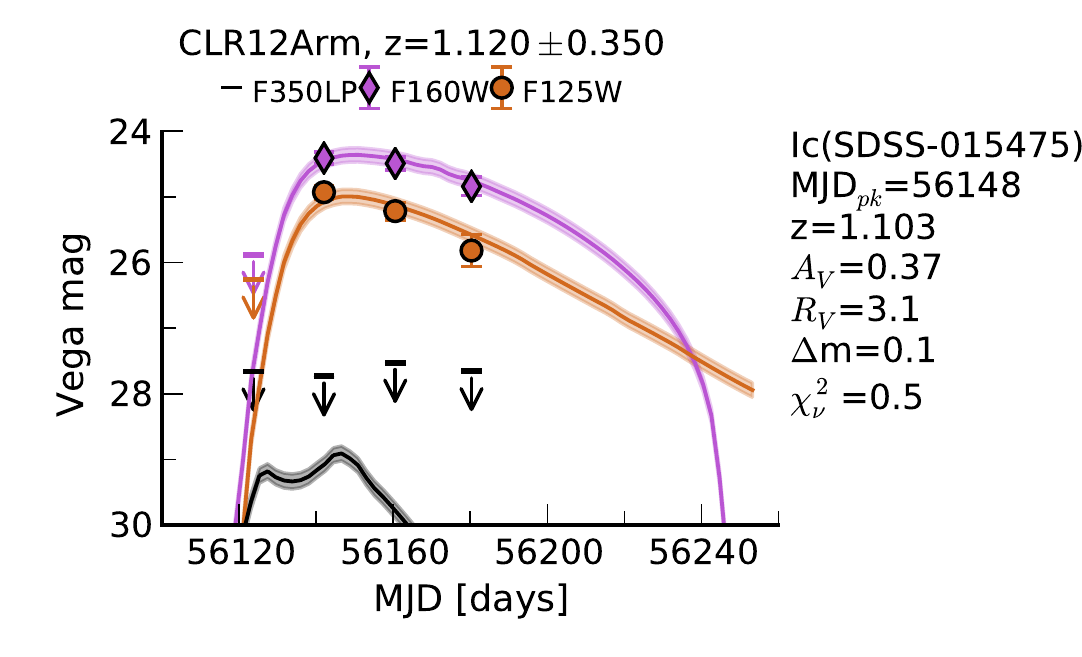} &
  \includegraphics[width=0.31\textwidth]{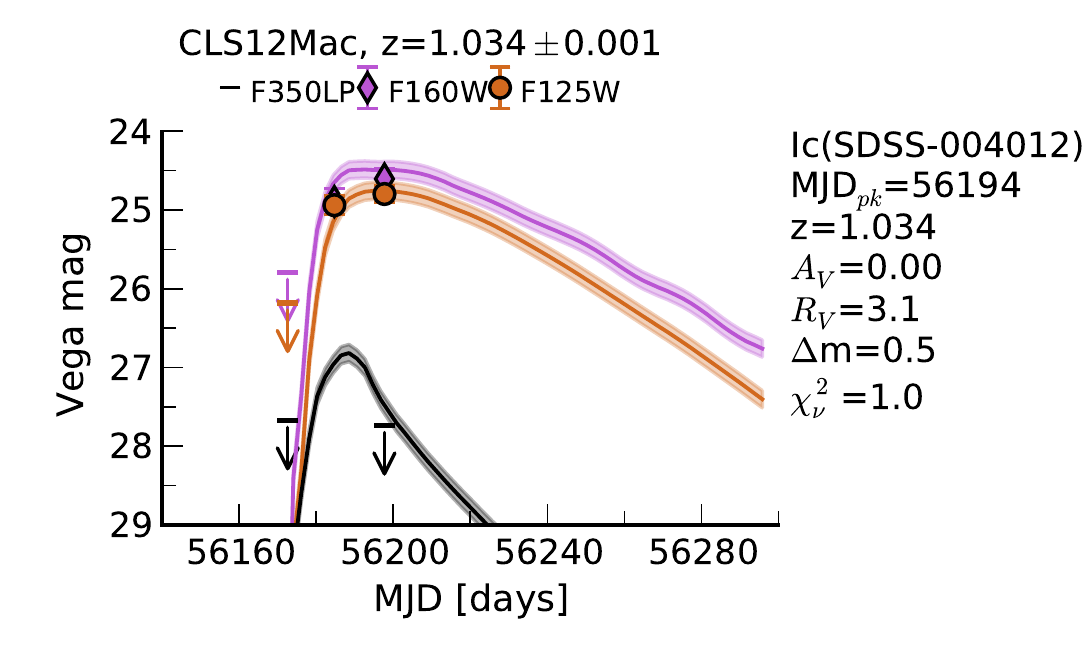} &
  \includegraphics[width=0.31\textwidth]{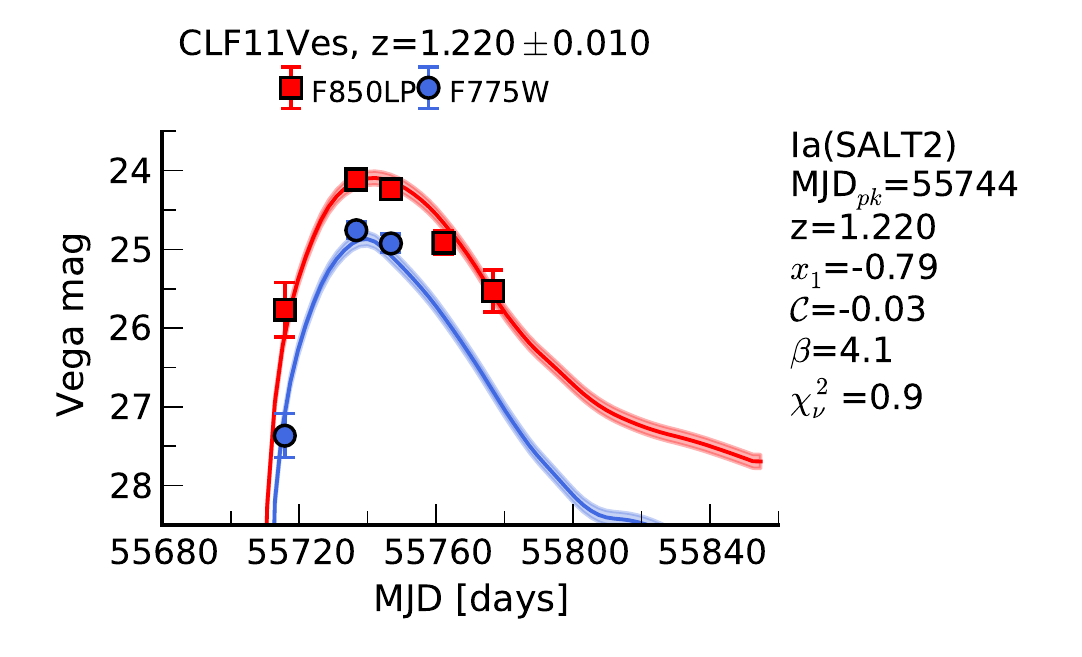} \\
  \includegraphics[width=0.31\textwidth]{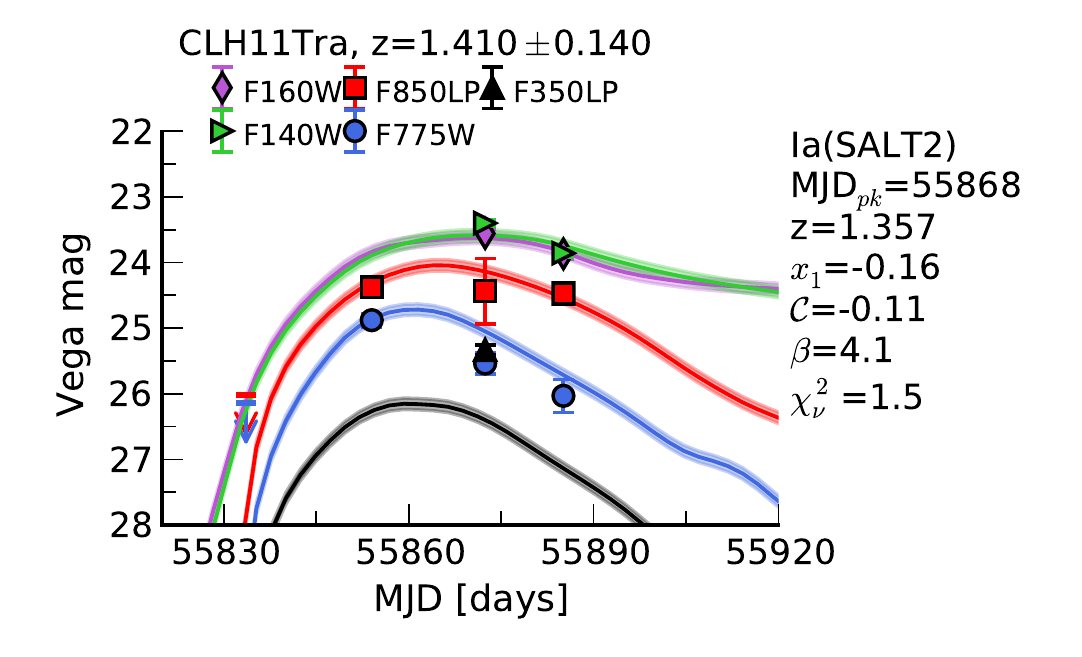} &
  \includegraphics[width=0.31\textwidth]{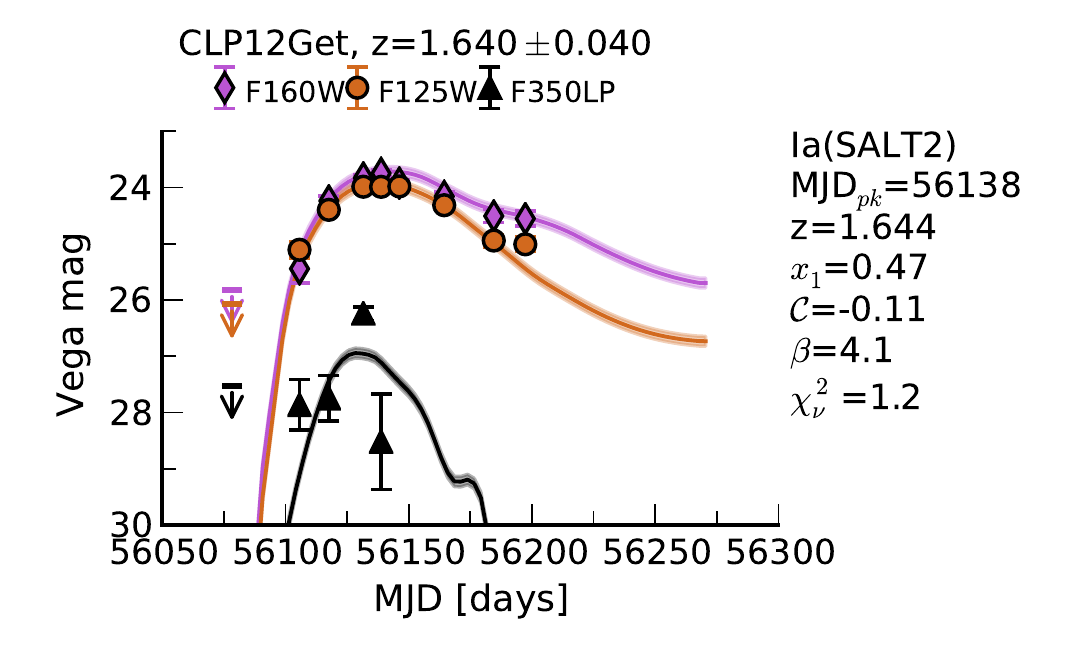} &
  \includegraphics[width=0.31\textwidth]{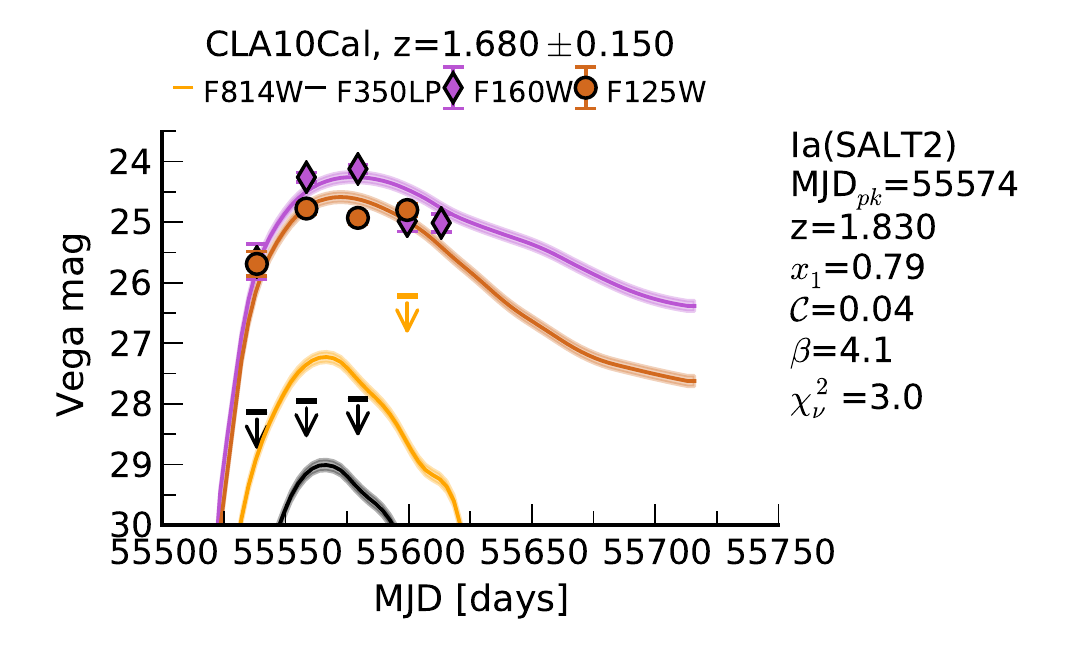} \\
 \end{tabular}
 \caption{Figure~\ref{fig:clash_classify_1}, continued, showing light-curve fits for the remaining 12 SN candidates.}
 \label{fig:clash_classify_2}
\end{figure*}

A further prior is placed on the fraction of each SN type as a function of redshift.
The distribution of SN type fractions has only been measured in the local universe \citep{li2011LF}, and it is expected to change with increasing redshift, once the SN~Ia rate starts to deviate from the star-formation rate.
However, assuming a prior on the evolution of the SN type fraction requires us to assume a prior on the CC~SN and SN~Ia rates as a function of redshift.
Such a prior might bias the SN~Ia rates measured in this work, and so we classify our SN sample twice: once using this prior, and once assuming that the fraction of SN types remains constant with redshift.
While the latter assumption is probably not the case in reality, it expresses our lack of concrete knowledge on the subject.
We report the classification probability, $P({\rm Ia})$, of each SN in our sample in Table~\ref{table:clash_SNe} both with ($P({\rm Ia})_{wp}$) and without ($P({\rm Ia})_{np}$) the SN type fraction prior.
The uncertainty reported for each $P({\rm Ia})_{wp}$ value takes into account uncertainties in both the extinction and SN-fraction priors, while the uncertainty of $P({\rm Ia})_{np}$ reflects only the uncertainty in the extinction prior.
For each of the SNe in our sample, the resultant $P({\rm Ia})$ values are consistent with each other, and while generally $P({\rm Ia})_{wp} < P({\rm Ia})_{np}$, the difference between the two values is small and has a negligible effect on the final SN~Ia rates.
The resultant light-curve fits, obtained without the SN-fraction prior, are presented in Figures~\ref{fig:clash_classify_1}--\ref{fig:clash_classify_2}.

SNe that were caught on the rise and were suspected of being SNe~Ia at $z>1$ were followed up with further \hst\ imaging in order to follow the evolution of their light curves.
In our sample, these include CLA10Cal, CLF11Ves, CLH11Tra, CLP12Get, CLR12Arm, and CLT12Ale.
Three SNe were caught sufficiently early, and were bright enough, to be followed up spectroscopically, either from the ground or with \hst.
These were CLI11Had, CLF11Ves, and CLY13Pup, whose spectra were obtained with Keck+LRIS, the ACS G800L grism, and VLT+FORS2, respectively.
Using the Supernova Identification code (\snid\footnote{http://marwww.in2p3.fr/$\sim$blondin/software/snid/index.html}; \citealt{Blondin2007}), we classify CLI11Had and CLY13Pup as SNe~Ia having the spec-$z$ measured from each of their host galaxies.
The best-fitting \snid\ templates are overlaid on the SN spectra in Figure~\ref{fig:clash_sn_spectra}.
Owing to its high redshift, the ACS G800L grisms caught CLF11Ves only in the rest-frame range $\sim 2500$--$4500$~\AA{}, and \snid\ fails to classify the spectrum as belonging to any type of SN at $z=1.22$.
When allowed to fit for the SN redshift, \snid\ classifies CLF11Ves as either a SN~Ia or SN~Ib at $z \approx 0.95$.
Consequently, we do not claim to have spectroscopic confirmation for CLF11Ves as a SN~Ia, although in the bottom panel of Figure~\ref{fig:clash_sn_spectra}, we show that the spectrum could be fit with the \citet{Hsiao2007} SN~Ia template at peak and at $z=1.22$.

Because declining SNe appear in all four epochs, their photometry cannot be measured from the subtraction images.
CLA10Ner, CLF11Dom, and CLL12Luc are well offset from their respective host galaxies, so we measure their photometry from the target images and assume any contamination by galactic light is minimal.
CLT12Ela is in a relatively faint (in the observed bands) area of its host galaxy.
Here, too, we assume that any contamination by galactic light is minimal.
As can be seen in Figure~\ref{fig:clash_sne}, CLK11Bur, CLV12Gor, and CLY13Gal exploded in relatively bright regions of their host galaxies, so contamination is a certainty.
However, all three of these SNe show signs of either a plateau or a slow decline in their light curves, and are classified as SNe~II, as shown in Figures~\ref{fig:clash_classify_1}--\ref{fig:clash_classify_2}.

\begin{table}
 \center
 \caption{SN Luminosity Functions Used for SN Classification}
 \begin{tabular}{l c c l}
  \hline
  \hline
  Type & $M_R$ & $\sigma$ & Source \\
  \hline
  Ia   & $-19.37$ & 0.47 & \citet{2006ApJ...641...50W} \\
  Ib   & $-17.90$ & 0.90 & \citet{2011ApJ...741...97D} \\
  Ic   & $-18.30$ & 0.60 & \citet{2011ApJ...741...97D} \\
  IcBL & $-19.00$ & 1.10 & \citet{2011ApJ...741...97D} \\
  II-P & $-16.56$ & 0.80 & \citet{li2011LF} \\
  II-L & $-17.66$ & 0.42 & \citet{li2011LF} \\
  IIn  & $-18.25$ & 1.00 & \citet{2012ApJ...744...10K} \\
  \hline
  \multicolumn{4}{l}{{\bf Notes.} The \citet{li2011LF} LFs have been} \\
  \multicolumn{4}{l}{corrected for host-galaxy extinction, as det-} \\
  \multicolumn{4}{l}{ailed in the text.}
 \end{tabular}
 \label{table:clash_LFs}
\end{table}

\subsection{Notes on Individual Supernovae}
\label{subsubsec:clash_individual}

We make the following more detailed notes on a number of cases with ambiguous host identification, redshift, or classification.

CLD11Cla has a broad $z$ probability distribution function (PDF) with a prominent peak at $z=2.66^{+0.07}_{-0.09}$ and a secondary peak at $z=0.24^{+0.07}_{-0.04}$.
We have attempted to obtain spectra of the host galaxy of this SN using several instruments (VLT+FORS2, Keck+LRIS, and the ACS+G800L grism).
However, the host galaxy appears to have an absorption spectrum with no discernible emission lines suitable for determining its redshift.
At $z \approx 2.7$, the SN would be too bright to be either a SN~Ia or a CC~SN.
While it could still be a high-$z$ superluminous SN \citep{2012Sci...337..927G}, it is more likely that this is a normal CC~SN at $z=0.24$, at which its colors agree with those of CC~SNe simulated using \snana.
In any case, we exclude this SN from the SN~Ia sample.

CLF11Dom was discovered on the decline, but actually peaked during the period of the survey, as evidenced by its photometry.
It has no immediately discernible host galaxy, the closest lying $\sim 4.7''$ away, with a photo-$z$ of $0.71^{+0.12}_{-0.09}$.
When classifying this SN, we chose a wide redshift prior of $0.7\pm0.6$ and found that the light curve was best fit as belonging to a SN~Ia at $z \approx 0.6$, consistent with the photo-$z$ of the nearby galaxy.

CLL12Luc is a declining SN with four possible host galaxies.
Of those, three have photo-$z$ values of 0.33, 0.59, and 1.13.
One has a spec-$z$ of 0.36 that is consistent with its photo-$z$ of 0.33.
At these redshifts, the SN would be $\sim45$, 30, and 70 kpc away from the core of each galaxy, respectively.
Based on the colors of the SN, and assuming it peaked 40--70 days before the first observation epoch, it could either be a CC~SN at $z=0.36$ or a SN~Ia at $z \approx 0.6$.
However, the SN colors measured from the last two observation epochs are inconsistent with the SN~Ia solution, so in this work we treat this SN as a CC~SN.

Like CLL12Luc, CLC12Thr has several potential host galaxies, for three of which we have measured the photo-$z$.
Two galaxies, a face-on spiral and an elliptical, are at $z=0.23$, while the third is an E/S0 galaxy at $z \approx 1.6$.
At either redshift, the colors of the SN are consistent with those of a CC~SN.
However, the $z=0.23$ spiral and elliptical galaxies are much closer to the SN than the $z \approx 1.6$ galaxy: $\sim 60$, 30, and 155 kpc, respectively.
Based on these data, we classify this SN as a CC~SN at $z=0.23$.
It would be interesting to inspect the elliptical galaxy, which is closer to the SN, for evidence of recent star formation.

We have discovered four SNe~Ia at $z>1.2$: CLA10Cal, CLF11Ves, CLH11Tra, and CLP12Get.
Although we have not yet succeeded in obtaining the spectroscopic redshift of the host galaxy of CLA10Cal, its photo-$z$ is based on both Subaru data and \hst\ NIR photometry in the \FJ\ and \FH\ bands, so we are confident of its redshift.
CLP12Get has a secure redshift from a photo-$z$ fit based on nine bands from the optical (Subaru) to the NIR (\hst).
The spectrum of its host galaxy, obtained with VLT+X-shooter, shows no emission or absorption lines, but the photometry extracted from its continuum is consistent with the photo-$z$ fit.
CLF11Ves, as noted above, has a secure, spectroscopic redshift of 1.22.
All four candidates have $P({\rm Ia})_{wp,np}>0.95$.

There are, at most, three more SNe with broad photo-$z$ PDFs that could potentially enter this redshift range: CLR12Arm, CLT12Ale, and CLY13Gal.
However, all three SNe are classified as CC~SNe according to their light curves, with posterior redshifts of $\sim 1.1, 1.4$, and $0.9$, respectively.
CLT12Ale exploded in a faint host galaxy (${\rm \FH}=26.4\pm0.1$~mag).
Based on the available Subaru photometry and additional \hst\ photometry in \FJ\ and \FH, the photo-$z$ of the galaxy can only be constrained to the very wide range $0.5$--$3.0$.
When considering the two-parameter space spanned by the host galaxy's redshift and type, we find that the most likely redshift solutions are either at $z \approx 1$ or $z>5$.
Although superluminous SNe have been observed out to redshift $\sim4$, they are exceedingly rare (\citealt{2012Natur.491..228C}).
It follows that the more likely redshift solution for this host galaxy is $\sim1$.
This is strengthened by the light-curve classification of this SN, which finds a posterior redshift of $\sim1.4$.
As there is some probability that CLT12Ale is a SN~Ia ($P({\rm Ia})_{np}=0.13\pm0.10$), we include it in our calculation of the SN~Ia rates, as detailed in Section \ref{sec:clash_rates}, below.


\section{The Type-Ia Supernova Rate}
\label{sec:clash_rates}

In this section, we use the SN~Ia sample from Section \ref{sec:clash_sne}, along with the detection efficiencies as a function of magnitude measured in Section \ref{subsec:clash_eff} and the classification probabilities measured in Section \ref{subsec:clash_colormag}, to measure the rates of SNe~Ia as a function of redshift, or lookback time.
So as not to bias our results, we use the SN classification without the assumption of the SN-fraction prior.

\begin{table*}
\center
\caption{SN~Ia Numbers and Rates}\label{table:clash_rates}
\begin{tabular}{lcccc}
\hline
\hline
{Subsample} & {$0.0<z<0.6$} & {$0.6<z<1.2$} & {$1.2<z<1.8$} & {$1.8<z<2.4$} \\
\hline
Total & 12 & 11 & 4 & 0 \\
\hline
SN host galaxies with spec-$z$ & 10 & 7 & 2 & 0 \\ 
Hostless SNe & 0 & 1 & 0 & 0 \\
\hline
SNe~Ia (raw) & \Nl & \Nm & \Nh & 0 \\
SNe~Ia (efficiency-corrected) & \Nld & \Nmd & \Nhd & 0 \\
SN~Ia rate without host-galaxy extinction$^a$ & \Rdl & \Rdm & \Rdh & $<\Rdul$ \\
SN~Ia rate [10$^{-4}$ yr$^{-1}$ Mpc$^{-3}$] & $\Rl$ & $\Rm$ & $\Rh$ & $<\Rul$\\
Effective redshift & 0.42 & 0.94 & 1.59 & 2.1 \\
\hline
\multicolumn{5}{l}{{\bf Notes.} The $1.8<z<2.4$ rate is a $2\sigma$ upper limit.} \\
\multicolumn{5}{l}{$^a$The errors, separated by commas, are respectively the 68\% Poisson statistical uncertainties on the number of SNe, and} \\
\multicolumn{5}{l}{systematic uncertainties due to possible misclassification and different host-galaxy extinction models, respectively.}
\end{tabular}
\end{table*}

The SNe~Ia in our sample can be divided among three categories, according to when they reached maximum light: before, during, or after the monitored interval of time spent on each field.
Each category will have a distinct detection efficiency as a function of redshift.
The date of maximum light can occur up to 40 days before and 20 days after the duration of the survey.
These values were chosen according to the approximate time when the SNe~Ia in our sample reached their peak, relative to the survey times in the fields where they were discovered, based on preliminary light-curve fits.
Accordingly, the visibility time of our survey is defined as the sum of the times each parallel field in each cluster was monitored (i.e., the time between the first and last epoch of that field), with the addition of 40 days before and 20 days after the observation period, in order to account for the SNe~Ia in our sample that were caught either in decline or on the rise.
Adding more or less time to the survey-extension times will either raise or lower, respectively, the number of SNe included in the rate calculation.
To within Poisson errors, the change in extension time should, in principle, cancel out the change in SN numbers, leaving the resultant SN~Ia rate unchanged.

We define the rate, $R_{\textrm{Ia}}$, in a redshift bin bound by redshifts $z_1$ and $z_2$, as
\begin{equation}\label{eq:clash_rate}
 R_{\textrm{Ia}}(z_1<z<z_2) = \frac{\sum\limits_i{N_i(z)/\eta_i(z)}}{\sum\limits_j{t_j A_j} \int_{z_1}^{z_2}{\frac{1}{(1+z)}\frac{dV}{dz}dz}},
\end{equation}
where $N_i$ is the number of SNe~Ia (see below); $\eta_i$ is that category's detection efficiency at the redshift, $z$, of each SN; $t_j$ is the visibility time, composed of the time between the first and last epoch of observation of a field $j$, plus $40$ days before the start of the survey and $20$ days after its end; $A_j$ is the solid angle of the searchable area of field $j$, divided by $4\pi$ steradians; $dV$ are thin volume elements behind each searchable area; and the $(1+z)$ factor converts the rates from the observer frame to the rest frame.
Although we classify a SN as a SN~Ia if $P({\rm Ia})\ge0.5$, we define $N_i$ as the sum of $P({\rm Ia})_{np}$ values of all the SNe in each subcategory (before, during, or after the monitored interval).
This is based on our treatment of $P({\rm Ia})$ as a measure of the probability of a SN being a SN~Ia.
Thus, for example, CLD11Cla has a $P({\rm Ia})_{np}=19$\% probability of being a SN~Ia, and is counted accordingly.
This approach allows us to take into account the uncertainty of our classifications, especially for SNe with sparse data.

To compute the detection efficiency of each SN category, $\eta$, we must first convert our detection efficiency from a function of magnitude to a function of redshift.
We do this by using the measured detection efficiency as a function of magnitude from Section \ref{subsec:clash_eff} to simulate the discovery process of $\sim$\,25,000 SNe~Ia. 
For each SN, we use the \citet{Hsiao2007} SN~Ia spectral templates to simulate a light curve.
The \citet{Hsiao2007} spectral templates are normalized so that an unredshifted, ``normal'' (i.e., with stretch $s=1$) SN~Ia at maximum light has a $B$-band apparent magnitude of $M_B=0$.
These templates are first redshifted and reddened using the \citet{1989ApJ...345..245C} extinction law.
We perform synthetic photometry on the redshifted and reddened spectral templates in the survey filters (\FZ\ and \FH) and construct light curves according to \citet{1999ApJ...517..565P}:
\begin{equation}\label{eq:LC}
m=m_{z,s=1}+M_B+\mu-\alpha(s-1),
\end{equation}
and
\begin{equation}\label{eq:LC1}
t_{s,z}=t_{z=0,s=1}\alpha(1+z),
\end{equation}
where $m_{z,s=1}$ is the apparent magnitude of a ``normal'' SN~Ia with $s=1$ at redshift $z$; $M_B$ is the absolute magnitude of the SN in the $B$ band at maximum light; $\mu$ is the distance modulus at redshift $z$; $\alpha=1.52\pm0.14$ (\citealt{2006A&A...447...31A}); and $t_{s,z}$ is the time axis of the light curve, stretched as a result of the stretch, $s$, and time dilation at redshift $z$.

\begin{table}
 \center
 \caption{SN~Ia Rate Uncertainty Percentages}\label{table:clash_errs}
 \begin{tabular}{lccc}
  \hline
  \hline
  Uncertainty & $0<z<0.6$ & $0.6<z<1.2$ & $1.2<z<1.8$ \\
  \hline
  Poisson        & $+92~/~-71$      & $+49~/~-44$     & $+75~/~-49$     \\
  Classification & $+17~/~-28$      & $+4.6~/~-9.4$   & $+1.9~/~-2.7$   \\
  Dust model     & $+5.4~/~-0.1$    & $+25~/~-3.3$    & $+8.6~/~-18$    \\
  Total          & $+114.4~/~-99.1$ & $+78.6~/~-56.7$ & $+85.5~/~-69.7$ \\
  \hline
  \multicolumn{4}{l}{{\bf Notes.} All uncertainties are reported as percentage of the rates.}
 \end{tabular}
\end{table}

\begin{figure}
 \center
 \includegraphics[width=0.5\textwidth]{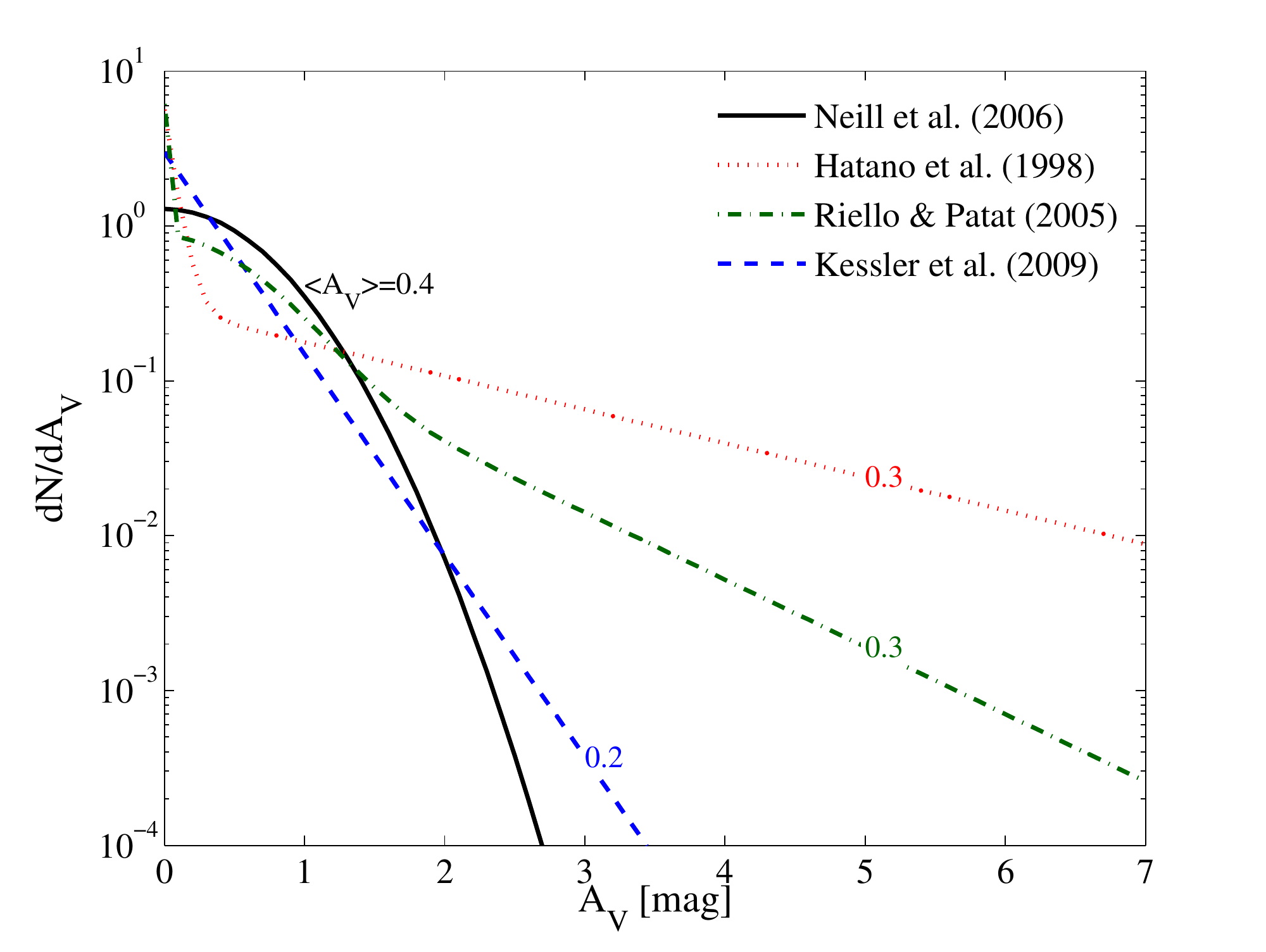}
 \caption[SN Ia host-galaxy dust extinction models]{SN Ia host-galaxy dust extinction models used in the derivation of the SN~Ia rates. We use the \citet{neill2006} model (solid black curve) as the fiducial model. While the models shown here go out to $A_V=7$ mag, we use values only to $A_V=3$ mag, as nearly all the objects produced by these models fall in the range $A_V=0$--3 mag.}
 \label{fig:clash_Av}
\end{figure}

Each SN is assigned a random cluster and field (WFC3 or ACS), redshift, $M_B$, stretch value, and host-galaxy extinction value ($A_V$).
The redshift values are drawn from a flat distribution in the range 0--3.
Following G11, the absolute $B$-band magnitude at maximum light is drawn from a Gaussian centered on $M_B=-19.37$ with a standard deviation of $\sigma_{M_B}=0.17$ (this standard deviation, smaller than the one used for the SN~Ia LF as it appears in Table~\ref{table:clash_LFs}, reflects the spread in SN~Ia intrinsic luminosity after correcting for the luminosity-stretch relation; \citealt{1995AJ....109....1H,1996AJ....112.2398H}; \citealt*{1996ApJ...473...88R}; \citealt{Phillips1999}).
The stretch values, following \citet{2006AJ....131..960S}, are drawn from a Gaussian centered on $s=1$ with a standard deviation of $\sigma_s=0.25$ and limited to the range $0.6<s<1.4$.
This distribution is wide enough to account for both subluminous and overluminous SNe~Ia.
The distribution of the amount of dust in the vicinities of SNe~Ia is as yet poorly constrained.
To gauge the systematic uncertainty of the SN~Ia rate caused by this, as in Section \ref{subsec:clash_colormag} above, we follow D08 and \citet{2012ApJ...745...31B} and use four different host-galaxy extinction models in our simulation: \citet*{1998ApJ...502..177H}, \citet{2005MNRAS.362..671R}, \citet{neill2006}, and \citet{2009ApJS..185...32K}, which have average extinctions of $\langle A_V \rangle = 0.3$, 0.3, 0.4, and 0.2 mag, respectively.
To remain consistent with G11, we choose the \citet{neill2006} model as our fiducial host-galaxy extinction model.
These models are illustrated in Figure~\ref{fig:clash_Av}.
Although we show these models (and specifically \citealt{1998ApJ...502..177H} and \citealt{2005MNRAS.362..671R}) going out to $A_V=7$ mag, we use values only out to $A_V=3$ mag, after which all models produce a negligible number of objects.

\begin{figure}[t]
 \center
 \includegraphics[width=0.5\textwidth]{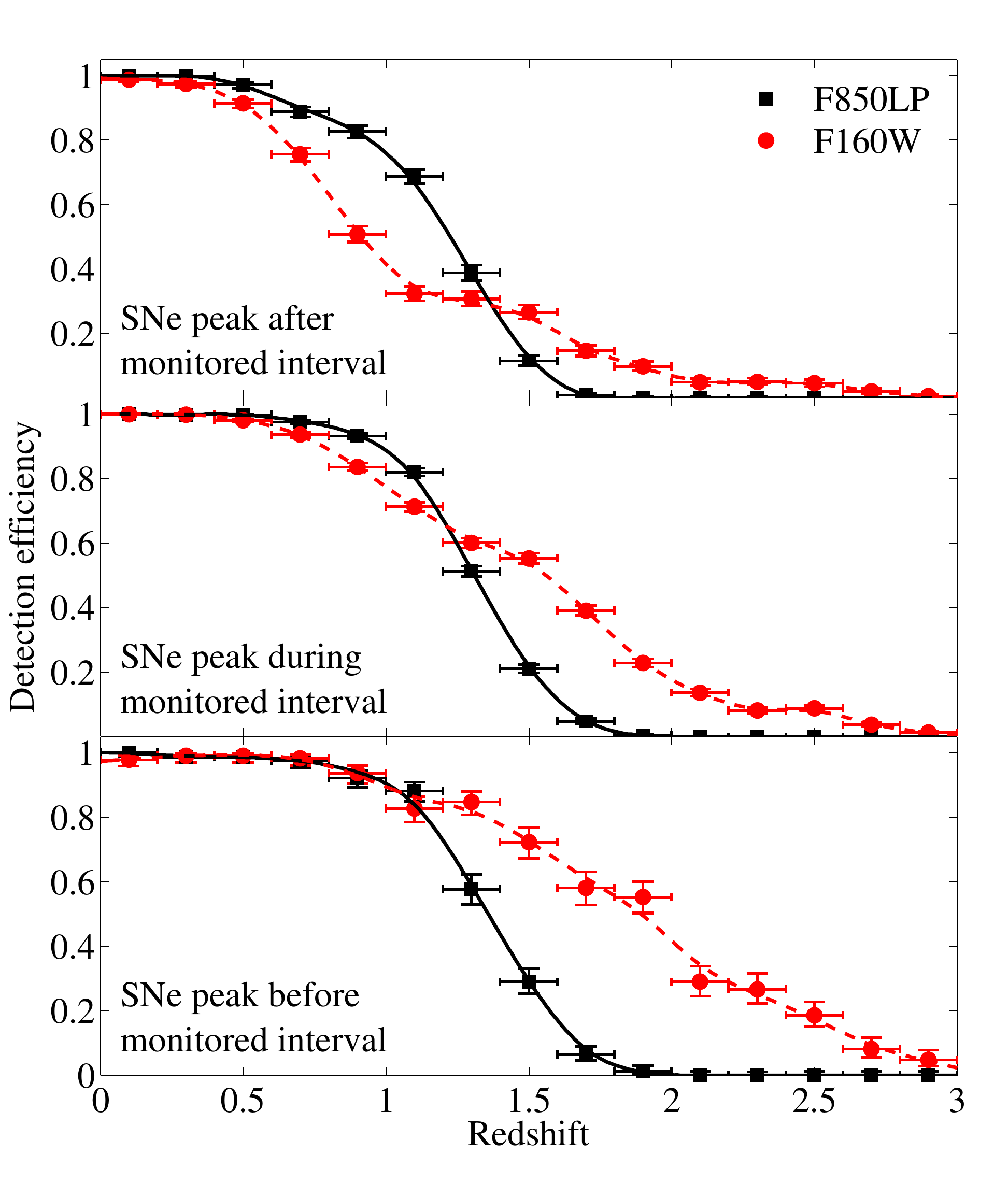}
 \caption[SN Ia detection efficiency vs. redshift in CLASH]{SN Ia detection efficiency as a function of redshift in the \FZ\ (solid black) and \FH\ (dashed red) bands for SNe~Ia that reached maximum light in the $B$ band before (bottom), during (center), and after (top) the monitored interval.}
 \label{fig:clash_effz}
\end{figure}

After choosing when the SN reached peak, the light curve is sampled according to the survey cadence in that particular cluster and field, and subtraction magnitudes are computed by subtracting the flux in each search epoch from the flux of the reference image.
Using the detection efficiency at the resultant magnitude in each search epoch, the SN is either discovered or missed. 
The resultant detection efficiency curves for SNe~Ia that reached maximum light before, during, or after the monitored interval are shown in Figure~\ref{fig:clash_effz}.

We divide the SNe in our sample into three redshift bins: $0<z<0.6$, $0.6<z<1.2$, and $1.2<z<1.8$.
In each redshift bin, we compute the effective redshift, $z_{\rm eff}$, as
\begin{equation}\label{eq:clash_zeff}
 z_{\rm eff}(z_1<z<z_2) = \frac{\int_{z_1}^{z_2}\frac{z}{(1+z)\eta(z)}dV}{\int_{z_1}^{z_2}\frac{1}{(1+z)\eta(z)}dV}.
\end{equation}

In each redshift bin, we take the minimal and maximal differences between the rate as computed with the fiducial dust model and with the models in Figure~\ref{fig:clash_Av} as lower and upper systematic uncertainties owing to dust extinction.
Since we express the number of SNe~Ia in our final sample as the sum of all the SN $P({\rm Ia})_{np}$ values, we also propagate the uncertainties in these values and add to the rates a systematic uncertainty due to our classification technique.
The systematic uncertainties from dust extinction and classification are then summed.
G11 also considered the systematic uncertainty due to the expected increase in extinction as a result of dust at high redshifts (e.g., \citealt*{2007MNRAS.377.1229M}; \citealt{2012ApJ...756..111M}).
However, G11 did not take into account the different extinction models used here. 
Specifically, the \citet{1998ApJ...502..177H} extinction model adds a $\sim+9$\% systematic uncertainty to the SN~Ia rate in the $1.2<z<1.8$ bin, similar to the $\sim+10$\% systematic uncertainty G11 added to their SN~Ia rate at $1.5<z<2.0$.

While we found no SNe at $z>1.8$, Figure~\ref{fig:clash_effz} shows that WFC3 is still sensitive to SNe~Ia out to $z \approx 2.5$.
Consequently, we add a fourth redshift bin, $1.8<z<2.4$, and compute a $2\sigma$ upper limit to the SN~Ia rate in that bin by taking the 95\% Poisson uncertainty in the number of SNe found in the bin (zero), and considering the detection efficiency of the different SN categories in the center of the bin at $z=2.1$.

\begin{table*}
 \center
 \caption{SN~Ia Rate Measurements}\label{table:clash_rates_lit_Ia}
 \begin{tabular}{ccll|ccll}
  \hline
  \hline
  Redshift & $N_{\textrm{Ia}}$ & Rate [$10^{-4}$ yr$^{-1}$ Mpc$^{-3}$]  & Reference & Redshift & $N_{\textrm{Ia}}$ & Rate [$10^{-4}$ yr$^{-1}$ Mpc$^{-3}$]  & Reference \\
  \hline
  
  0.01 & 70 & $0.183\pm 0.046$ & \citet{cappellaro1999}$^a$ & 0.55 & 72 & $0.55^{+0.07,+0.05}_{-0.07,-0.06}$ & \citet{perrett2012}$^e$ \\

  $<0.019$ & 274 & $0.265^{+0.034,+0.043}_{-0.033,-0.043}$ & {\citet{li2011rates}}$^b$ & 0.552 & 41 & $0.63^{+0.10,+0.26}_{-0.10,-0.27}$ & \citet{neill2007}$^d$ \\

  0.0375 & 516 & $0.278^{+0.112,+0.015}_{-0.083,-0.000}$ & \citet{dilday2010a}$^c$ & 0.62 & 7 & $1.29^{+0.88,+0.27}_{-0.57,-0.28}$ & \citet{melinder2012} \\

  0.09 & 17 & $0.29^{+0.09}_{-0.07}$ & \citet{dilday2008} & 0.65 & 23 & $1.49\pm 0.31$ & \citet{2006ApJ...637..427B}$^d$ \\

  0.098 & 19 & $0.24^{+0.12}_{-0.12}$ & \citet{Madgwick2003}$^{a,d}$ & 0.65 & 10.09 & $0.49^{+0.17,+0.14}_{-0.17,-0.08}$ & \citet{2010ApJ...723...47R} \\

  0.1 & 516 & $0.259^{+0.052,+0.018}_{-0.044,-0.001}$ & \citet{dilday2010a}$^c$ & 0.65 & 91 & $0.55^{+0.06,+0.05}_{-0.06,-0.07}$ & \citet{perrett2012}$^e$ \\

  0.1 & 52 & $0.569^{+0.098,+0.058}_{-0.085,-0.047}$ & \citet{Krughoff2011}$^d$ & 0.714 & 42 & $1.13^{+0.19,+0.54}_{-0.19,-0.70}$ & \citet{neill2007}$^d$ \\

  0.11 & 90 & $0.247^{+0.029,+0.016}_{-0.026,-0.031}$ & \citet{GraurMaoz2013} & 0.74 & 5.5 & $0.43^{+0.36}_{-0.32}$ & \citet{poznanski2007sdf}$^d$ \\

  0.13 & 14 & $0.158^{+0.056,+0.035}_{-0.043,-0.035}$ & \citet{blanc2004}$^a$ & 0.74 & 20.3 & $0.79^{+0.33}_{-0.41}$ & \citet{Graur2011} \\

  0.14 & 4 & $0.28^{+0.22,+0.07}_{-0.13,-0.04}$ & \citet{hardin2000}$^a$ & 0.75 & 28 & $1.78\pm 0.34$ & \citet{2006ApJ...637..427B}$^d$ \\

  0.15 & 516 & $0.307^{+0.038,+0.035}_{-0.034,-0.005}$ & \citet{dilday2010a}$^c$ & 0.75 & 14.29 & $0.68^{+0.21,+0.23}_{-0.21,-0.14}$ & \citet{2010ApJ...723...47R} \\

  0.15 & 1.95 & $0.32^{+0.23,+0.07}_{-0.23,-0.06}$ & \citet{2010ApJ...723...47R} & 0.75 & 110 & $0.67^{+0.07,+0.06}_{-0.07,-0.08}$ & \citet{perrett2012}$^e$ \\

  0.16 & 4 & $0.16^{+0.10,+0.07}_{-0.10,-0.14}$ & \citet{perrett2012}$^e$ & 0.80 & 14 & $1.57^{+0.44,+0.75}_{-0.25,-0.53}$ & \citet{dahlen2004}$^d$ \\

  0.2 & 17 & $0.189^{+0.042,+0.018}_{-0.034,-0.015}\pm 0.42$ & \citet{horesh2008} & 0.80 & 18.33 & $0.93^{+0.25}_{-0.25}$ & \citet{kuznetsova2008goods}$^d$ \\

  0.2 & 516 & $0.348^{+0.032,+0.082}_{-0.030,-0.007}$ & \citet{dilday2010a}$^c$ & 0.807 & 5.25 & $1.18^{+0.60,+0.44}_{-0.45,-0.28}$ & \citet{2012ApJ...745...31B} \\

  0.25 & 1 & $0.17\pm 0.17$ & \citet{2006ApJ...637..427B}$^d$ & 0.83 & 25 & $1.30^{+0.33,+0.73}_{-0.27,-0.51}$ & \citet{dahlen2008} \\

  0.25 & 516 & $0.365^{+0.031,+0.182}_{-0.028,-0.012}$ & \citet{dilday2010a}$^c$ & 0.85 & 15.43 & $0.78^{+0.22,+0.31}_{-0.22,-0.16}$ & \citet{2010ApJ...723...47R} \\

  0.26 & 16 & $0.32^{+0.08,+0.07}_{-0.08,-0.08}$ & \citet{perrett2012}$^e$ & 0.85 & 128 & $0.66^{+0.06,+0.07}_{-0.06,-0.08}$ & \citet{perrett2012}$^e$ \\

  0.3 & 31.05 & $0.34^{+0.16,+0.21}_{-0.15,-0.22}$ & \citet{botticella2008}$^f$ & \textbf{0.94} & \textbf{\Nm} & $\mathbf{\Rm}$ & \textbf{CLASH (this work)} \\

  0.3 & 516 & $0.434^{+0.037,+0.396}_{-0.034,-0.016}$ & \citet{dilday2010a}$^c$ & 0.95 & 13.21 & $0.76^{+0.25,+0.32}_{-0.25,-0.26}$ & \citet{2010ApJ...723...47R} \\

  0.35 & 5 & $0.530\pm 0.024$ & \citet{2006ApJ...637..427B}$^d$ & 0.95 & 141 & $0.89^{+0.09,+0.12}_{-0.09,-0.14}$ & \citet{perrett2012}$^e$ \\

  0.35 & 4.01 & $0.34^{+0.19,+0.07}_{-0.19,-0.03}$ & \citet{2010ApJ...723...47R} & 1.05 & 11.01 & $0.79^{0.28,+0.36}_{-0.28,-0.41}$ & \citet{2010ApJ...723...47R} \\ 

  0.35 & 31 & $0.41^{+0.07,+0.06}_{-0.07,-0.07}$ & \citet{perrett2012}$^e$ & 1.05 & 50 & $0.85^{+0.14,+0.12}_{-0.14,-0.15}$ & \citet{perrett2012}$^e$ \\

  0.368 & 17 & $0.31^{+0.05,+0.08}_{-0.05,-0.03}$ & \citet{neill2007}$^d$ & 1.187 & 5.63 & $1.33^{+0.65,+0.69}_{-0.49,-0.26}$ & \citet{2012ApJ...745...31B} \\

  0.40 & 3 & $0.69^{+0.34,+1.54}_{-0.27,-0.25}$ & \citet{dahlen2004}$^d$ & 1.20 & 6 & $1.15^{+0.47,+0.32}_{-0.26,-0.44}$ & \citet{dahlen2004}$^d$ \\

  0.40 & 5.44 & $0.53^{+0.39}_{-0.17}$ & \citet{kuznetsova2008goods}$^d$ & 1.20 & 8.87 & $0.75^{+0.35}_{-0.30}$ & \citet{kuznetsova2008goods}$^d$ \\

  \textbf{0.42} & \textbf{\Nl} & $\mathbf{\Rl}$ & \textbf{CLASH (this work)} & 1.21 & 20 & $1.32^{+0.36,+0.38}_{-0.29,-0.32}$ & \citet{dahlen2008} \\

  0.442 & 0 & $0.00^{+0.50,+0.00}_{-0.00,-0.00}$ & \citet{2012ApJ...745...31B} & 1.23 & 10.0 & $1.05^{+0.45}_{-0.56}$ & \citet{poznanski2007sdf}$^d$ \\

  0.45 & 9 & $0.73\pm 0.24$ & \citet{2006ApJ...637..427B}$^d$ & 1.23 & 27.0 & $0.84^{+0.25}_{-0.28}$ & \citet{Graur2011} \\

  0.45 & 5.11 & $0.31^{+0.15,+0.12}_{-0.15,-0.04}$ & \citet{2010ApJ...723...47R} & 1.535 & 1.12 & $0.77^{+1.07,+0.44}_{-0.54,-0.77}$ & \citet{2012ApJ...745...31B} \\

  0.45 & 42 & $0.41^{+0.07,+0.05}_{-0.07,-0.06}$ & \citet{perrett2012}$^e$ & 1.55 & 0.35 & $0.12^{+0.58}_{-0.12}$ & \citet{kuznetsova2008goods}$^d$ \\

  0.46 & 8 & $0.48\pm 0.17$ & \citet{2003ApJ...594....1T} & \textbf{1.59} & \textbf{\Nh} & $\mathbf{\Rh}$ & \textbf{CLASH (this work)} \\

  0.467 & 73 & $0.42^{+0.06,+0.13}_{-0.06,-0.09}$ & \citet{neill2006}$^d$ & 1.60 & 2 & $0.44^{+0.32,+0.14}_{-0.25,-0.11}$ & \citet{dahlen2004}$^d$ \\

  0.47 & 8 & $0.80^{+0.37,+1.66}_{-0.27,-0.26}$ & \citet{dahlen2008} & 1.61 & 3 & $0.42^{+0.39,+0.19}_{-0.23,-0.14}$ & \citet{dahlen2008} \\

  0.55 & 38 & $0.568^{+0.098,+0.098}_{-0.088,-0.088}$ & \citet{2002ApJ...577..120P}$^a$ & 1.67 & 3.0 & $0.81^{+0.79}_{-0.60}$ & \citet{poznanski2007sdf}$^d$ \\

  0.55 & 29 & $2.04\pm 0.38$ & \citet{2006ApJ...637..427B}$^d$ & 1.69 & 10.0 & $1.02^{+0.54}_{-0.37}$ & \citet{Graur2011} \\
  
  0.55 & 6.49 & $0.32^{+0.14,+0.07}_{-0.14,-0.07}$ & \citet{2010ApJ...723...47R} & \textbf{2.1} & \textbf{0} & $\mathbf{<\Rul}$ & \textbf{CLASH (this work)}$^g$ \\

  \hline

  \multicolumn{8}{l}{{\bf Notes.} Redshifts are means over the redshift intervals probed by each survey. $N_{\rm{Ia}}$ is the number of SNe~Ia used to derive the rate. Where necessary, rates} \\
  \multicolumn{8}{l}{have been converted to $h=0.7$. Where reported, the statistical errors are followed by systematic errors, and separated by commas. Rates from this work} \\
  \multicolumn{8}{l}{are shown in bold.} \\
  \multicolumn{8}{l}{$^a$Rates have been converted to volumetric rates using Equation~\ref{eq:clash_Ldense}.} \\
  \multicolumn{8}{l}{$^b$\citet{li2011rates} consider SNe~Ia within 80 Mpc.}  \\
  \multicolumn{8}{l}{$^c$\citet{dilday2010a} compute their rates using 516 SNe~Ia at $z<0.5$.} \\
  \multicolumn{8}{l}{$^d$These measurements have been superseded by more recent results, as detailed in Section \ref{sec:clash_dtd}.} \\
  \multicolumn{8}{l}{$^e$\citet{perrett2012} do not include SN1991bg-like SNe~Ia in their rates. Here, their measurements are scaled up by 15\% (see their Section 6).} \\
  \multicolumn{8}{l}{$^f$\citet{botticella2008} found a total of 86 SN candidates of all types. See their Section 5.2 for details on their various subsamples and classification techniques.} \\  
  \multicolumn{8}{l}{$^g2\sigma$ upper limit on the SN~Ia rate, as derived in Section \ref{sec:clash_rates}.}

 \end{tabular}
\end{table*}

The resultant SN~Ia rates, including both statistical and systematic uncertainties, are shown in Figure~\ref{fig:clash_rates_dtd}.
Table~\ref{table:clash_rates} summarizes the SN~Ia rates, with and without correction for host-galaxy dust extinction, and Table~\ref{table:clash_errs} shows the complete error budget of our SN~Ia rates.
Table~\ref{table:clash_rates_lit_Ia} compares the rates from this work to previous rates from the literature.
Where necessary, the measurements have been corrected to reflect the value of $h=0.7$ used in this work.
As \citet{perrett2012} did not take into account low-stretch, SN1991bg-like SNe~Ia, we scale up their SN~Ia rates by 15\% (see their Section 6).
As in G11, in instances where rates were originally reported in units of SNuB (SNe per century per $10^{10}~L_{\odot,B}$; \citealt{cappellaro1999, hardin2000, 2002ApJ...577..120P, Madgwick2003, blanc2004}), we have converted them to volumetric rates using the \citet{botticella2008} redshift-dependent luminosity density function,
\begin{equation}\label{eq:clash_Ldense}
j_B(z)=(1.03+1.76\, z)\times 10^8~L_{\odot,B}~\textrm{Mpc}^{-3}.
\end{equation}


\section{The Type-Ia supernova delay-time distribution}
\label{sec:clash_dtd}

In this section, we test different models of the DTD by convolving them with various cosmic SFHs and fitting the resultant SN~Ia rate histories to the SN~Ia rate measurements from the previous section, along with rates from the literature.
We include all the rate measurements from Table~\ref{table:clash_rates_lit_Ia} except for \citet{neill2006,neill2007}, which have been superseded by \citet{perrett2012}; \citet{dahlen2004} and \citet{kuznetsova2008goods}, which have been superseded by D08; \citet{2006ApJ...637..427B}, which has been superseded by \citet{2010ApJ...723...47R}; \citet{poznanski2007sdf}, which has been superseded by G11; and \citet{Madgwick2003} and \citet{Krughoff2011}, which have been superseded by \citet{GraurMaoz2013}.
We do not use the $z>2$ upper limit from the previous section as it is too high to affect the DTD fits.
In total, we use 50 SN~Ia rate measurements, of which 41 are at $z<1$ and 9 are at $z>1$.

\begin{figure}
 \center
 \includegraphics[width=0.5\textwidth]{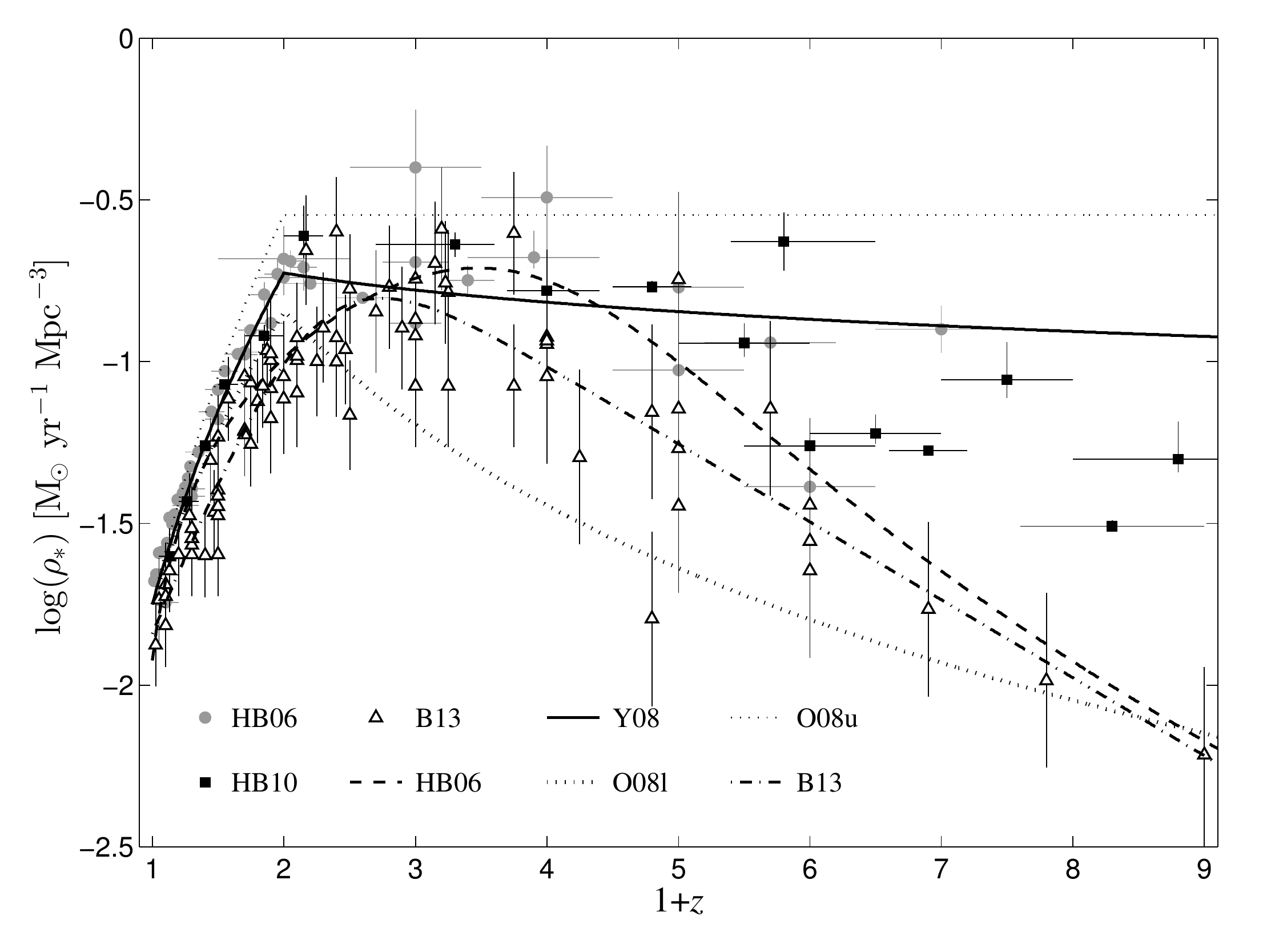}
 \caption[Star-formation history measurements and parameterizations]{SFH measurements and parameterizations. SFH measurement compilations up to 2006 from HB06 are shown in filled circles, and additional measurements up to 2010 compiled by \citet{horiuchi2010}, here marked as HB10, are shown as filled squares. The most up-to-date compilation from B13 is shown as open triangles. The different parameterizations are shown as curves and include the \citet{2001MNRAS.326..255C} fit to the HB06 data (dashed); the Y08 (solid) and O08u/O08l (thin/thick dotted) power-law fits; and the B13 parameterization (dot-dashed). All data and parameterizations have been rescaled to the \citet{2003ApJS..149..289B} ``diet'' Salpeter IMF.}
 \label{fig:clash_SFHs}
\end{figure}

Following G11, we test different SFHs, including the \citet{2001MNRAS.326..255C} parameterization fit to the data collected by \citet[HB06]{2006ApJ...651..142H}; the SFH presented by \citet[Y08]{2008ApJ...683L...5Y} and upper (O08u) and lower (O08l) limits from \citet[O08]{2008PASJ...60..169O} which can be approximated as broken power laws with a break at $z=1$ and with varying indices before and after the break; and the recent \citet[B13]{Behroozi2012} SFH.
These SFHs, and the data they are based on, are presented in Figure~\ref{fig:clash_SFHs}.

\begin{figure}
 \center
 \includegraphics[width=0.5\textwidth]{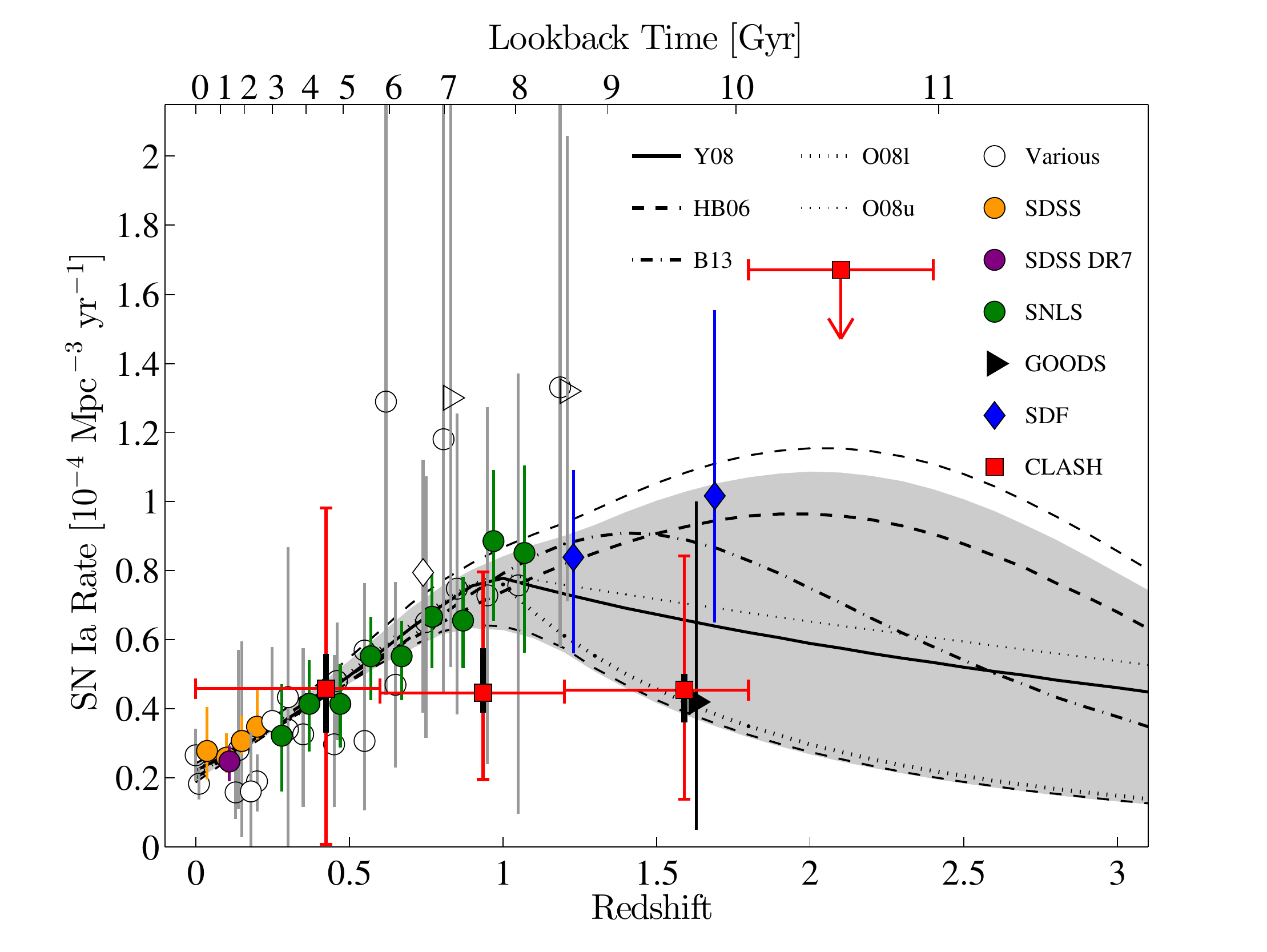}
 \caption[SN~Ia rates from CLASH]{SN~Ia rates from CLASH (filled, red squares) compared to rates from the literature and best-fitting SN~Ia rate evolutions derived by convolving a power-law DTD with different SFHs. Circles denote data from surveys with measurements out to $z \approx 1$ from \citet{cappellaro1999}, \citet{hardin2000}, \citet{2002ApJ...577..120P}, \citet{2003ApJ...594....1T}, \citet{blanc2004}, \citet{botticella2008}, \citet{horesh2008}, \citet{2010ApJ...723...47R}, \citet{li2011rates}, \citet{2012ApJ...745...31B}, and \citet{melinder2012}. Filled circles denote the most accurate and precise measurements at $z<1$ and are from the SDSS Stripe 82 survey (\citealt[orange]{dilday2010a}), SNLS (\citealt[green]{perrett2012}), and SDSS DR7 (\citealt[purple]{GraurMaoz2013}). The GOODS rates from \citet{dahlen2008} are shown as triangles and the SDF rates from \citet{Graur2011} are shown as diamonds. The $z>1.5$ rates from these two surveys are colored in black and blue, respectively. 
 The thick curves are convolutions of several SFHs (dashed, \citealt{2006ApJ...651..142H}; solid, \citealt{2008ApJ...683L...5Y}; thin/thick dotted, \citealt{2008PASJ...60..169O}; dot-dashed, \citealt{Behroozi2012}) with the best-fitting power-law DTDs.
 The shaded area is the confidence region resulting from the combined 68\% statistical uncertainties in the values of the power-law index fit with the above SFHs.
 The thin dashed lines indicate the 68\% statistical uncertainty region obtained without the new CLASH measurements.
 All vertical error bars are sums of the statistical and systematic uncertainties. The CLASH vertical error bars are composed of the systematic uncertainty, shown as black thick lines, and the statistical uncertainty, shown as red thin lines. The horizontal error bars delineate the CLASH redshift bins.
 The \citet{perrett2012} and $z>1.5$ \citet{dahlen2008} SN~Ia rates have been shifted by $\Delta z=+0.02$ to disentangle them from other results.}
 \label{fig:clash_rates_dtd}
\end{figure}

When deriving SFH measurements, various authors use different versions of the initial-mass function (IMF), leading to different scalings of the SFH.
In order to maintain consistency across the different SFHs, we must choose one IMF and re-scale the SFHs accordingly.
As in G11 and \citet{GraurMaoz2013}, we assume a ``diet'' Salpeter IMF \citep{2003ApJS..149..289B}, which is similar to the \citet{1955ApJ...121..161S} IMF with lower and upper mass limits of 0.1 and 125~$M_\odot$, respectively, but with a stellar mass-to-light ($M/L$) ratio that is scaled down by a factor of 0.7 in order to fit the $M/L$ ratios measured in disks (\citealt{2001ApJ...550..212B}).
The choice of this IMF requires us to scale down the SFHs of HB06, O08, and Y08, who assumed a \citet{1955ApJ...121..161S} IMF, by a factor of 0.7.
The B13 SFH, where a \citet{2003PASP..115..763C} IMF was assumed, is scaled up by a factor of 0.7.
To allow comparisons with G11, we use the Y08 SFH as the fiducial model in our DTD recoveries and the other SFHs to estimate a systematic uncertainty in the values of the parameters of the DTD model tested below.

We test a power-law DTD of the form $\Psi(t)=\Psi_{\rm 1Gyr}(t/{\rm 1~Gyr})^\beta$, setting its index, $\beta$, and scaling, $\Psi_{\rm 1Gyr}$, as free parameters, leaving 48 degrees of freedom for the fit.
The DTD is set to zero before $40$~Myr, to allow for 8~$M_{\odot}$ stars to evolve into CO WDs.
The Y08 SFH yields a best-fit index value of $\beta=-1.00^{+0.06(0.09)}_{-0.06(0.10)}$ with a reduced $\chi^2$ ($\chi^2_\nu$) of 0.7, where the statistical uncertainties are the 68\% and 95\% (in parentheses) confidence regions, respectively.
The other SFHs yield a systematic uncertainty of $^{+0.12}_{-0.08}$, with $\chi^2_\nu$ values in the range 0.7--0.8, yielding a final value of $\beta=$\plind.
This value is consistent with those obtained by G11 and \citet{GraurMaoz2013} and in a variety of different SN surveys and using different DTD recovery techniques (see \citealt{Maoz2012review}).
Integrating the DTD over a Hubble time, we find that the number of SNe~Ia per formed mass, $N/M_*$, lies in the range (0.5--1.3) $\times 10^{-3}~{\rm SNe}~M_\odot^{-1}$, similar to the ranges found by G11 and \citet{GraurMaoz2013}.
The best-fitting SN~Ia rate histories derived from each SFH, along with the 68\% uncertainty region, are shown in Figure~\ref{fig:clash_rates_dtd}.

\begin{figure}
 \center
 \begin{tabular}{c}
  \includegraphics[width=0.5\textwidth]{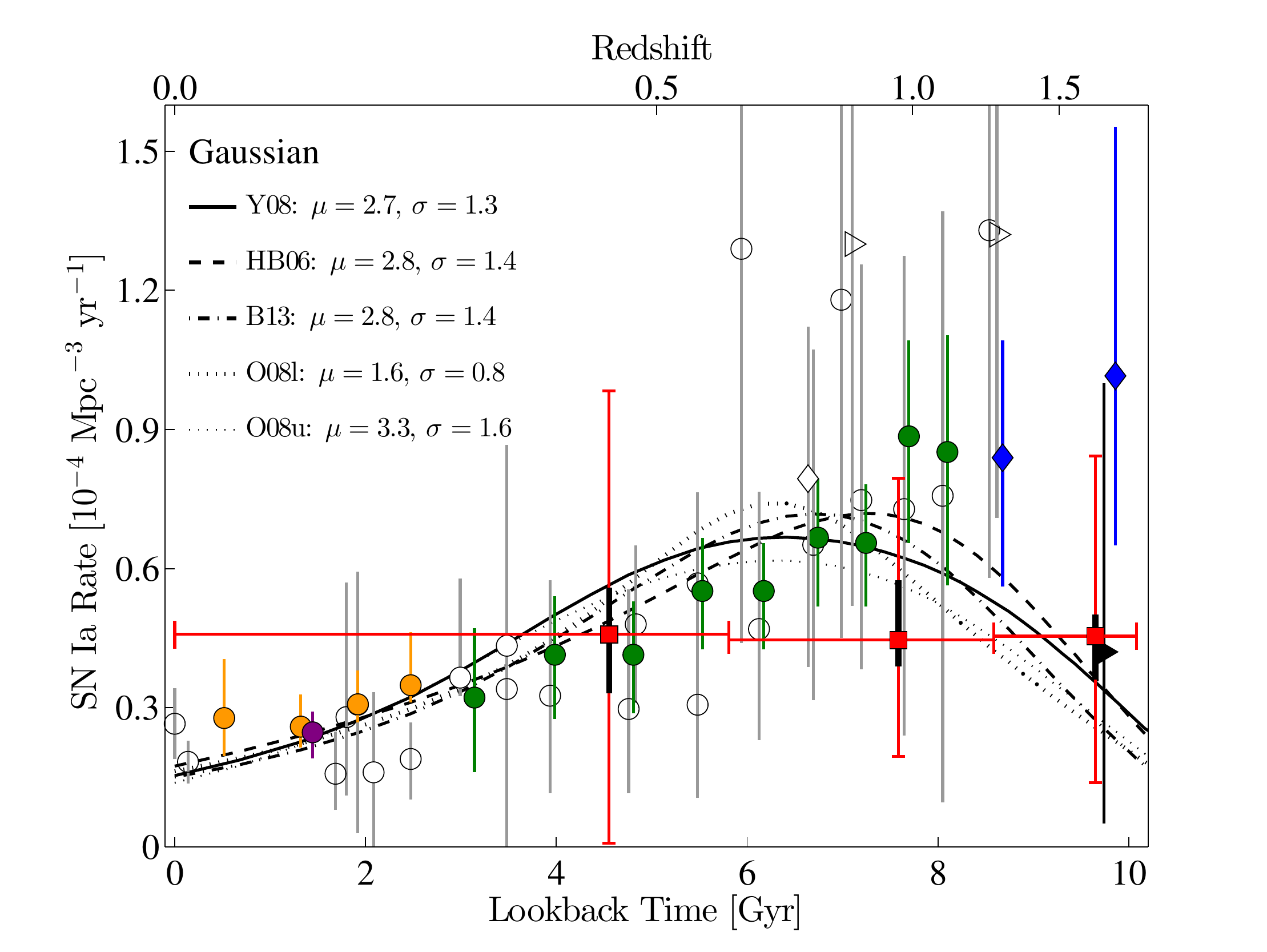} \\
  \includegraphics[width=0.5\textwidth]{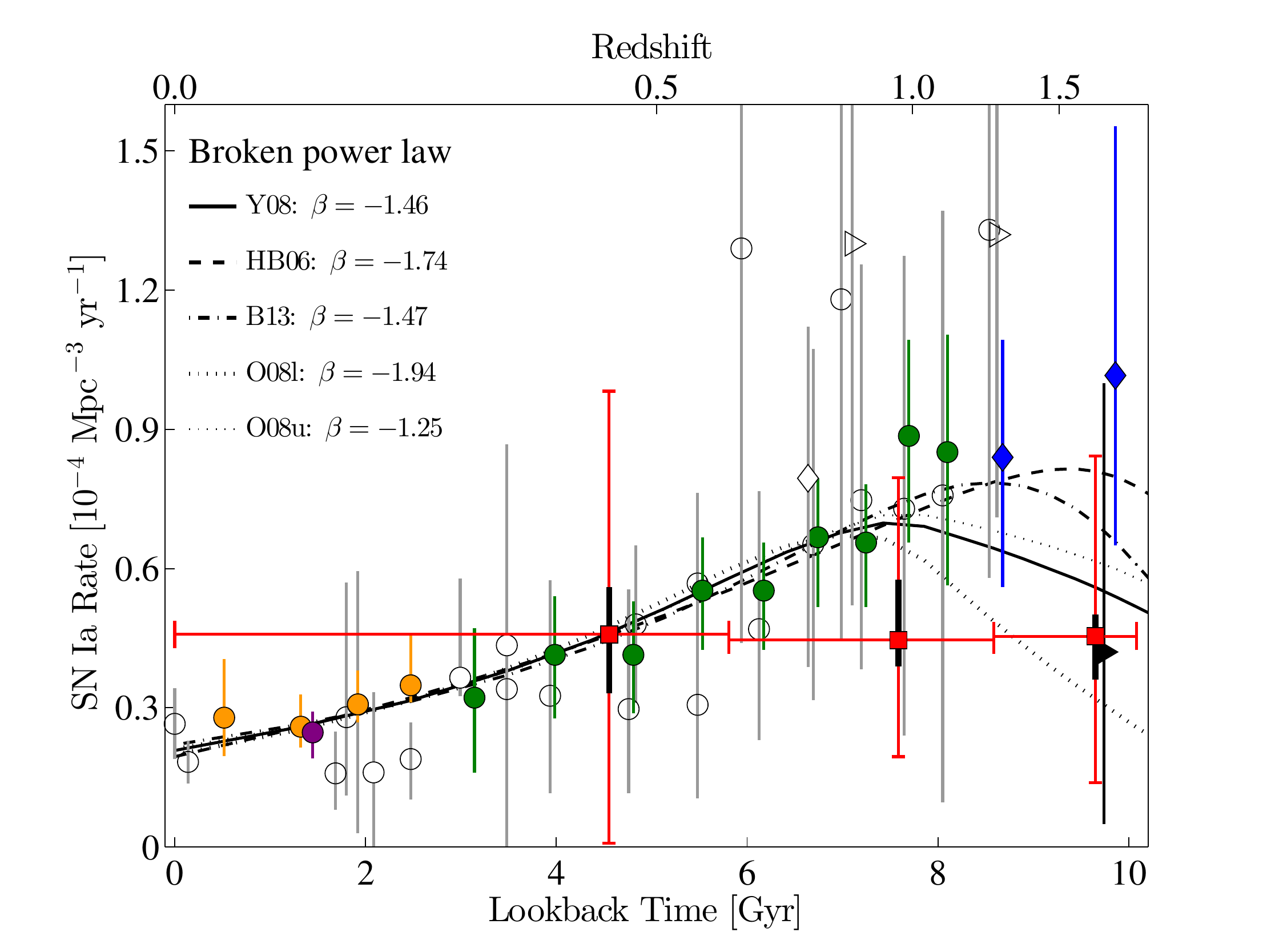} \\
 \end{tabular}
 \caption{Observed SN~Ia rates compared to predicted SN~Ia rate evolution tracks from the convolution of different SFHs with a best-fitting (top) Gaussian DTD and (bottom) broken power-law DTD of the form $\Psi(t) = \Psi(t/{\rm 1~Gyr})^{-1/2}$ up to $t_c=1.7$~Gyr, and $\Psi(t) \propto t^{\beta}$ afterward. Symbols are as in Figure~\ref{fig:clash_rates_dtd}. The SFHs used for each fit are listed in the top panel along with best-fitting parameter values: the mean, $\mu$, and standard deviation, $\sigma$, of the Gaussian DTD, in Gyr; and the slope of the second power law, $\beta$, of the broken power-law DTD. The CLASH upper limit at $z>1.8$, which was not used in the fits, has been removed.}
 \label{fig:clash_alternate_DTDs}
\end{figure}

We investigate also the viability of a Gaussian DTD fit \citep{Strolger2004,Strolger2010} to the SN~Ia rates.
We start by testing the Gaussian DTD proposed by D08, with a mean delay time of $3.4$~Gyr and a standard deviation of $0.68$~Gyr.
As in G11, we allow the scaling of this DTD to vary as a free parameter.
Coupled with the SN~Ia rates, the only SFH that does not disfavor this DTD is the HB06 SFH, with a reduced $\chi^2$ value of $1.1$.
All other SFHs result in SN~Ia rate evolutions that are excluded at a $>95$\% significance level, with the O08 and B13 SFHs specifically excluded at a $>99$\% significance level.
We next test a general Gaussian DTD, where we allow the mean delay time, standard deviation, and scaling to vary as free parameters, while requiring that 95\% of the area under the DTD remain above a delay time of 40~Myr (thus ensuring the resultant DTD retains a Gaussian shape). 
The B13 and lower-limit O08 SFHs result in Gaussians that are excluded at a $>95$\% significance level.
The Y08, upper-limit O08, and HB06 SFHs, on the other hand, result in Gaussians with means in the range $\mu = 2.7$--$3.3$~Gyr with standard deviations of $\sigma = 0.8$--$1.6$~Gyr and reduced $\chi^2$ values of $0.9$--1.3.
These Gaussian DTDs are centered at slightly lower mean delay times, but are wider, than the D08 Gaussian DTD.
The resultant fits to the SN~Ia rate evolution are shown in Figure~\ref{fig:clash_alternate_DTDs}. 
Although at first sight, it might appear that the $z>1.5$ SDF rate measurement is driving the exclusion of the Gaussian DTDs, most of the fitting power actually comes from the accurate $z<1$ SDSS and SNLS measurements.

As in G11, we test the further possibility that at early times the DTD is dominated by the production efficiency of double WD systems, which is described by a power law of the form $t^{-0.5}$ (\citealt*{2008ApJ...683L..25P}), until some cutoff time, $t_c$, when a second physical process takes over, described as a power law having a different slope, $t^{\beta}$.
Whereas in G11, we set the slope of the second power law to $\beta=-1$ and fit for $t_c$, here we set $t_c=1.7$~Gyr, the lifetime of a 2~$M_\odot$ star, the least massive star expected to produce a $\sim 0.7$~$M_\odot$ CO WD (see Figure~4 of \citealt{2000A&AS..141..371G}), and find that the best-fitting slope of the second power law is $\beta = -1.46^{+0.16(0.26)}_{-0.13(0.22)}$~(statistical)~$^{+0.48}_{-0.21}$~(systematic) with reduced $\chi^2$ values in the range $0.7$--$0.9$.

\begin{figure}
 \center
 \begin{tabular}{c}
  \includegraphics[width=0.5\textwidth]{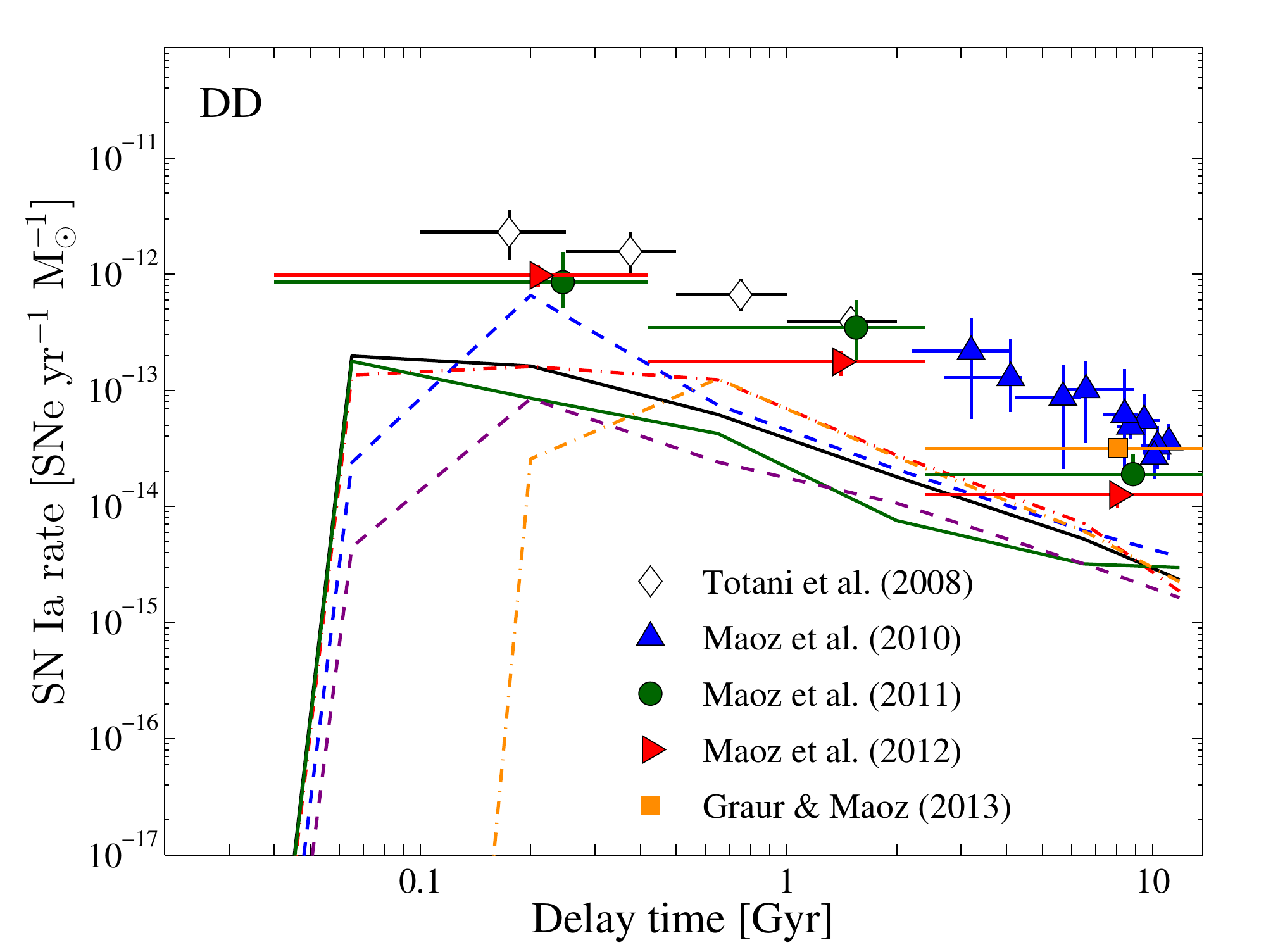} \\
  \includegraphics[width=0.5\textwidth]{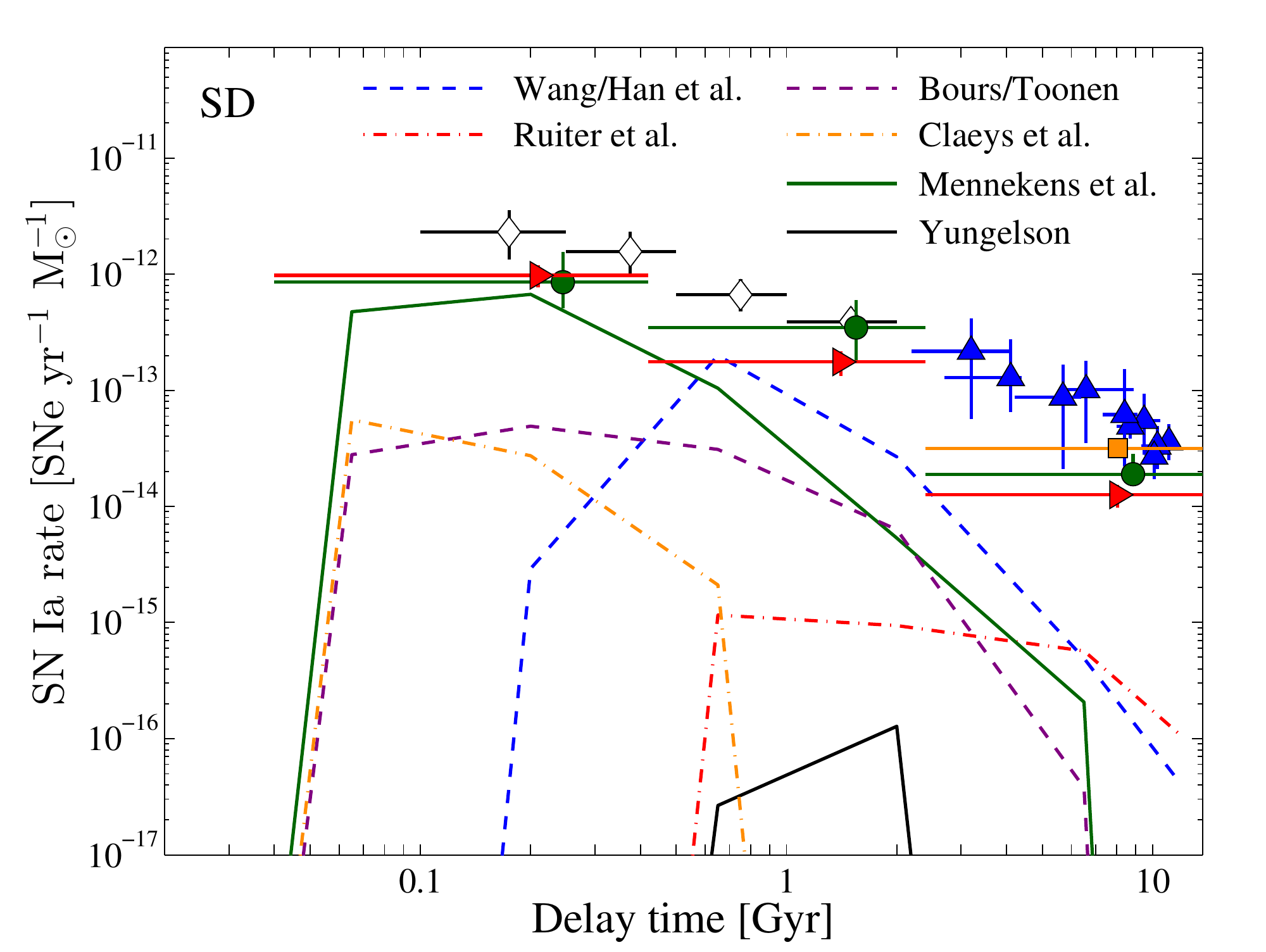} \\
 \end{tabular}
 \caption{Binary-population synthesis DTD models for the DD (top) and SD (bottom) scenarios, compared to observations. The BPS DTD models, shown here as different-colored curves, are updated versions of the ones that appear in \citet[G. Nelemans, private communication]{2013IAUS..281..225N} and are listed according to the groups that computed them: Yungelson (e.g., \citealt{2010AstL...36..780Y}; black solid), the Yunnan group (Wang/Han et al.; e.g., \citealt{2010MNRAS.401.2729W}; blue dashed), the StarTrack code (Ruiter et al.; e.g., \citealt{2009ApJ...699.2026R}; red dot-dashed), the Brussels group (Mennekens et al.; e.g., \citealt{2010A&A...515A..89M}; green solid), the Utrecht group (Claeys et al.; e.g., \citealt{2013IAUS..281..236C}; orange dot-dashed), and the SeBa code (Bours/Toonen; e.g., \citealt{2012A&A...546A..70T}; purple dashed). For comparison, we also show reconstructed components of the DTD from observations of SNe Ia in $0.4<z<1.2$ elliptical galaxies (\citealt{2008PASJ...60.1327T}; white diamonds), galaxy clusters (\citealt*{Maoz2010clusters}; blue triangles), LOSS-SDSS galaxies (\citealt{Maoz2010loss}; green circles), SDSS-II galaxies (\citealt*{Maoz2012sdss}; red right-pointing triangles), and SDSS DR7 spectra (\citealt{GraurMaoz2013}; orange squares).} 
 \label{fig:clash_BPS_models}
\end{figure}

Finally, we test both DD and SD DTDs resulting from binary population synthesis (BPS) simulations.
Here, we use updated versions of the scaled models presented in Figures 2--3 of \citet*[for color versions of these DTD figures, see \citealt{2012NewAR..56..122W}; updated versions of the models courtesy of G. Nelemans, private communication]{2013IAUS..281..225N}.
As in \citet{2013IAUS..281..225N}, we designate the BPS DTD models by the groups that computed them: Yungelson (e.g., \citealt{2010AstL...36..780Y}), the Yunnan group (Wang/Han et al.; e.g., \citealt{2010MNRAS.401.2729W}), the StarTrack code (Ruiter et al.; e.g., \citealt{2009ApJ...699.2026R}), the Brussels group (Mennekens et al.; e.g., \citealt{2010A&A...515A..89M}), the Utrecht group (Claeys et al.; e.g., \citealt{2013IAUS..281..236C}), and the SeBa code (Bours/Toonen; e.g., \citealt{2012A&A...546A..70T}).
We present the updated versions of the scaled BPS DTD models in Figure~\ref{fig:clash_BPS_models}.
As \citet{2013IAUS..281..225N} scaled the different DTDs to a \citet{1993MNRAS.262..545K} IMF, we rescale them to the diet Salpeter IMF by multiplying them by a factor of $0.7$.
For comparison with the volumetric SN~Ia rates, we convolve the different BPS DTDs with the B13 SFH, as parametrized by their Equation F1, which we reproduce here as
\begin{equation}\label{eq:clash_B13}
 {\rm CSFR}(z) = \frac{C}{10^{A(z-z_0)}+10^{B(z-z_0)}},
\end{equation}
where CSFR$(z)$ is the cosmic SFH as a function of redshift, and the constants $A$, $B$, $C$, and $z_0$ are given in Table 7 of B13 as $A=-0.997$, $B=0.241$, $C=0.180$, and $z_0=1.243$.

\begin{figure}
 \center
 \begin{tabular}{c}
  \includegraphics[width=0.5\textwidth]{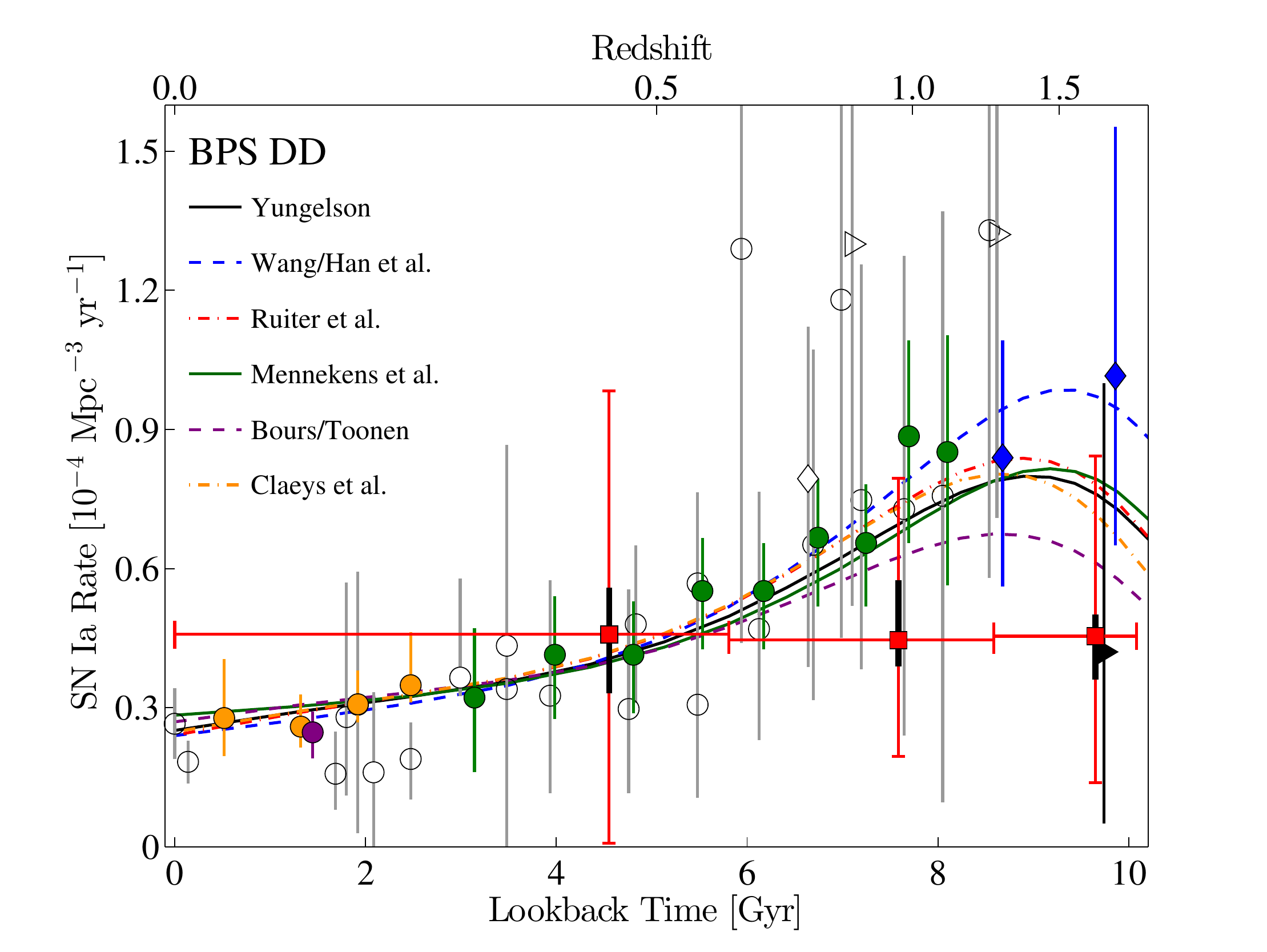} \\
  \includegraphics[width=0.5\textwidth]{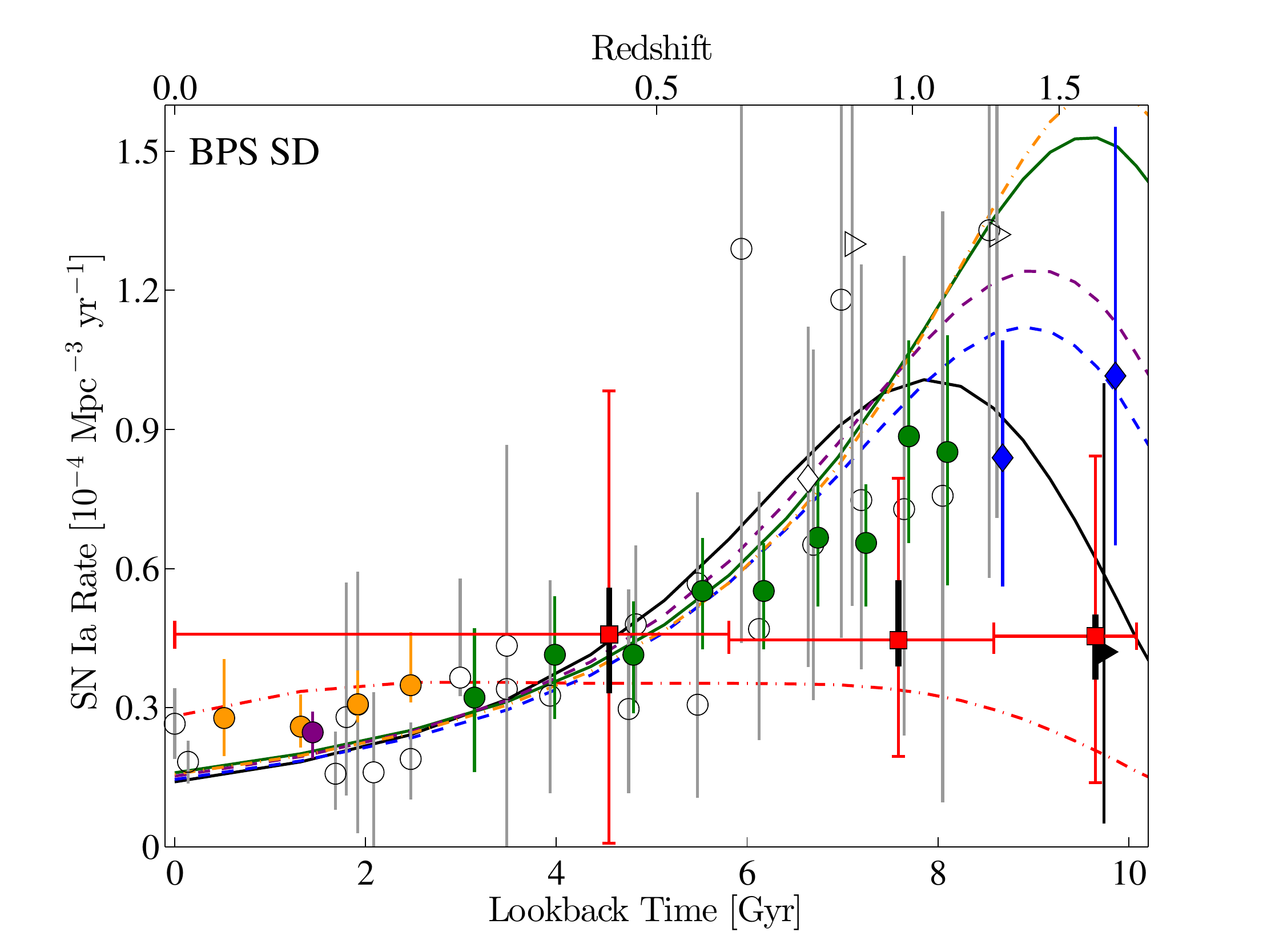} \\
 \end{tabular}
 \caption{Observed SN~Ia rates compared to predicted SN~Ia rate evolution tracks from the convolution of the \citet{Behroozi2012} SFHs with BPS DD (top) and SD (bottom) DTD models from the literature \citep{2013IAUS..281..225N}. Symbols are as in Figure~\ref{fig:clash_alternate_DTDs}, except for the SN~Ia rate evolution curves, which are labeled in the top panel according to the DTD used in each fit, as shown in Figure~\ref{fig:clash_BPS_models}.}
 \label{fig:clash_BPS_DTDs}
\end{figure}

As has been commented elsewhere (e.g., \citealt{Maoz2012review,2013IAUS..281..225N}), the BPS DTDs, both for DD and SD scenarios, fail to produce the number of observed SNe~Ia.
Here, we test the shape of the BPS DTDs by treating their scaling as a free parameter.
The resultant SN~Ia rate evolutions are presented in Figure~\ref{fig:clash_BPS_DTDs}.
The BPS DD models require scalings by small factors of 3--9 and result in reduced $\chi^2$ values of $\chi^2_\nu\ltsim1$, consistent with the SN~Ia rates.
The BPS SD models, on the other hand, require large scaling factors of $>10$ (except for the Wang/Han and Mennekens DTDs, which require scaling factors of $\sim4$) and result in reduced $\chi^2$ values of $\chi^2_\nu>1.8$, thus excluding all BPS SD models at a $>99$\% significance level.

The poor fits of the SD models are the result of their low DTD amplitudes at long delay times.
The accurate and precise $z<1$ SN~Ia rates, which are most sensitive to the long delay-time component of the DTD, have the most leverage on the scalings of the BPS SD DTDs. 
Because the SD models have low amplitudes at long delay times, the $z<1$ SN~Ia rates force scalings of large factors that then cause the resultant SN~Ia rate evolutions to overshoot the $z>1$ rates.

It is instructive to compare some of the SD DTD models, and their resultant SN~Ia rate evolutions, in detail.
The Claeys model has the lowest amplitude at long delay times, which is why its resultant SN~Ia rate evolution overshoots all the other models in the bottom panel of Figure~\ref{fig:clash_BPS_DTDs}.
On the other hand, although the Yungelson model only has an intermediate delay-time component, that component is at longer delay times than the Claeys model, so it results in a SN~Ia rate evolution with less amplitude than the rate evolution produced by the Claeys model.
Finally, the Ruiter model has the highest amplitude at long delay times, after the Wang/Han model, which results in a low scaling. 
However, the Ruiter model has lower amplitude at short delay times, compared to all other DTD models besides Yungelson, which is why the low scaling, forced by the long-delay component, results in a SN~Ia rate evolution that undershoots the $z>1$ SN~Ia rates.


\section{Conclusions}
\label{sec:clash_conclude}

In this work, we have presented a sample of 27 SNe discovered in the parallel fields of the 25 CLASH galaxy clusters.
Of these, $\sim13$ were classified as SNe~Ia, four of which are at $z>1.2$.
Using the SN~Ia sample, we measured the SN~Ia rate out to $z \approx 1.8$ and obtained an upper limit on the rate in the redshift range $1.8<z<2.4$.
Within the uncertainties of all the measurements, these rates are consistent with both the \hst/GOODS and the Subaru/SDF SN~Ia rates.
Based on these rates, along with previous rates from the literature, we have shown that when convolved with different cosmic SFHs, a power-law DTD with an index of \plind\ is consistent with the data.
The systematic uncertainty derives from the wide range of possible SFHs considered.

We have also shown that the overall shape of DTDs from BPS DD models are consistent with the SN~Ia rate measurements, as long as the models are scaled up by factors of 3--9, while all BPS SD models are ruled out at a $>99$\% significance level.

The SN~Ia rates at $z<1$ require a DTD with a significant delayed component, such as the power-law DTD tested here.
The high-redshift SN~Ia rates provide a probe of the early times of the DTD, where the DTD could either continue with an index of $\sim -1$, as found here, or perhaps transit to a lower index of $-0.5$, as proposed by \citet{2008ApJ...683L..25P}.
If the SD scenario contributes significantly to the SN~Ia rate, as claimed by some recent work (e.g., \citealt{2011Sci...333..856S} and \citealt{Dilday2012}), its main effect would be on the high-redshift rates.
However, to have any real discriminatory power on the different DTD models, the SN~Ia rates at $z>1$ must be more accurate and precise than they currently are.
To make the most efficient use of the CLASH SN sample, we will combine it with the final CANDELS sample in a future paper.
Together, the two samples will contain a similar number of SNe~Ia as the Subaru/SDF sample from G11.
However, their systematic uncertainties will be lower, as they will make use of light curves and spectroscopy, where available, as done in this work.
Finally, the upcoming \hst\ Frontier Fields program\footnote{http://www.stsci.edu/hst/campaigns/frontier-fields/} (PI: M.~Mountain) will observe six pairs of galaxy clusters and blank fields containing field galaxies, using 140 orbits of ACS and WFC3 for each pair of galaxy cluster / blank field during Cycles 21--23.
Based on our work on CLASH and CANDELS, we expect that this survey, which will go deeper than either of the previous surveys, will discover $\sim 20$ SNe, including five $z>1.5$ SNe~Ia.
Once this sample is added to the combined CLASH+CANDELS SN sample, we may finally have high-redshift SN~Ia rates accurate enough to probe the early part of the DTD.


\begin{acknowledgements}
We thank Gijs Nelemans for sharing the BPS DTD models with us, and Ori Fox, Patrick Kelly, Isaac Shivvers, Brad Tucker, and WeiKang Zheng for assistance with some of the Keck observations.
Financial support for this work was provided by NASA through grants HST-GO-12060 and HST-GO-12099 from the Space Telescope Science Institute (STScI), which is operated by Associated Universities for Research
in Astronomy, Inc., under NASA contract NAS 5-26555.
O.G. and D.M. acknowledge support by the I-CORE program of the PDC and the ISF, Grant 1829/12, and by a grant from the Israel Science Foundation.
Support for S.R.  was provided by NASA through Hubble Fellowship grant \#HF-51312.01 awarded by STScI.
This work was supported by NASA Keck PI Data Awards (to Rutgers University, PI: S.W.J.), administered by the NASA Exoplanet Science Institute. Supernova research at Rutgers University is additionally supported by NSF CAREER award AST-0847157 to S.W.J.
A.M. and N.B. acknowledge support from AYA2010-22111-C03-01 and PEX/10-CFQM-6444, and from the Spanish Ministerio de Educaci\'{o}n y Ciencia through grant AYA2006-14056 BES-2007-16280.
M.N. is supported by PRIN INAF-2010.
A.V.F. is also grateful for the support of NSF grant AST-1211916, the TABASGO Foundation, and the Christopher R. Redlich Fund.
The Dark Cosmology Centre is funded by the Danish National Research Foundation.
J.M.S. is supported by an NSF Astronomy and Astrophysics Postdoctoral Fellowship under award AST-1302771.
K.J., C.L., K.L., J.N., A.O., and H.E.R. were supported by the American Museum of Natural History's Science Research Mentoring Program under NASA grant award NNX09AL36G.

This work is based, in part, on data collected at the Subaru Telescope, which is operated by the National Astronomical Observatory of Japan. Additional data presented here were obtained at the W. M. Keck Observatory, which is operated as a scientific partnership among the California Institute of Technology, the University of California, and NASA; the Observatory was made possible by the generous financial support of the W. M. Keck Foundation. We wish to recognize and acknowledge the very significant cultural role and reverence that the summit of Mauna Kea has always had within the indigenous Hawaiian community. We are most fortunate to have the opportunity to conduct observations from this mountain.

Part of the research presented here is based on observations obtained at the Gemini Observatory, which is operated by the Association of Universities for Research in Astronomy, Inc., under a cooperative agreement with the NSF on behalf of the Gemini partnership: the NSF (United States), the National Research Council (Canada), CONICYT (Chile), the Australian Research Council (Australia), Minist\'{e}rio da Ci\^{e}ncia, Tecnologia e Inova\c{c}\={a}o (Brazil) and Ministerio de Ciencia, Tecnolog\'{i}a e Innovaci\'{o}n Productiva (Argentina) [Programs GS-2011A-Q-16, GN-2011A-Q-14, and GN-2012A-Q-32].

This research is based in part on observations made with ESO telescopes at the La Silla Paranal Observatory under program IDs 086.A-0660, 088.A-0708, and 089.A-0739; also, on observations collected at the European Southern Observatory, Chile (ESO Programmes 086.A-0070, 087.A-0295, 089.A-0438 and 091.A-0067).

We used data obtained with the MODS spectrographs built with funding from NSF grant AST-9987045 and the NSF Telescope System Instrumentation Program (TSIP), with additional funds from the Ohio Board of Regents and the Ohio State University Office of Research.

This research has made use of NASA's Astrophysics Data System (ADS) Bibliographic Services and of the NASA/IPAC Extragalactic Database (NED), which is operated by the Jet Propulsion Laboratory, California Institute of Technology, under contract with NASA.

\end{acknowledgements}

{\it Facilities:} \facility{Gemini:South (GMOS-S)} \facility{Gemini:Gillett (GMOS-N)} \facility{HST (ACS, WFC3)} \facility{Keck:I (LRIS)} \facility{Keck:II (DEIMOS)} \facility{LBT (MODS)} \facility{Subaru (Suprime-Cam)} \facility{VLT:Antu (FORS2)} \facility{VLT:Kueyen (X-shooter)} \facility{VLT:Melipal (VIMOS)}
\smallskip



\end{document}